\documentclass{elsart}
\usepackage{epsfig,multirow}
\usepackage{amssymb}

\begin{document}

\begin{frontmatter}

\title{The Monte Carlo Event Generator \\
 AcerMC version 1.0\\
with interfaces to \\ PYTHIA 6.2 and  HERWIG 6.3}

\author{B. P. Ker\v sevan}
\address{Jozef Stefan Institute, Jamova 39, SI-1000 Ljubljana, Slovenia; \\
 Faculty of Mathematics and Physics, University of Ljubljana, 
 Jadranska 19, SI-1000 Ljubljana, Slovenia.}
\author{E. Richter-W\c{a}s\thanksref{KBN}}
\address{Institute of Computer Science, Jagellonian University\\
 30-072 Krakow, ul. Nawojki 11, Poland;\\
 Institute of Nuclear Physics,  30-055 Krakow, ul. Kawiory 26a, Poland.}
\thanks[KBN]{ Supported in part by
  Polish Government grant KBN 2P03B11819, 
  by the European Commission 5-th framework contract HPRN-CT-2000-00149  and
  by Polish-French Collaboration with IN2P3.}

\begin{abstract} 
The {\bf AcerMC} Monte Carlo Event Generator is dedicated for the generation
of Standard Model background processes at $pp$ LHC collisions. 
The program itself provides a library of the massive matrix elements 
and phase space modules for generation of a set of  
selected processes:  $gg, q \bar q \to t \bar t b \bar b$, 
$ q \bar q W(\to \ell \nu) b \bar b$, 
$q \bar q W(\to \ell \nu) t \bar t$, 
$gg, q \bar q \to Z/\gamma^*(\to \ell \ell) b \bar b$,
$gg, q \bar q \to Z/\gamma^*(\to \ell \ell, \nu \nu, b \bar b) t \bar t$ and
complete electroweak $gg \to (Z/W/\gamma^* \to) b \bar b t \bar t$ process. 
The hard process event, 
generated with one of these modules, can be completed by the initial and final 
state radiation, hadronisation and decays, simulated with either 
{\tt PYTHIA} or {\tt HERWIG} Monte Carlo event generator. 
Interfaces to both of 
these generators are provided in the distribution version. 
The matrix element codes have been derived with the help of the {\tt MADGRAPH} package.
The phase-space generation is based on the multi-channel self-optimising approach as
proposed in {\tt NEXTCALIBUR} event generator.
Eventually, additional
smoothing of the phase space was obtained by using a modified {\tt ac-VEGAS}
routine in order to improve the generation efficiency.  
\end{abstract}


\end{frontmatter}

\tableofcontents

\newpage

{\scriptsize
\begin{verbatim}      
  -----------------------------------------------------------------------------
                                   ._                           
                                 .j%3]:,                        
                               ~!%%%%%%% ,._.                   
                            _|xx%xxxx%%%%%+`                    
                             :~]%xxxx]xx%x_,+_x_%`              
                      -__||x||xx]+]]]+]]]]x|xxx]`               
                       -+%%xxxx]]]]+]]++]+]x]|>- .;..;.:_/`     
                          -+x]]]]|+]+]+]++]++]]+|+|]+|]]+-      
              ,.   .  ., |x]+]+||=++]+]=++++++=|++]=+]|~-       
              -|%x]]];x]]||=++++++++|];|++++++++++=+]=];   ..   ..  :..,; 
                -/]|]+]|]++++++++||||==|:;:=|==++==+;|,   :;;. :;=;;===|` 
                 _|]|+>+++]+]]|+|||=|=;|;;:|;=:===|==;::.;;:;,;:;;;;;;:-  
                     -++]+++:+|x+::||=||:=:::::;;::::::::;:;:;:;:;;;=::   
                   .|x+|+|]++:,-..:::|=;=:-.:.:-::.:.:.::-::::::::;:-     
         .,  .:||_   --||;:|:::.-.:.-:|;||::.:...-.-.-....::::-:::;::.    
     __._;++;;;|=|;. -:::::::---:.--;;|==|:;:.:-:-:.-.-.:::.--:::;:;--    
     -+++=+=======;==:::::-:.::...:.|+=;;===:-..:.:.:::::-...--:  -       
     :|:-:|===;:;:;::::::--::.:-::::.|||===:::.:.:::::.-......--:: .      
        -;|===;;:;;;;::;::.:-:: -:=;;|+||=;==|=::::::::...-.:..-.::..     
         ---|;:,::::.::::=;=:=::  -++===|======:--:   ...-...-:-:.-       
             ---;:::::;:--:===;;;:|-|+=+|+|==;:`    ...-....-..           
             .::;;:::;:;;:--- -- --.|=||=+|=:=,      ..... .              
            .;;=;;:-:::;:::.        -+;,|]| ;;=`                          
              - -:- --:;;:- -         - :|; -                             
                       -                 :`                               
                                          :                               
                                          :                               
                                          .                               
                                                                  
   40000L,                                           |0000i   j000&  .a00000L#0
   --?##0L        .aaaa aa     .aaaa;     aaaa, _aaa, -000A  _0001- _d0!`  -400
     d0 40,     _W0#V9N0#&    d0#V*N#0,   0##0LW0@4#@' 00j#; J0|01  d0'      40
    J0l -#W     #0'    ?#W   ##~    -#0;    j##9       00 4#|01|01  00     
   _00yyyW0L   :0f      ^-  :000###00001    j#1        00 ?#0@`|01  #0      
   ##!!!!!#0;  -0A       _  -0A             j#1        00  HH< |01  j0L       _
 ad0La,  aj0Aa  4#Aaa_aj#0`  ?0Laa_aaa0L  aaJ0Laaa,  _a00aa  _aj0La  *0Aaa_aad
 HHHRHl  HHHHH   `9##009!     `9NW00@!!`  HHHHRHHHl  :HHHHH  ?HHHRH   ?!##00P!`
      
  -----------------------------------------------------------------------------
      
        AcerMC 1.0 (February 2002),  B. P. Kersevan, E. Richter-Was
      
  -----------------------------------------------------------------------------
\end{verbatim}
}
\vspace{0.5cm}
\begin{flushleft}
{\it Available from web page: {\bf http://cern.ch/Borut.Kersevan} }\\
\vspace{0.10cm}
{\it Contact author email: {\bf Borut.Kersevan@cern.ch}}
\end{flushleft}

\newpage
\boldmath
\section{PROGRAM SUMMARY}
\unboldmath

{\it Title of the program:} {\bf AcerMC version 1.0}\\
{\it Operating system:} Linux\\
{\it Programming language:} FORTRAN 77 with popular extensions.\\
{\it External libraries:} CERNLIB.\\
{\it Size of the compressed distribution directory:} about 3.6 MB. The distribution includes modified
versions of {\tt PYTHIA~6.2}, {\tt HERWIG 6.3} and {\tt HELAS} libraries.\\
{\it Key words:} Standard Model backgrounds at LHC, massive matrix elements, Monte Carlo generator,
heavy flavor production, multi-channel phase-space generation.\\

{\it Nature of physical problem:}
Despite a large repertoire of processes implemented for generation in event
generators like {\tt PYTHIA} [1] or {\tt HERWIG} [2] a number of background processes, 
crucial for studying expected physics potential of the LHC experiments is missing.
For some of these processes the matrix element expressions are rather lengthly and/or
to achieve a reasonable generation efficiency it is necessary to tailor the
phase-space selection procedure to the dynamics of the process. 
That is why it is not practical to imagine that any of the above general
purpose generators will contain {\it every}, or even only {\it observable}, processes which
will occur at LHC collisions. A more practical solution
can be found in  a library of dedicated  matrix-element-based generators,
with the standardised interfaces  like that proposed in [3],
 to the more universal one which is used to complete the event generation.\\
{\it Method of solution:}
The {\bf AcerMC} Event Generator provides itself library of the matrix-element-based 
generators for a few example 
processes. The initial- and final- state showers, beam remnants and underlying events,
fragmentation and remaining 
decays are supposed to be performed by the other universal generator
to which this one is interfaced. We will call it {\it supervising
generator}. 
The interfaces to {\tt PYTHIA~6.2} and {\tt HERWIG 6.3}, as such  generators, 
are provided. At present, the following  matrix-element-based processes have
 been implemented: 
$gg, q \bar q \to t \bar t b \bar b$, 
$q \bar q \to W (\to \ell \nu) b \bar b$;
$q \bar q \to W (\to \ell \nu) t \bar t$,
$gg, q \bar q \to Z/\gamma^*(\to \ell \ell) b \bar b$;
$gg, q \bar q \to Z/\gamma^*(\to \ell \ell, \nu \nu, b \bar b) t \bar t$ and
complete EW $gg \to (Z/W/\gamma^* \to) t \bar t b \bar b$.
Both interfaces allow the use of the {\tt PDFLIB} library of  parton density functions.\\
{\it Restriction on the complexity of the problem:}
The package is dedicated for the 14 TeV $pp$ collision simulated in the LHC environment.
In particular, using it for the 2 TeV $p \bar p$ collision although technically possible
might not be efficient and would require dedicated optimisation of the phase space generation.
The consistency between results of the complete generation using {\tt PYTHIA~6.2} 
or {\tt HERWIG~6.3} interfaces is technically limited by the different
approaches taken in both these generators for evaluating $\alpha_{QCD}$ 
and $\alpha_{QED}$ couplings and by the different models
for fragmentation/hadronisation.
For the consistency check, in the {\bf AcerMC} library  contains native coded 
definitions of the $\alpha_{QCD}$ and  $\alpha_{QED}$. Using these native
definitions leads to the same total cross-sections both with 
 {\tt PYTHIA~6.2} or {\tt HERWIG~6.3} interfaces. 
{\it Typical running time:} On an PIII 800 MHz PC it amounts to $\sim 0.05 \to 1.1$ events/sec, depending
on the choice of process. 
  
[1]. T. Sjostrand et al., {\it High energy physics generation with PYTHIA~6.2},
eprint hep-ph/0108264, LU-TP  01-21, August 2001.

[2]. G. Marchesini et al., Comp. Phys. Commun. {\bf 67} (1992) 465,
G. Corcella et al., JHEP {\bf 0101} (2001) 010.

[3]. E. Boos at al., {\it Generic user process interface for event generators}, hep-ph/0109068.

\newpage
\boldmath
\section{Introduction}
\unboldmath

Despite a large repertoire of processes implemented for generation in the
universal generators like {\tt PYTHIA} \cite{Pythia62} or {\tt HERWIG}
\cite{Herwig6.3} a number of Standard Model background processes for studying
expected physics potential of the LHC experiments is still missing.  For some of these
processes the matrix element expressions are rather lengthy and/or to achieve a reasonable
generation efficiency it is necessary to tailor the phase-space selection procedure to the
dynamics of the process. Due to this fact it cannot be expected that any of the universal
purpose generators will contain {\it every} proces that is expected to participate at LHC
pp collisions with an observable rate.  A more practical solution could come in form of
dedicated matrix-element-based generators with standardised interfaces, like the one
proposed in \cite{Boo01}, to the more general ones which are used to complete event
generation.

The {\bf AcerMC} Monte Carlo Event Generator follows up on this idea.
It is dedicated for the simulation of the specific Standard Model
background processes at LHC collisions: the $gg, q \bar q \to t \bar t b \bar b$, 
$q \bar q \to W (\to \ell \nu) b \bar b$;
$q \bar q \to W (\to \ell \nu) t \bar t$,
$gg, q \bar q \to Z/\gamma^*(\to \ell \ell) b \bar b$;
$gg, q \bar q \to Z/\gamma^*(\to \ell \ell, \nu \nu, b \bar b) t \bar t$ and
complete EW $gg \to (Z/W/\gamma^* \to) t \bar t b \bar b$.
They are characterised by the
presence of the heavy flavour jets and multiple isolated leptons in
the final state. For the Higgs boson searches,  the $t \bar t H$, $ZH, WH$ 
with $H \to b \bar b$, the $gg \to H$ with $H\to ZZ^* \to 4 \ell$,
the $b\bar b h/H/A$ with $h/H/A \to \tau \tau, \mu \mu $ are the most obvious
examples of signals where the implemented processes would contribute
to the dominant irreducible backgrounds. The same background processes
should also be considered for e.g. estimating the observability of SUSY events
with a signature of multi-b-jet and multi-lepton production.

The program itself provides library of the massive matrix elements 
and phase space modules for the generation of a few selected $2 \to 4$ processes.
The hard process event, generated with these modules, can be completed by the 
initial and final state radiation, hadronisation and decays, simulated with either 
{\tt PYTHIA~6.2} or {\tt HERWIG 6.3} Monte Carlo Event Generators. 
These will subsequently be called the
{\it Supervising Generators}. Interfaces of {\bf AcerMC} to both,
 {\tt PYTHIA~6.2} or {\tt HERWIG 6.3}
generators, are provided in the distribution version. The {\bf AcerMC} also uses
several other external libraries: {\tt CERNLIB}, {\tt HELAS} \cite{HELAS}, {\tt VEGAS} \cite{vegas}.
The matrix element codes have been derived with the help of {\tt MADGRAPH} 
\cite{Madgraph} package.
The achieved typical efficiency for the generation of unweighted events 
is of {\bf 20\%~-~30\%}, rather high given a complicated topology of the 
implemented processes.
 
The very first version of this library, interfaced to {\tt PYTHIA 6.1 } within the
standard of the so-called {\it external processes} (i.e. stand-alone implementations of
hard processes interfaced to {\tt PYTHIA} for further treatment of ISR/FSR and
hadronisation, c.f \cite{Pythia62} ), was already available and used by ATLAS
Collaboration for physics simulation studies since several months. The documentation, for
the processes implemented in this early version: $q \bar q \to W(\to \ell \nu) b \bar b$,
$gg, q \bar q \to Z/\gamma^*(\to \ell \ell) b \bar b$ and $gg, q \bar q \to t \bar t b
\bar b$, can be found respectively in \cite{ATLCOMP013}, \cite{ATLCOMP014},
\cite{ATLCOMP025}.  Since then, when upgrading to the {\bf AcerMC} standard, the
efficiency has been significantly improved thanks to the additional optimisation step in
the phase space generation. Also, the interface standard was changed from {\tt PYTHIA 6.1
} to {\tt PYTHIA~6.2 } conventions, an interface to {\tt HERWIG~6.3} generator was
introduced and the native {\bf AcerMC} calculations of the $\alpha_{QED}$ and
$\alpha_{QCD}$ couplings were coded to allow for consistent benchmarking between results
obtained with {\tt PYTHIA} and {\tt HERWIG} as supervising generators. As a significant
extension, the: $q\bar q \to W(\to \ell \nu) t \bar t$; $gg, q \bar q \to Z/\gamma^*(\to
\ell \ell, \nu \nu, b \bar b) t \bar t$ and complete electroweak $gg \to (Z/W/\gamma^*
\to) t \bar t b \bar b$ processes were added, which have been implemented for the first
time in the {\bf AcerMC} library.

The outline of this paper is as follows. In Section 3, we describe physics
motivation for implementing each of the above processes and we collect some
numerical results (plots, tables) which can be used as benchmarks.  In Section 4
we describe the overall Monte Carlo algorithm. Section 5 gives details on the
structure of the program.  Section 6 collects information on how to use this
program and existing interfaces to {\tt PYTHIA~6.2} and {\tt HERWIG
6.3}. Summary, Section 7, closes the paper. Appendix A documents sets of Feynman
diagrams used for calculation of the matrix element for each subprocess,
Appendices B and C give examples of the input/output of the program.


\boldmath
\section{Physics content}
\unboldmath

The physics programme of the general purpose LHC experiments, ATLAS \cite{ATL-PHYS-TDR}
and CMS \cite{CMS-Documents}, focuses on the searches for the {\it New Physics} with the
distinctive signatures indicating production of the Higgs boson, SUSY particles, exotic
particles, etc.  The expected environment will in most cases be very difficult, with the
signal to background ratio being quite low, on the level of a few
percent after final selection in the signal window.

Efficient and reliable Monte Carlo generators, which allow one to understand and predict
background contributions, are becoming the key point to the discovery. As the
cross-section for signal events is rather low, even rare Standard Model processes might
become the overwhelming background in such searches. In several cases, generation of such
a process is not implemented in the general purpose Monte Carlo generators, when the
complicated phase space behaviour requires dedicated (and often rather complex)
pre-sampling, whilst the general purpose Monte Carlo generators due to a large number of
implemented processes tend to use simpler (albeit more generic) phase space sampling
algorithms.  In addition, the matrix element for these processes is often lengthy and thus
requiring complicated calculations.  Only recently, with the appearance of modern
techniques for automatic computations, their availability {\it on demand} became feasible
for the tree-type processes. With the computation power becoming more and more easily
available, even very complicated formulas can now be calculated within a reasonable time
frame.

The physics processes implemented in {\bf  AcerMC} library represent such a set of cases.
They are all being key background processes for the  discovery in the channels
characterised by the presence of the heavy flavour jets and/or 
multiple isolated leptons. For the Higgs boson searches,  the $t \bar t H$, $ZH, WH$ 
with $H \to b \bar b$, the $gg \to H$ with $H\to ZZ^* \to 4 \ell$,
the $b\bar b h/H/A$ with $h/H/A \to \tau \tau, \mu \mu $ are the most obvious
examples of such channels. 

It is not always the case that the matrix element calculations in the lowest
order for a given topology represent the total expected background of a given
type. This particularly concerns the heavy flavour content of the event.  The
heavy flavour in a given event might occur in the hard process of a much simpler
topology, as the effect of including higher order QCD corrections (eg. in the
shower mechanism). This is the case for the b-quarks present in the inclusive
Z-boson or W-boson production, which has a total cross-section orders of
magnitude higher than the discussed matrix-element-based $Wb \bar b$ or $Zb \bar
b$ production.  Nevertheless, the matrix-element-based calculation is a very
good reference point to compare with parton shower approaches in different
fragmentation/hadronisation models.  It also helps to study matching
procedures between calculations in a fixed $\alpha_{QCD}$ order and parton
shower approaches.  For very exclusive hard topologies matrix-element-based
calculations represent a much more conservative approximation than the parton
shower ones
\cite{ATLCOMP032}.

Let us shortly discuss the motivation for these few Standard Model background 
processes which are implemented in the {\bf AcerMC 1.0} library. 

\boldmath {\bf The $t \bar t b \bar b$ production  } \unboldmath
at LHC is a dominant irreducible background for the Standard Model (SM) and
Minimal Supersymmetric Standard Model (MSSM) Higgs boson search in the
associated production, $ t \bar t H$, followed by the decay $H \to b \bar b$.
The potential for the observability of this channel has been carefully studied
and documented in \cite{ATL-PHYS-TDR} and \cite{ATL-PHYS-98-132}.  Proposed
analysis requires identifying four b-jets, reconstruction of both top-quarks in
the hadronic and leptonic mode and visibility of the peak in the invariant mass
distribution of the remaining b-jets.  The irreducible $t \bar t b \bar b$
background contributes about 60-70\% of the total background from the $t \bar t$
events ($t \bar t b \bar b$, $t \bar t b j$, $t \bar t j j$).

\boldmath {\bf The $W b \bar b$ production} \unboldmath at LHC is
recognised as a substantial irreducible background for the Standard Model (SM)
and Minimal Supersymmetric Standard Model (MSSM) Higgs boson search in the
associated production, $WH$, followed by the decay $H \to b \bar b$.  The
massive matrix element for $q \bar q \to W g^*(\to b \bar b)$ process has been
calculated \cite{Wbb93} and interfaced with {\tt HERWIG 5.6} Monte Carlo
\cite{Herwig5.6} already a few years ago.  A more recent implementation of the
$Wbb$ + multi-jet final states is available from \cite{Wbbgen}. Recently, the
massless matrix element has been implemented in the general purpose Monte Carlo
program {\tt MCFM} \cite{RKEllis2000}, where the radiative corrections to this
process are also addressed. Another implementation of the $q \bar q \to W ( \to
\ell \nu)g^* (\to b \bar b)$ massive matrix elements, with the interface to {\tt
PYTHIA 6.1} became available in \cite{ATLCOMP013}.  The {\bf AcerMC}
library discussed here includes even more efficient implementation of the
algorithm presented in \cite{ATLCOMP013}.

\boldmath {\bf The $W t \bar t$ production } \unboldmath 
at LHC has to our knowledge not been implemented in the publicly available code
so far\footnote{ We thank M. L. Mangano for bringing this process to our
attention and for providing benchmark numbers for verifying the total
cross-section.}. It is of interest because it contributes an overwhelming
background \cite{Note-Federica} for the measurement of the Standard Model Higgs
self-couplings at LHC in the most promising channel $pp \to HH \to WWWW$.

\boldmath {\bf The  $Z/\gamma^*(\to \ell \ell)  b \bar b$ production} \unboldmath at LHC 
has since several years been recognised as one of the most substantial
irreducible (or reducible) backgrounds for the several Standard Model (SM) and
Minimal Supersymmetric Standard Model (MSSM) Higgs boson decay modes as well as
for observability of the SUSY particles.  There is a rather wide spectrum of
{\it regions of interest} for this background.  In all cases the leptonic
$Z/\gamma^*$ decay is asked for, but events with di-lepton invariant mass around
the mass of the Z-boson mass or with the masses above or below the resonance
peak could be of interest.  The presented process enters an analysis either by
the accompanying b-quarks being tagged as b-jets, or by the presence of leptons
from the b-quark semi-leptonic decays in these events, in both cases thus
contributing to the respective backgrounds.

Good understanding of this background, and having a credible Monte Carlo
generator available, which allows studying of expected acceptances for different final states
topologies, is crucial. Despite a very large effort taken in time of the preparation of the
Aachen Workshop \cite{Aachen1990}, such well established Monte Carlo generator was missing
for several years.
The matrix element for the $gg \to Z b \bar b \to b \bar b \ell \ell$ production has been
published already in \cite{Kleiss1990} and in time of Aachen Workshop implemented into
{\tt EUROJET} Monte Carlo \cite{EUROJET}.  But that generator was not giving the
possibility for having fully generated hadronic event with modelled initial and final state
radiation and hadronisation, for analyses presented in \cite{DellaNegra1990} it was
interfaced to {\tt PYTHIA 5.6} \cite{Pythia57} Monte Carlo.  This is no longer
supported and  available at present.
The same matrix element has been directly implemented into {\tt PYTHIA 5.7}
\cite{Pythia57}. However, with this implementation the algorithm for the phase space
generation never working credibly, it has finally  been removed from the version
{\tt PYTHIA 6.1} \cite{Pythia61}.
A year ago, the massless matrix elements for $gg, q \bar q \to Z b \bar b$ processes have been
implemented in the general purpose Monte Carlo program {\tt MCFM} \cite{RKEllis2000}.  In
that implementation radiative corrections to this process are addressed as well.  The massive
matrix element implementation is also present in the very recent  
version of {\tt HERWIG 6.3} \cite{Herwig6.3}.
At that time the $gg, q \bar q \to Z/\gamma^* b \bar b \to \ell \ell b \bar b$
massive matrix elements, with the interface to {\tt PYTHIA 6.1} became 
available \cite{ATLCOMP014}.
The {\bf AcerMC} library discussed here includes  more efficient implementation of
the algorithm presented in \cite{ATLCOMP014}.

\boldmath {\bf The  $Z/\gamma^*(\to \ell \ell, \nu \nu, b \bar b)  t \bar t$ production}
 \unboldmath at LHC is an irreducible background to the Higgs search in the
invisible decay mode (case of $Z \to \nu \nu)$ in the production with association
to the top-quark pair \cite{Gunion94}. With the  $Z/\gamma^*(\to b \bar b)$ it is also
an irreducible resonant background to the Higgs search in the $t \bar t H$  production channel
but with the Higgs boson decaying to the b-quark pair \cite{ATL-PHYS-98-132}. 

The complete \boldmath {\bf EW production}  of the
\boldmath {\bf $gg \to (Z/W/\gamma^* \to)  b \bar b t \bar t$} \unboldmath  final
state is also provided. It can be considered as a benchmark for the previous
process, where only the diagrams with resonant $gg \to (Z/\gamma^* \to) b \bar b
t \bar t$ are included. It thus allows the verification of the question, whether
the EW resonant contribution is sufficient in case of studying the $t \bar t b
\bar b$ background away from the Z-boson peak, like for the $t \bar t H$ with
Higgs-boson mass of 120~GeV.

This completes the list of the native {\bf AcerMC} processes implemented so far.
Having all these different production processes implemented in the consistent framework, 
which can also be directly used for generating standard subprocesses implemented in either
{\tt PYTHIA} or {\tt HERWIG} Monte Carlo, represents a very convenient environment 
xor several phenomenological studies dedicated to the LHC physics.

For the cases, where radiative photon emission from final state leptons is important 
the package {\tt PHOTOS} \cite{PHOTOS} can be used in the chain of event generation.
In similar way also package {\tt TAUOLA} \cite{TAUOLA} can be interfaced directly to
the generation chain and used for events generation in cases where
more detailed treatment of the tau-lepton decay and including spin
correlations effects is relevant.

In the following subsections we discuss in more detail implementation of each
subprocess. We also give benchmark Tables with the total
cross-sections obtained with {\tt AcerMC} processes but  different 
implementations and setting of 
$\alpha_{\rm QCD}(Q_{QCD})$: the native {\tt AcerMC}, {\tt PYTHIA} and
{\tt HERWIG} ones. For a more detailed discussion on this
topic the reader is referred to Section \ref{s:alphas}. If the  native
{\bf AcerMC} definition is used, the same cross-section is obtained
either  with {\tt PYTHIA} or {\tt HERWIG} generation chains.

\boldmath
\subsection{The  $g g, q \bar q \to t \bar t b \bar b$ processes}
\unboldmath

In the implementation discussed here, the matrix element was derived using the
{\tt MADGRAPH} package \cite{Madgraph}. These matrix elements are not covering
the decay of the top-quarks, the latter are considered as massive final states
of the process. The top-quark decays is than performed by the supervising
generator.  Rather important spin effects (spin correlations) in the top decays
are therefore not yet included. The similar solution, like for tau decay in
the Z-boson production process discussed in \cite{Acta2001}, is planned to be
implemented here in the near future.

As a benchmark, the processes $g g, q \bar q \to t \bar t b \bar b$ have been
simulated for pp collisions with 14~TeV centre-of-mass energy and CTEQ5L \cite{cteq5l}
parton density functions, using event generation with massive $2 \to 4$ matrix
element implemented as an external process to {\tt PYTHIA 6.2} (see Section 4
and 5).  The decays of the top-quarks have been left under control of {\tt
PYTHIA 6.2} generator.  The $q \bar q \to t \bar t b \bar b$ subprocess
contributes less than 10\% of the total cross-section.

The total cross-section is very sensitive to the choice of the QCD energy scale
used for calculation of that process, thus indicating potentially large
contributions from higher order corrections. The same definition for the
factorisation and renormalisation scale is used. The example values of the total
cross-section for implemented choices of the QCD energy scale are given in
Table~\ref{T1:1}.

As a cross-check, the processes $gg, q \bar q \to t \bar t b \bar b$ have been
coded independently using the {\tt COMPHEP} package \cite{COMPHEP}.  The same
set of diagrams was selected and only the integrating part of the package was
used to calculate total cross-section.  The choices for the QCD energy scale
were kept consistent.  A very good agreement between the cross-sections obtained
with two independent calculation streams prepared for this study has been
achieved \cite{ATLCOMP025}.

\begin{table}
\newcommand{\lstrut}{{$\strut\atop\strut$}}
  \caption {\em {\bf AcerMC} cross-sections  for the $gg, q \bar q \to t \bar t b \bar b$
production at different choices of the QCD energy scale and $\alpha_{QCD}$ implementations. 
The 14 TeV centre-of-mass
energy and CTEQ5L parton density functions were used for the simulation with interfaces
to {\tt PYTHIA 6.2} and {\tt HERWIG 6.3}.  The $m_H~=~120$~GeV and $m_t~=~175$~GeV 
were used for calculating the $Q^2_{QCD}$ in the last row of this table. The default 
settings of $\alpha_{QCD}$ as implemented in {\bf AcerMC}, {\tt PYTHIA 6.2} and {\tt HERWIG 6.3} 
were used. \label{T1:1}}\vspace{0.5cm}
\begin{center}
\scriptsize
\begin{tabular}{|c|c|c|c|} \hline \hline
Factorisation scale  &  $\alpha_{QCD}$ (1L)  &  $\alpha_{QCD}$ (1L) &  $\alpha_{QCD}$ (2L)  \\ 
       & native AcerMC & as in {\tt PYTHIA 6.2} & as in {\tt HERWIG 6.3}  \\
\hline \hline
\cline{2-4} \multicolumn{1}{|c|}{}& \multicolumn{3}{|c|}{$gg \to t \bar t b \bar b$} \\
\hline 
$Q^2_{QCD}~=~\hat{s}$ &  4.2 [pb]     &  3.9 [pb]     & 2.7 [pb]     \\
\hline
$Q^2_{QCD}~=~\sum{({p^i_T}^2 + m_i^2)}/4$ &  10.3 [pb]      & 10.2 [pb]    & 6.4 [pb]    \\
\hline
$Q^2_{QCD}~=~\sum{({p^i_T}^2)}/4$ &  17.0 [pb]      &  16.9 [pb]   & 10.1 [pb]    \\
\hline
$Q^2_{QCD}~=~(m_t + m_H/2)^2$ & 8.2  [pb]     & 8.1 [pb]   & 5.2 [pb]      \\
\hline \hline
\cline{2-3} \multicolumn{1}{|c|}{}& \multicolumn{3}{|c|}{$q \bar q \to t \bar t b \bar b$} \\
\hline 
$Q^2_{QCD}~=~\hat{s}$ &  0.30 [pb]    &  0.29 [pb]     & 0.20 [pb]    \\
\hline
$Q^2_{QCD}~=~\sum{({p^i_T}^2 + m_i^2)}/4$ & 0.61  [pb]   & 0.60 [pb]    &  0.38 [pb]    \\
\hline
$Q^2_{QCD}~=~\sum{({p^i_T}^2)}/4$ &  0.91 [pb]       &  0.90 [pb]     & 0.54 [pb]   \\
\hline
$Q^2_{QCD}~=~(m_t + m_H/2)^2$ & 0.52  [pb]     &  0.51  [pb]     & 0.33 [pb]     \\
\hline \hline 
\end{tabular}
\end{center}
\end{table}

One can observe a very strong scale dependence of the cross-section for the $gg, q \bar q
\to t \bar t b \bar b$ process (c.f. Table \ref{T1:1}).  Factor four (!!) can be expected
on the predicted cross-section when changing from the scale $Q^2_{QCD}~=~\hat{s}$ to the
scale $Q^2_{QCD}~=~<p_T^2>$.  This very strong dependence on the energy scale is also
observed in the case of the $t \bar t H$ production, for recent discussion see
\cite{NLOttH}. There, the recommended {\it central} factorisation and renormalisation
energy scale is $\mu_0~=~(m_t + m_H/2)$.  Having in mind that the primary interest of
evaluating this background is the Higgs search in the $t \bar t H$ production, i.e.  with
the b-quark system being produced with the invariant mass of the expected Higgs boson, we
have also introduced this {\it central} energy scale, with $m_H~=~120$~GeV as one of the
possible choices.  

\begin{figure}[ht]
\begin{center}
\mbox{
     \hspace{-0.2cm}
     \epsfxsize=5.0cm
     \epsffile{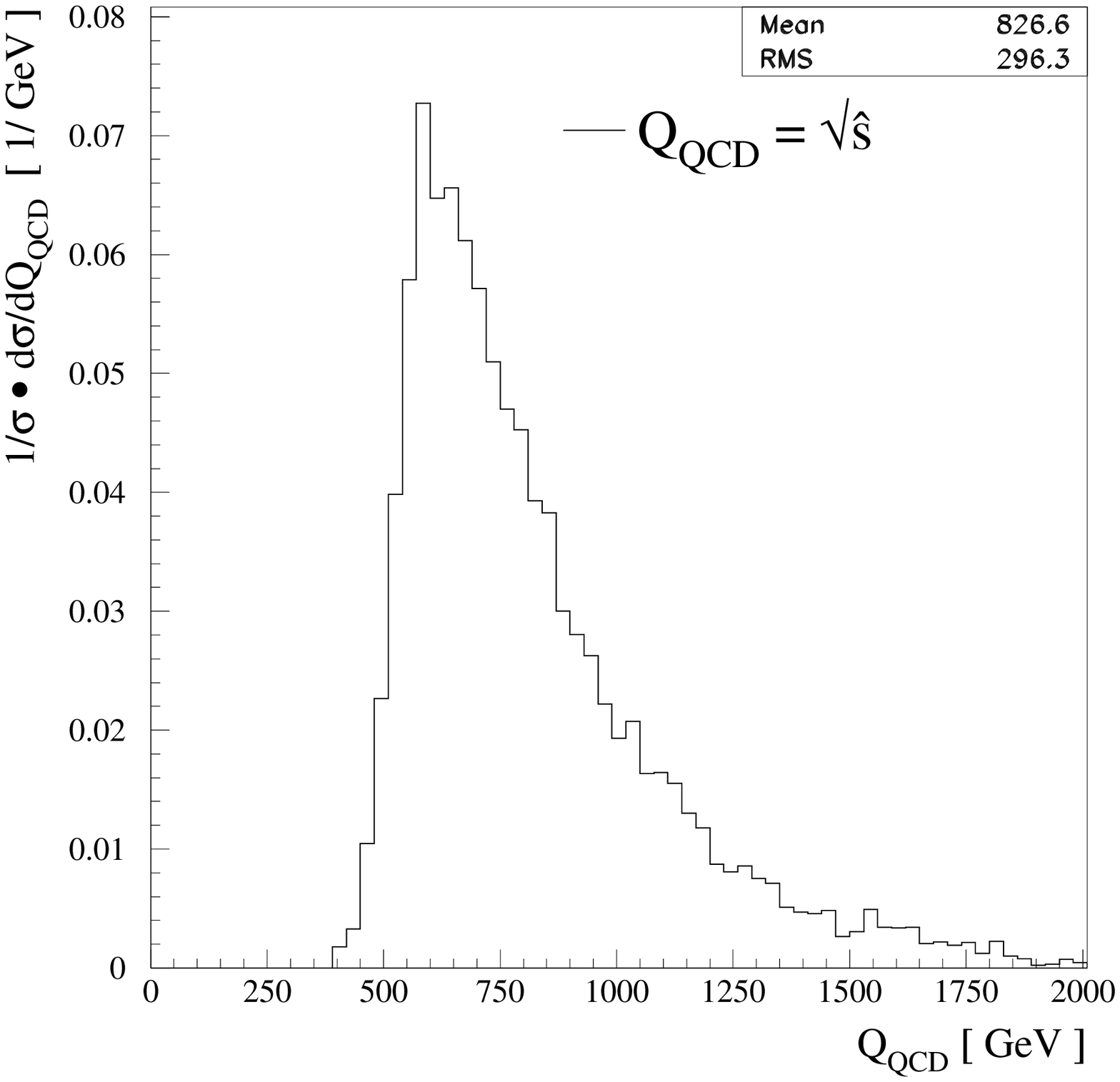} 
     \hspace{-0.7cm}
     \epsfxsize=5.0cm
     \epsffile{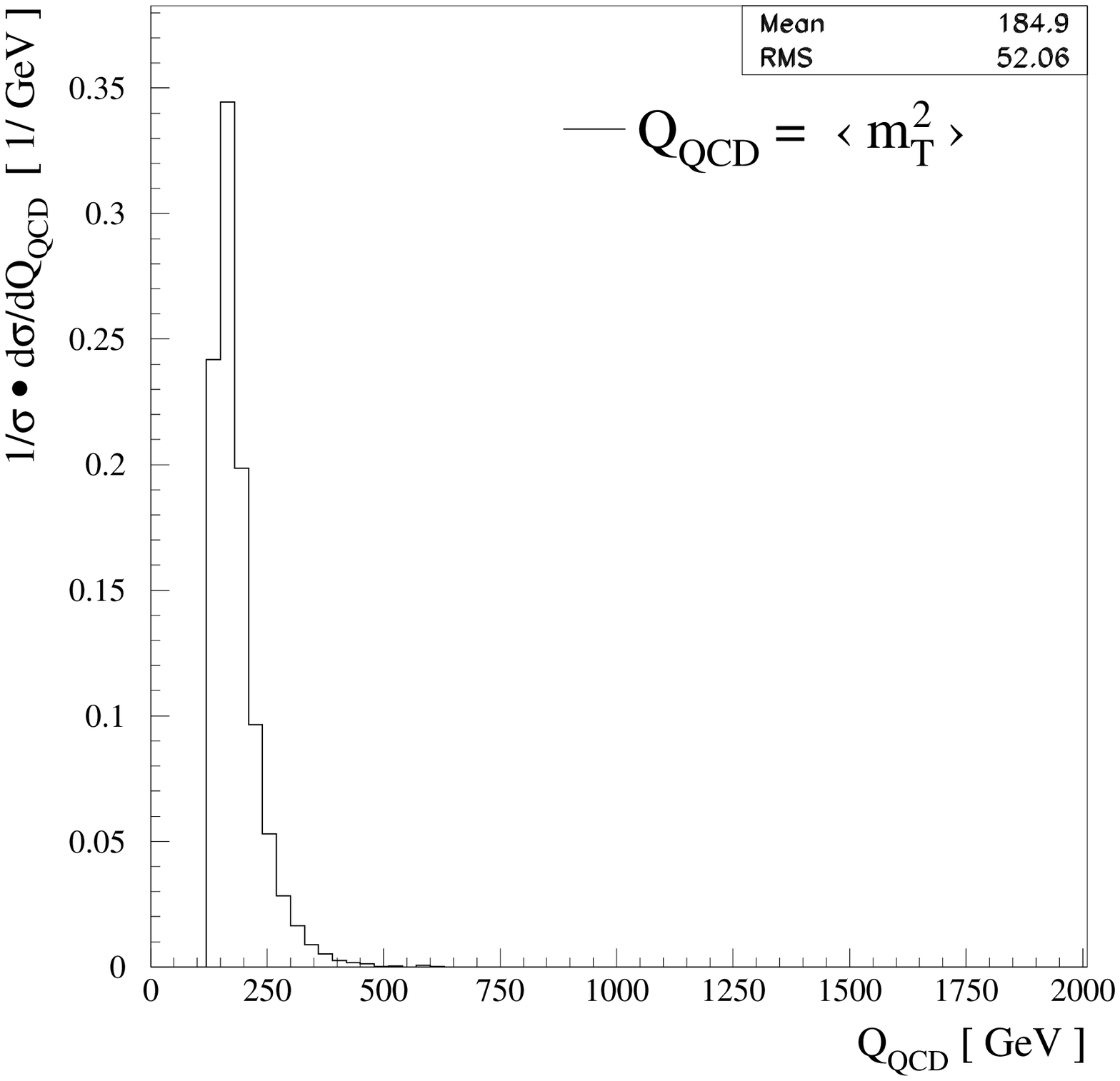}
     \hspace{-0.7cm}
     \epsfxsize=5.0cm
     \epsffile{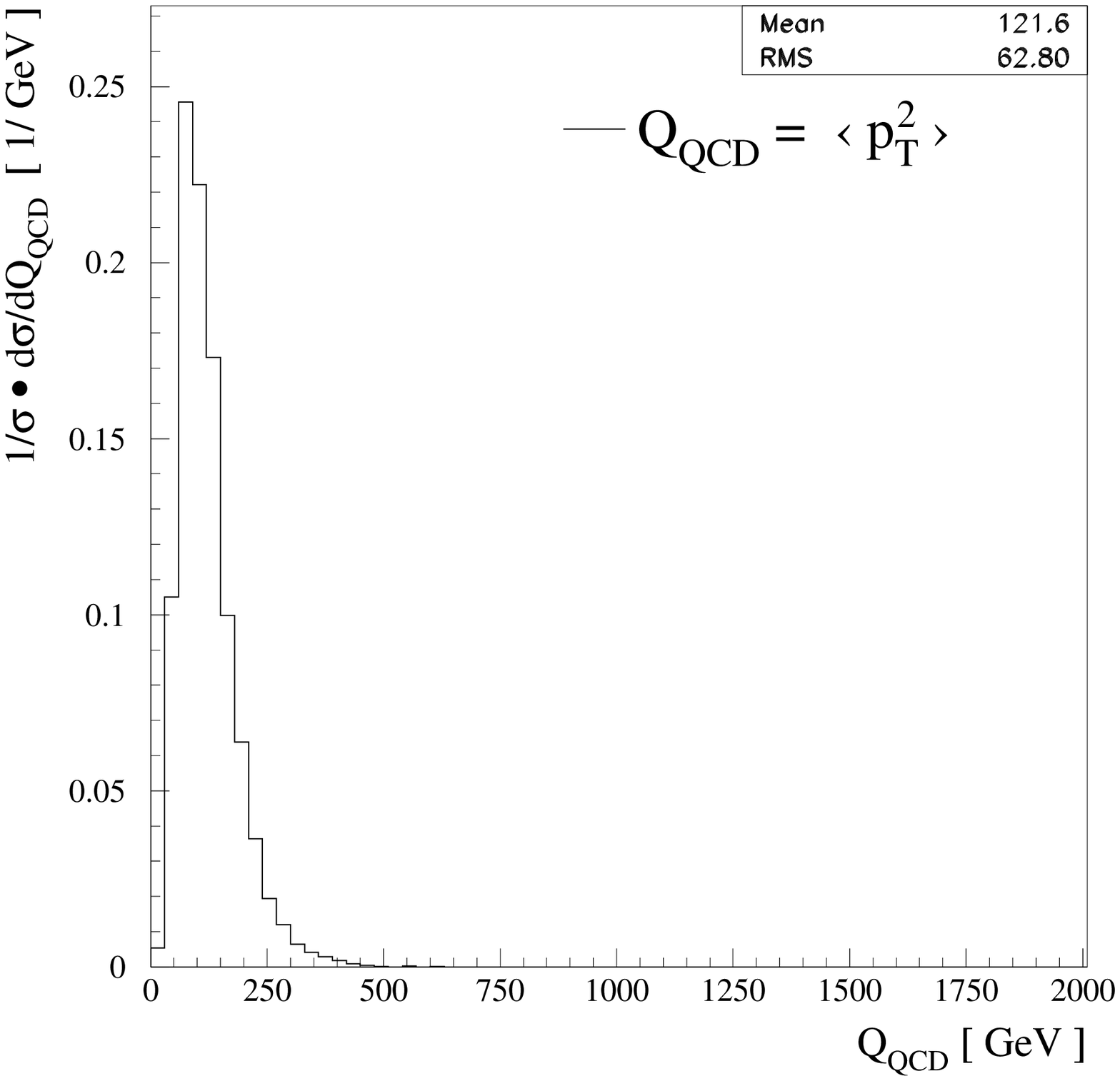}
}
\vglue -0.3cm
\mbox{
     \epsfxsize=5.0cm
     \epsffile{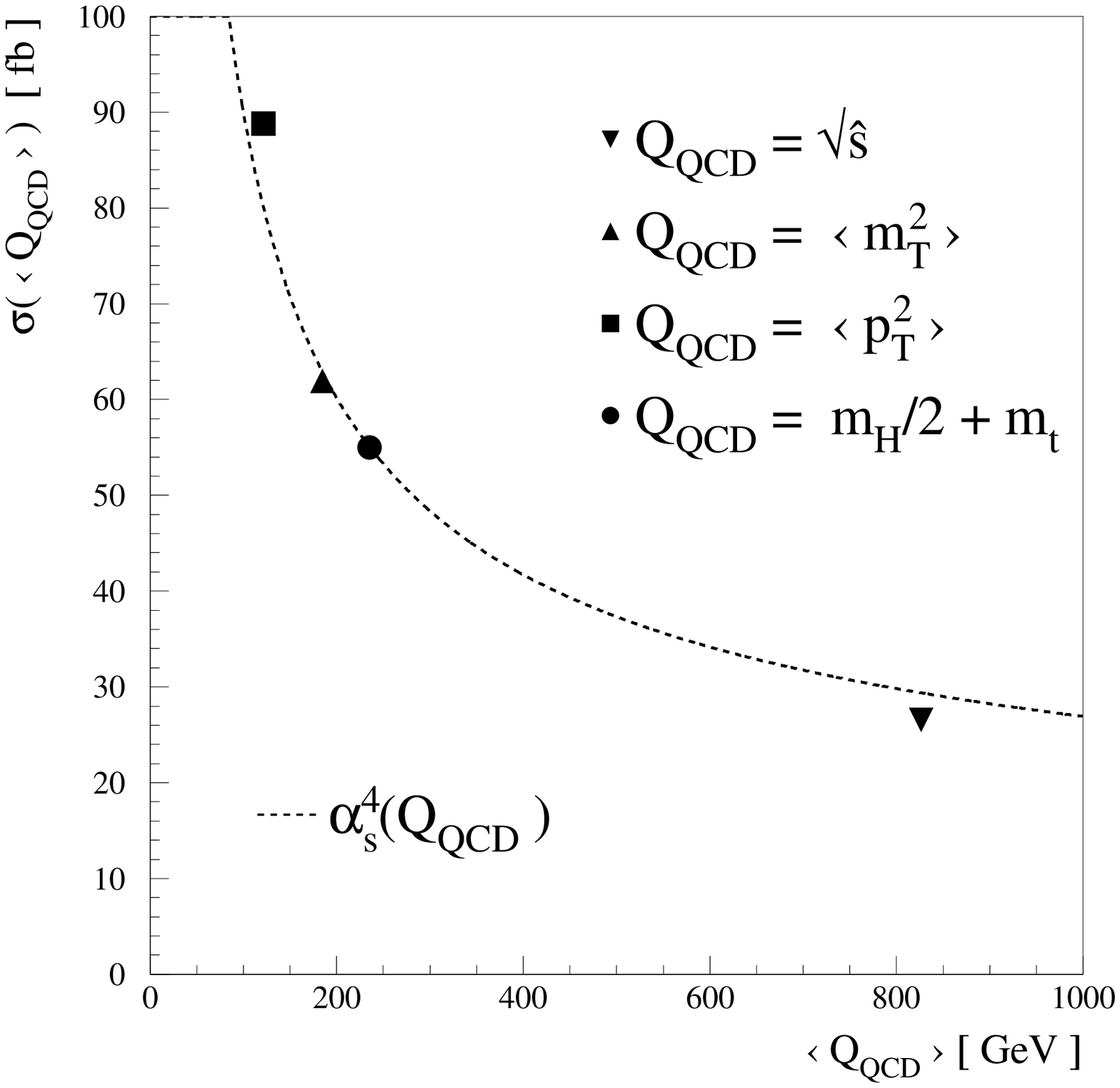}
}
\end{center}
\caption{\em
Top: the $Q^2_{QCD}$ distributions for $t \bar t b \bar b$ events with 
the invariant mass of the b-jets system $m_{bb-jets}=120~\pm~30$~GeV.
Bottom: the total cross-section of accepted
events as a function of the averaged $Q^2_{QCD}$ (for these events). 
\label{FS2a:a}}
\end{figure}

Fig.~\ref{FS2a:a} shows the distributions of the $Q_{QCD}=\sqrt{Q^2_{QCD}}$ (distributions
have been normalised to one) for the $t \bar t b \bar b$ events with the invariant mass
of the b-jets system, calculated using the default {\tt PYTHIA} (LO) $\alpha_{\rm QCD}$ 
implementation\footnote{ This would makes distributions directly relevant for the $t
\bar tH$ analysis. For details on the jet reconstruction see \cite{ATL-PHYS-98-131}.}.
$m_{bb-jets}~=~120~\pm~30$~GeV. The distribution is well collimated around the
average value when $Q^2_{QCD}$ is defined as $<m_T^2>$ or $<p_T^2>$ while it is
much broader when $Q^2_{QCD}$ is defined as $\hat{s}$. The kinematic
distributions are very similar in shape for separate $gg \to t \bar t b \bar b$
and $q \bar q \to t \bar t b \bar b$ contributions.  The total cross-section for
accepted events as a function of the averaged $Q^2_{QCD}$ (for these events) is
shown in the bottom plot. It can be noted that the cross-section decreases
rather fast with the  increasing value of the average $<Q^2_{QCD}>$.  Also shown is
the $\alpha_s^4(Q_{QCD})$ dependence scaled to match the cross-section at
$Q_{QCD}=(m_H/2 + m_t)$ with $ m_H=120\;$ GeV, it being the only calculated
cross-section point with a fixed scale. The other cross-sections are shown to
follow the expected $\alpha_s^4(Q_{QCD})$ dependence rather well, while the
deviations are induced by the parton density function dependence on the $Q^2_{QCD}$
scale, most notably at $Q^2_{QCD}=\hat{s}$ value. The deviations induced by
the parton density functions dependence on the $Q^2_{QCD}$ scale are different for the
$gg$ and $q \bar q$ contributions, as can be concluded from results given in
Table~\ref{T1:1}.

The series of plots illustrating the most relevant differential distributions for the
top-quarks and b-quarks can be found in \cite{ATLCOMP025}.

\boldmath
\subsection{The  $q \bar q \to W (\to  \ell \nu) g^*(\to b \bar b)$ 
process}
\unboldmath

The matrix element for the implemented process was again coded by using the {\tt
MADGRAPH} package \cite{Madgraph}.  This process is represented by only two
Feynman diagrams, with quark exchange in the t-channel, leading to the
production of the $W$-boson and virtual gluon splitting into $ b \bar b$
pair. Only the $u, d, s, c$ quarks were considered in this implementation, the
possibility of the b-quark in the initial state was omitted as expected to be
negligible numerically (e.g.. $|V_{bc}/V_{ud}|^2 \sim 0.002$) but leading to
several additional diagrams which would have to be included.  The massive matrix
element takes into account spin correlations in the W-boson decay and angular
correlations between leptons and quarks.  Due to the massive treatment of the
final state fermions the amplitude has no singularities; the total cross-section
is well defined.  The effect from the $W$-boson natural width and the $W$-boson
propagator are also properly included.

As a benchmark, the process $ q \bar q \to W (\to \ell \nu) g^*(\to b \bar b)$ has
been simulated for pp
collision with 14~TeV centre-of-mass energy.  The total cross-section, including branching
ratio for $W \to \ell \nu$ (single flavour) is 36.5~pb (CTEQ5L parton density functions, $Q^2 =
M_W^2$, {\tt PYTHIA 6.2} interface)\footnote{This can be compared with the matrix element implementation to {\tt HERWIG 5.6},
used in \cite{ATL-PHYS-94-043},\cite{ActaB31}, where originally this cross-section was estimated
to 19.8~pb (CTEQ2L parton density functions) but, when implementing CTEQ5L parton density functions
and setting kinematic parameters to be in approximate accordance with {\tt PYTHIA} defaults, rises to
36.0~pb, which is consistent with the AcerMC implementation by taking into account the remaining 
differences in the two calculations (e.g. the former implementation uses an on-shell W boson in the ME
calculation).}.

\vspace{0.2cm}
\begin{table}[hb]
\newcommand{\lstrut}{{$\strut\atop\strut$}}
  \caption {\em {\bf AcerMC} production cross-sections for the $q \bar q \to W b \bar b$ with
  $W \to e \nu~$ decay (single flavour). The 14 TeV centre-of-mass energy and
  CTEQ5L parton density functions were used with different definitions of
  $\alpha_{QED}$, $\alpha_{QCD}$ ( as in default {\tt PYTHIA 6.2} and {\tt
  HERWIG 6.3}) and several choices of the factorisation scale, $\alpha_{QED}$ and $\alpha_{QCD}$ 
  implementations.
\label{T2:1c}}
\vspace{0.5cm}
\begin{center}
\scriptsize
\begin{tabular}{|c|c|c|c|} \hline \hline
Factorisation scale  &  $\alpha_{QED}$, $\alpha_{QCD}$ (1L)  & 
       $\alpha_{QED}$, $\alpha_{QCD}$ (1L) &  $\alpha_{QED}$, $\alpha_{QCD}$ (2L)  \\ 
       & native AcerMC & as in {\tt PYTHIA 6.2} & as in {\tt HERWIG 6.3}  \\
\hline 
 $Q^2 = M_W^2$   &  36.5 [pb]   &  36.4 [pb]   & 29.5 [pb]  \\
\hline 
 $Q^2 = s^*_{b \bar b}$   &  44.1 [pb]   &  44.0 [pb]   & 34.8 [pb]  \\
\hline 
 $Q^2 = M_W^2+pT_W^2$   &  36.0 [pb]   &  36.0 [pb]   & 29.1 [pb]  \\
\hline 
 $Q^2 = (s^*_W + s^*_{b \bar b})/2 +pT_W^2$     &  37.2 [pb]   &  37.1 [pb]   & 30.0 [pb]  \\
\hline \hline
\end{tabular} 
\end{center}
\vspace{0.5cm}
\end{table}

The dependence on the choice of the factorisation scale is rather modest (c.f. Table
\ref{T2:1c}) and does not exceed 20\% for the choices implemented in {\bf AcerMC}
library. The variation of the cross-section due to different $\alpha_{QED}$ and
$\alpha_{QCD}$ implementations and default settings is again evident; as one can expect the two-loop 
$\alpha_{QCD}$ implementation given in {\tt HERWIG} gives a $\sim$20 \% lower cross-section when compared to
the cases when native {\bf AcerMC} and {\tt PYTHIA} one-loop $\alpha_{QCD}$ were used\footnote{While 
performing further comparisons of native {\bf AcerMC} and {\tt PYTHIA} processes we discovered a 
misinterpretation of our CKM matrix implementation. This correction efectively changes the cross-section for 
$q \bar q \to W b \bar b$ and $q \bar q \to W t \bar t$ processes by $\sim$10\% compared to the draft 
versions of this paper, which is nevertheless  still well within the physics precision of the program. The 
affected tables in this paper are already updated.}.

The differential distributions of the $q \bar q \to Wb \bar b$ events turn out to be interesting
when compared to the corresponding ones of the $q \bar q \to Z b \bar b$ and $g
g \to Z b \bar b$ events (generated with pure Z-boson exchange). Such comparison
is well documented in
\cite{ATLCOMP014}.

\boldmath
\subsection{The  $q \bar q \to W (\to  \ell \nu) g^*(\to t \bar t)$ 
process}
\unboldmath
\vspace{-0.2cm}

The matrix elements, coded by the {\tt MADGRAPH} package \cite{Madgraph}, are not covering
the decay of the top-quarks; the latter are considered as massive final states of the
process. The top decay is than performed by the supervising generator. As in the case of
$gg, q \bar q \to t \bar t b \bar b$ process spin effects in the top decays are therefore
not yet included. This process, although rare, contributes an overwhelming irreducible
background to possible measurement of the Higgs-boson self-coupling in the $HH \to WWWW$
decay mode \cite{Note-Federica}.

Table~\ref{T2:1d} shows the expected {\bf AcerMC} cross-sections for different choices of the energy
scale and coupling ($\alpha_{QED}$, $\alpha_{QCD}$) definitions. One should notice the
effect of almost a factor two from different choices of the energy scale.
\vspace{0.5cm} 

\begin{table}[hb]
\newcommand{\lstrut}{{$\strut\atop\strut$}}
  \caption {\em {\bf AcerMC} production cross-sections for the $q \bar q \to W t \bar t$ with primary
  $W \to e \nu~$ decay (single flavour). The 14 TeV centre-of-mass energy, CTEQ5L parton
  density functions with different factorisation scales and different definitions of the
  $\alpha_{QED}$ and $\alpha_{QCD}$ were used in the matrix element calculations.
\label{T2:1d}}
\vspace{2mm} 
\begin{center}
\scriptsize
\begin{tabular}{|c|c|c|c|} \hline \hline
Factorisation scale  &  $\alpha_{QED}$, $\alpha_{QCD}$ (1L)  & 
       $\alpha_{QED}$, $\alpha_{QCD}$ (1L) &  $\alpha_{QED}$, $\alpha_{QCD}$ (2L)  \\ 
       & native AcerMC & as in {\tt PYTHIA 6.2} & as in {\tt HERWIG 6.3}  \\
\hline 
$Q^2_{QCD}~=~ M_W^2$     &  69.3 [fb]   &  69.1 [fb]   &  56.0 [fb] \\
\hline
$Q^2_{QCD}~=~ s^*_{t \bar t}$ & 40.9 [fb]    & 39.9 [fb]   & 33.9  [fb]     \\
\hline
$Q^2_{QCD}~=~ M_W^2+pT_W^2$ & 59.7 [fb]    &  59.5  [fb]    &  48.8 [fb]   \\
\hline
$Q^2_{QCD}~=~(s^*_W + s^*_{t \bar t})/2 +pT_W^2$  & 43.7[fb]   &  42.8 [fb]    &  36.0 [fb]     \\
\hline \hline 
\end{tabular} 
\end{center}
\end{table}

\boldmath
\subsection{The  $gg, q \bar q \to  Z/\gamma^* (\to \ell \ell) b \bar b$ 
processes}
\unboldmath 

The matrix elements, derived using the {\tt MADGRAPH} package \cite{Madgraph},
properly take into account spin correlations in the Z-boson decay and angular
correlations between leptons and quarks.  Thank to keeping non-zero b-quark
masses the amplitude has no singularities; the total cross-section is well
defined.

The full $Z/\gamma^*$ exchange proves to be important: For events well below the
Z-boson resonance the contribution from $\gamma^*$ becomes dominant; the
$\gamma^*$ contribution is also sizeable in the high mass tail and increases
proportionally with the effective mass of the di-lepton system.

As a benchmark result, the process has been simulated for pp collisions at
14~TeV centre-of-mass energy.  The total cross-sections, including the branching
ratio for $Z/\gamma^* \to \ell \ell$ (single flavour) are given in
Table~\ref{T2:1a} for different definitions of  $\alpha_{QED}$, $\alpha_{QCD}$ couplings.

Several differential benchmark distributions for leptons and b-quarks
originating from the hard process has been collected and discussed in
\cite{ATLCOMP013}.

\begin{table}[ht]
\newcommand{\lstrut}{{$\strut\atop\strut$}}
  \caption {\em {\bf AcerMC} production cross-sections for the $gg, q \bar q \to Z/\gamma^* b
  \bar b$ with $Z/\gamma^* \to e e$ decay (single flavour). The 14 TeV
  centre-of-mass energy, CTEQ5L parton density functions and different
  definitions for the $\alpha_{QED}$, $\alpha_{QCD}$ (as in default {\tt PYTHIA
  6.2} and {\tt HERWIG 6.3})  were used in the matrix element calculations.  The threshold
  $m_{\ell \ell}~\geq~10$~GeV was used in the event generation.
\label{T2:1a}}
\vspace{0.2cm} 
\begin{center}
\begin{tabular}{|c|c|c|c|} \hline \hline
Factorisation scale  &  $\alpha_{QED}$, $\alpha_{QCD} (1L)$ & 
 $\alpha_{QED}$, $\alpha_{QCD} (1L)$ &  $\alpha_{QED}$, $\alpha_{QCD} (2L)$  \\ 
 & native AcerMC     & as in {\tt PYTHIA 6.2} & as in {\tt HERWIG 6.3}  \\
\hline \hline 
\cline{2-3} \multicolumn{1}{|c|}{}& \multicolumn{3}{|c|}{$gg \to Z/\gamma^* b \bar b$} \\
\hline 
 $Q^2 = m_Z^2$   &  26.4 [pb]  &  26.4 [pb]   & 20.5 [pb]  \\
\hline \hline
\cline{2-3} \multicolumn{1}{|c|}{}& \multicolumn{3}{|c|}{$q \bar q \to Z/\gamma^* b \bar b$} \\
\hline 
\hline 
 $Q^2 = m_Z^2 $   &  4.3 [pb]   &  4.3 [pb]  &  3.3 [pb]   \\
\hline \hline
\end{tabular}
\end{center}
\end{table}

\boldmath
\subsection{The  $gg, q \bar q \to  Z/\gamma^* (\to \ell \ell, \nu \nu, b \bar b) t \bar t$ 
processes}
\unboldmath 
\vspace{-0.2cm}
 
This process, in spite of having a very small cross-section at LHC energies,
contributes as irreducible background to the $t \bar t H$ production at low
masses. In case the Higgs boson is searched within the $H \to b \bar b$ mode, this
contribution becomes less and less important with the Higgs boson mass moving
away from the Z-boson mass. In case of the Higgs-boson search in the invisible
decaying mode, the $Z \to \nu \nu$ might be more relevant also for the
higher masses, as the mass peak cannot be reconstructed for signal events.  The
$Z/\gamma^* \to \ell \ell$ decay is of less interest, as the expected observability
 at LHC is very low (c.f. Table \ref{T2:1b}). 
\vspace{0.2cm}
\begin{table}[ht]
\newcommand{\lstrut}{{$\strut\atop\strut$}}
  \caption {\em {\bf AcerMC} production cross-sections for the $gg, q \bar q \to Z/\gamma^* t
  \bar t$ with $Z \to \nu_e \nu_e $ decay (single flavour). The 14 TeV
  centre-of-mass energy, CTEQ5L parton density functions and different
  definitions for the $\alpha_{QED}$, $\alpha_{QCD}$ (as in native {\bf AcerMC},
  default {\tt PYTHIA 6.2} and {\tt HERWIG 6.3}) were used.  The threshold $m_{\nu_e
  \nu_e}~\geq~60$~GeV was used in the event generation.\vspace{0.5cm}
\label{T2:1b}}
\begin{center}
\begin{tabular}{|c|c|c|c|} \hline \hline
Factorisation scale  &  $\alpha_{QED}$, $\alpha_{QCD}(1L)$ &  $\alpha_{QED}$, $\alpha_{QCD}(1L)$ &  $\alpha_{QED}$, $\alpha_{QCD}(2L)$  \\ 
   & native AcerMC    & as in {\tt PYTHIA 6.2} & as in {\tt HERWIG 6.3}  \\
\hline \hline
\cline{2-4} \multicolumn{1}{|c|}{}& \multicolumn{3}{|c|}{$gg \to Z/\gamma^*(\to \nu_e \nu_e) t \bar t$} \\
\hline 
 $Q^2 = m_Z^2$   &  41.3  [pb]   & 41.3  [pb]   & 32.1  [pb]     \\
\hline \hline
\cline{2-4} \multicolumn{1}{|c|}{}& \multicolumn{3}{|c|}{$q \bar q \to Z/\gamma^*(\to \nu_e \nu_e) t \bar t$} \\
\hline  
 $Q^2 = m_Z^2 $   &  21.2 [pb]   &  21.2 [pb]    &  16.5 [pb]     \\
\hline \hline
\end{tabular}
\vspace{-0.7cm}
\end{center}
\end{table}

\boldmath
\subsection{The electroweak  $gg \to  (Z/W/\gamma^* \to) b \bar b t \bar t$ 
process}
\unboldmath 

One should be well aware, that the $gg, q \bar q \to  Z/\gamma^* t \bar t$ with 
$Z/\gamma^* \to b \bar b$ does not represent a complete electroweak production of the 
$t \bar t b \bar b$ final state. Consequently, a separate implementation for generation
of the complete set of such diagrams (including as well W-boson exchange) was addressed.
In fact this final state leads to complicated pattern of the 72 Feynman diagrams.

\begin{figure}[hb]
\begin{center}
\mbox{
     \epsfxsize=6.6cm
     \epsffile{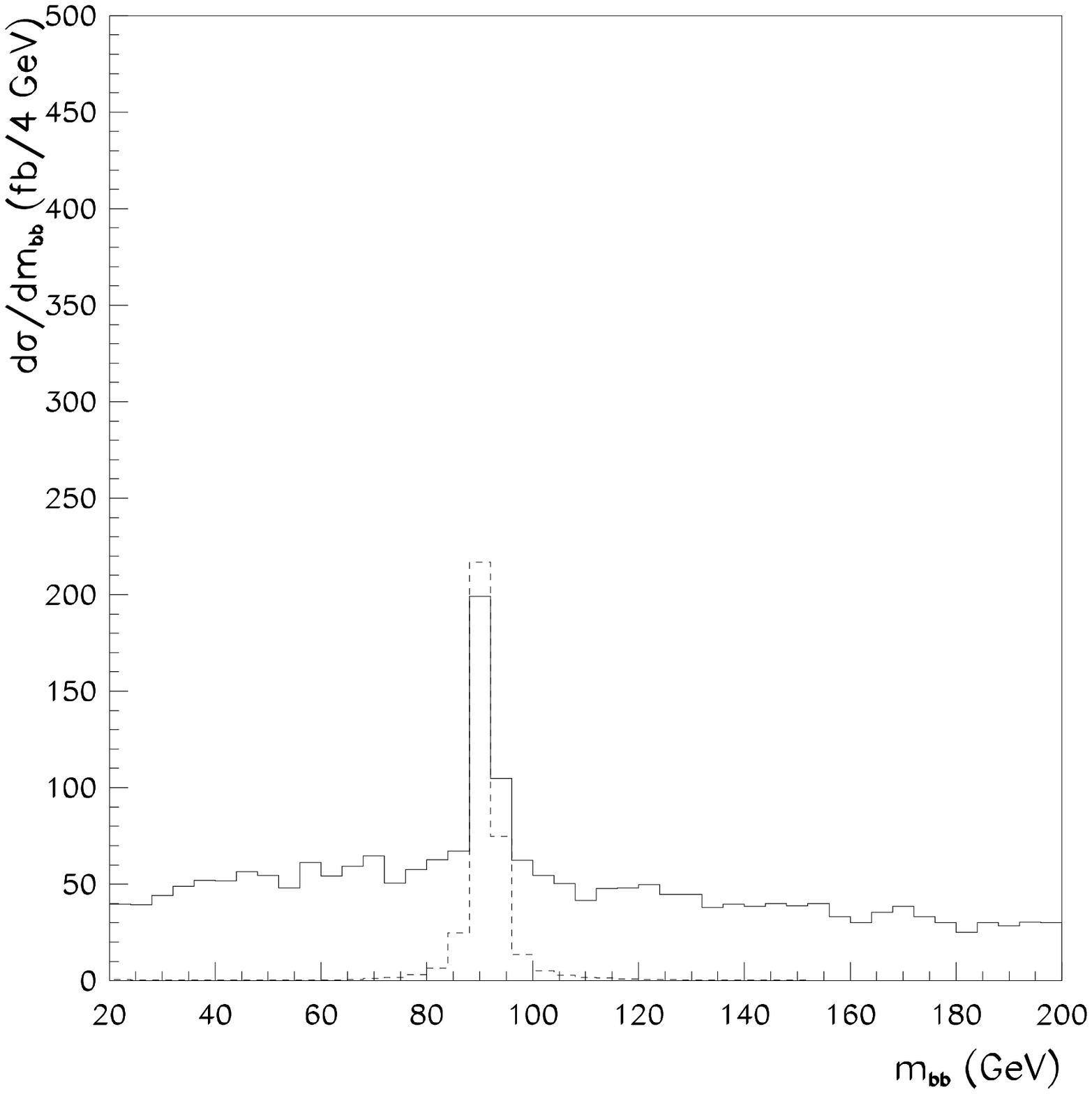}
     \hspace{-0.5cm}
     \epsfxsize=6.6cm
     \epsffile{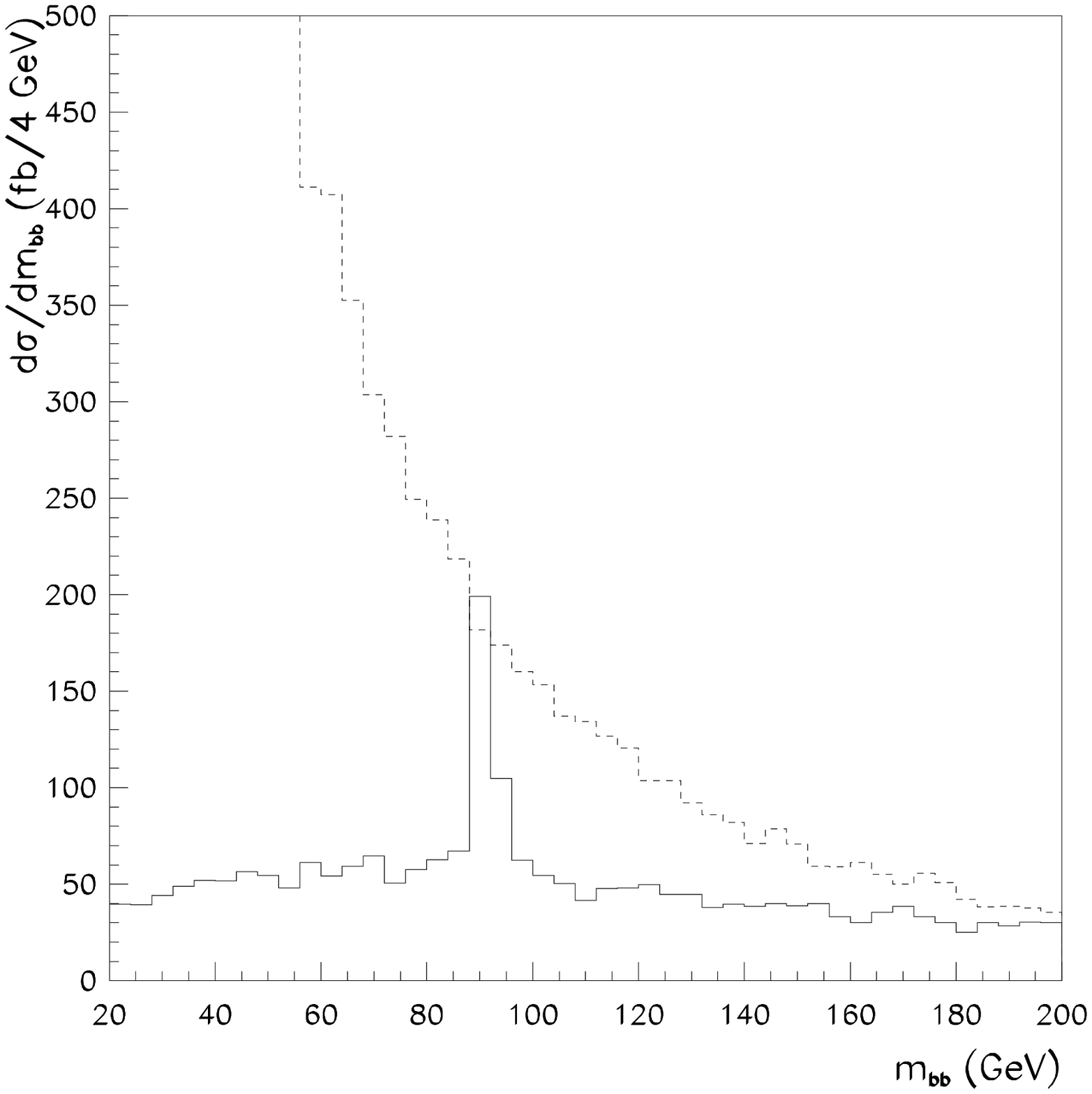}
}
\end{center}
\caption{\em
The invariant mass $m_{bb}$ distribution of the b-quark system: {\bf Left:}
EW $gg \to (Z/W/\gamma^* \to) t \bar t b \bar b$ (solid) and only resonant
$gg \to (Z/\gamma^* \to b \bar b) t \bar t $ (dashed); {\bf Right:}
EW $gg \to (Z/W/\gamma^* \to) t \bar t b \bar b$ (solid) and QCD
$gg \to  t \bar t b \bar b $  (dashed). The energy scale
$Q^2_{QCD}~=~(m_t + m_H/2)^2$ with $m_H~=~120$~GeV was used for generating QCD and 
full electro-weak production. 
\label{FS3.6:a}}
\end{figure}

The contribution from all non-resonant channels is a dominant one for the
inclusive cross-section, see Table~\ref{T2:1e}.  An almost factor 10 higher
cross-section is calculated with the full electroweak $gg \to (Z/W/\gamma^* \to)
b \bar b t \bar t$ with respect to calculated with the $gg \to (Z/\gamma^* \to b \bar b) t \bar
t$ process only.  The interesting region for the background estimates to the
Higgs searches is the one with the mass of the $b \bar b$ system around 120 GeV, see Fig.~\ref{FS3.6:a}.
This contribution was not included in results discussed in \cite{ATL-PHYS-TDR},\cite{ATL-PHYS-98-132},
so these analyses will require revisiting.
One should also note that the electroweak $gg \to (Z/W/\gamma^* \to) b \bar b t
\bar t$ inclusive cross-section is on the level of 10\% of the QCD $gg \to b
\bar b t \bar t$ cross-section, see Table~\ref{T1:1}, for the same choice of the
energy scale. But in the mass range around 120~GeV it is on the level of 50\% of 
the QCD contribution, as clearly indicated in Fig.~\ref{FS3.6:a}.

The implementation of the  EW  $q \bar q \to (Z/W/\gamma^* \to) b \bar b t
\bar t$  is still lacking but will be added in the near future. One could expected,
on the base of  cross-sections in Table~\ref{T2:1b}, that the contribution from the
quark-antiquark anihilation will be also significant.

\begin{table}[t]
\newcommand{\lstrut}{{$\strut\atop\strut$}}
  \caption {\em {\bf AcerMC} [roduction cross-sections for the electroweak $gg \to
  (Z/W/\gamma^* \to) b \bar b t \bar t$ and only for the resonant $gg \to
  (Z/\gamma^* \to b \bar b) t \bar t$.  The 14 TeV centre-of-mass energy and
  CTEQ5L parton density functions were used along with different definitions for the
  $\alpha_{QED}$, $\alpha_{QCD}$ (as in native {\bf AcerMC}, default {\tt PYTHIA 6.2}
  and {\tt HERWIG 6.3}).  The threshold $m_{b \bar b}~\geq~10$~GeV was used for
  generation in the resonant case. The $m_H~=~120$~GeV was used for caluculation
  of the energy scale.
\label{T2:1e}}
\vspace{0.5cm} 
\begin{center}
\scriptsize
\begin{tabular}{|c|c|c|c|} \hline \hline
Factorisation scale  &  $\alpha_{QED}$, $\alpha_{QCD}(1L)$ &  $\alpha_{QED}$, $\alpha_{QCD}(1L)$ &  $\alpha_{QED}$, $\alpha_{QCD}(2L)$  \\ 
   & native AcerMC    &  as in {\tt PYTHIA 6.2} & as in {\tt HERWIG 6.3}  \\
\hline \hline
\cline{2-4} \multicolumn{1}{|c|}{}& \multicolumn{3}{|c|}{$gg \to (Z/W/\gamma^* \to) b \bar b t \bar t$} \\
\hline 
$Q^2_{QCD}~=~\hat{s}$ &  0.58 [pb]     &  0.56 [pb]     & 0.71 [pb]     \\
\hline
$Q^2_{QCD}~=~\sum{({p^i_T}^2 + m_i^2)}/4$ &  1.10 [pb]     &  1.05 [pb]   & 0.84 [pb]    \\
\hline
$Q^2_{QCD}~=~\sum{({p^i_T}^2)}/4$ &  1.50 [pb]      & 1.50  [pb]    & 1.16 [pb]    \\
\hline
$Q^2_{QCD}~=~(m_t + m_H/2)^2$ &  0.90 [pb]     & 0.89 [pb]   & 0.71 [pb]     \\
\hline \hline
\cline{2-4} \multicolumn{1}{|c|}{}& \multicolumn{3}{|c|}{$gg \to (Z/\gamma^* \to b \bar b) t \bar t$} \\
\hline 
$Q^2_{QCD}~=~M_Z^2$ &  0.092 [pb]    & 0.092  [pb]    & 0.071 [pb]    \\
\hline \hline 
\end{tabular}
\end{center}
\end{table}

\boldmath 
\section{Monte Carlo algorithm} 
\unboldmath

The conceptual motivation leading to the present implementation of {\bf AcerMC}
was to exploit the possibility of dedicated matrix-element-based generation
interfaced to a more general event generator, called {\it supervising} event
generator, which is subsequently used to complete the event generation procedure.

The goal of the dedicated matrix-element-based part is to efficiently generate
complicated event topologies using native (multi-channel based) phase space
generation procedures.  The strategy is based on the understanding that a
case-by-case optimisation is in complex cases of phase space topologies
preferable to an universal algorithm.  Given that phase-space is optimised on a
case-by-case basis, an user-defined pre-selection for the generated regions of
the phase-space is not implemented. Due to the fact that the $2 \to 4$ matrix
elements, provided by the {\tt MADGRAPH/HELAS} \cite{Madgraph} package, contain
full massive treatment of the final state particles, there are no explicit
divergences present for implemented processes and {\bf AcerMC} can indeed cover
the {\it full} (kinematically allowed) phase space of the processes at hand.

The matrix-element-based part uses $\alpha_{QCD}(Q^2)$ and $\alpha_{QED}(Q^2)$
couplings and mass spectra, as calculated by the supervising event generator, to
insure the full internal consistency in treatment of the event itself.
Optionally, the native $\alpha_{QCD}(Q^2)$ and $\alpha_{QED}(Q^2)$ definitions
can also be invoked.

The generation chain is built from the following steps:

\begin{itemize}
\item
The {\tt PYTHIA 6.2} or {\tt HERWIG 6.3} interfaces to the library of the structure
functions {\tt PDFLIB 8.04} \cite{PDF} are used to calculate convolution of the partonic density.
\item
{\bf AcerMC} modules produce unweighted hard-process events with colour flow information and
pass them to the supervising  generator {\tt PYTHIA 6.2} or {\tt HERWIG 6.3} as an external
event.
\item 
The generated events are then further treated within {\tt PYTHIA 6.2} or
{\tt HERWIG 6.3} event generators, where the fragmentation and hadronisation procedures,
as well as the initial and final state radiation are added and final unweighted events are
produced.
\end{itemize}

The {\bf AcerMC} efficiency\footnote{ Note that efficiency is energy scale
dependent and phase-space optimisation is done individually for each choice.  So
it might vary for the same process but different choices of the energy scale
definition.}  for generating unweighted events, using the implementation of the
phase-space generation discussed below, is summarised in Table~\ref{T3:1}. A
certain (very small) fraction of events is further rejected in the
showering/fragmentation procedures of the supervising generators.

In the following we will briefly describe the key points of the implemented {\bf
AcerMC} modules and developed algorithms: matrix element calculations,
four-fermion phase-space generation, the issue of the s-dependent width for
resonances, unitary generation of multipheral topology, and finally, the
modification of the {\tt VEGAS} algorithm.
\begin{table}[h]
\newcommand{\lstrut}{{$\strut\atop\strut$}}
  \caption {\em Efficiency  for the generation of unweighted events with the default
definition of the energy scale (see Section \ref{s:scdef} for details).
 For generation of the $q \bar q, gg \to Z/\gamma^*(\to \ell \ell) b \bar b$  and 
 $q \bar q, gg \to Z/\gamma^*(\to \ell \ell) t \bar t$ events 
 threshold  $m_{\ell \ell}~\geq~60$~GeV has been used. For generating 
 $gg  \to (Z/W/\gamma^* \to) t \bar t b \bar b $ central energy scale
 was used. The $f=e,\mu,\tau,b$. 
\label{T3:1}}
\vspace{2mm}
\begin{center}
\begin{tabular}{|c|c|c|} \hline \hline
Process   & Description & Internal AcerMC  efficiency \\
\hline \hline
 [1] & $gg \to t \bar t b \bar b$           &  20.2 \%    \\
\hline 
 [2] & $q \bar q \to t \bar t b \bar b$           &  26.3 \%    \\
\hline 
 [3] & $q \bar q \to W(\to \nu \ell) b \bar b$   &  33.0 \%    \\
\hline 
 [4] &$q \bar q \to W(\to \nu \ell) t \bar t$   &  25.2\%    \\
\hline 
 [5] &$gg \to Z/\gamma^*(\to \ell \ell) b \bar b$ & 33.0 \%     \\
\hline 
 [6] &$q \bar q \to Z/\gamma^*(\to \ell \ell) b \bar b$ & 29.7 \%   \\
\hline 
 [7] &$gg \to Z/\gamma^*(\to f \bar f, \nu \nu) t \bar t$ & 28.2 \%     \\
\hline 
 [8] &$q \bar q \to Z/\gamma^*(\to f \bar f, \nu \nu) t \bar t$ & 34.6 \%   \\
\hline
 [9] & $gg  \to (Z/W/\gamma^* \to) t \bar t b \bar b $ & 11.2 \% \\
\hline \hline
\end{tabular}
\end{center}
\end{table}

\boldmath
\subsection{The Matrix Element Calculation}
\unboldmath

The squared matrix elements of the processes were obtained by using the {\tt
MADGRAPH/HELAS} \cite{Madgraph} package. They take properly into account the
masses and helicity contributions of final states particles, incoming quarks are
considered as massless. The particle masses, charges and coupling values that
are passed to the code derived with the {\tt MADGRAPH} package are calculated
from functions consistent with the ones used in supervising generators ({\tt
PYTHIA/HERWIG}). This allows to preserve the internal consistency of the event
generation procedure.  In particular, the (constant) coupling values of
$\alpha_s$ and $\alpha_{\rm QED}$ were replaced with the appropriate running
functions that were either taken from the interfaced generators or provided by
the {\bf AcerMC} code according to user settings.  Slightly modified {\tt
MADGRAPH/HELAS} allowed for obtaining colour flow information of the implemented
processes.

The sets of the  {\tt MADGRAPH/HELAS} coded diagrams, for each of the  implemented 
processes, are collected in Appendix A.

\boldmath
\subsection{The Four Fermion Phase Space Generation}
\unboldmath

\vspace{-0.2cm}
The four-fermion phase space for the implemented processes  was modelled
using the importance sampling technique based on the procedures implemented in
the $e^+e^-$ event generators {\tt FERMISV} \cite{fermisv}, {\tt EXCALIBUR}
\cite{excal} and {\tt NEXTCALIBUR} \cite{nextcalibur}. For each implemented
process a sequence of different kinematic diagrams ({\it channels}) modelling
the expected event topologies was constructed and the relative weights between
contributions of each sampling channel were subsequently obtained by using the
multi-channel self-optimising approach \cite{Kleiss1994}. Eventually, additional
smoothing of the phase space was obtained by using a modified {\tt VEGAS}
routine to improve the generation efficiency. 

The procedure of multi-channel importance sampling used in the event generation can
briefly be outlined as follows. An analytically integrable function
$g(\vec{\Phi})$, which aims to approximate the peaking behaviour of the differential
cross-section dependence on various kinematic quantities is introduced into the
differential cross-section equation as:
\begin{equation}
d\sigma = s(\vec{\Phi})\, d\vec{\Phi} = \frac{s(\vec{\Phi})}{g(\vec{\Phi})} \cdot
g(\vec{\Phi})\, d\vec{\Phi} = w(\vec{\Phi})\,g(\vec{\Phi})\, d\vec{\Phi},
\end{equation}
where the $d\vec{\Phi}$ denotes the (four-)particle phase space and the
$s(\vec{\Phi})$ summarises the matrix element, flux and parton density functions,
which all depend on the chosen phase space point.  The function $g(\vec{\Phi})$
has to be a normalised probability density:
\begin{equation}
\int g(\vec{\Phi}) d\vec{\Phi} = 1.
\end{equation}
\vspace{-0.2cm}
Since the peaking behaviour of $s(\vec{\Phi})$ can be very complex due to several possible
topologies introduced by a large number of contributing Feynman diagrams, 
the function $g(\vec{\Phi})$
is composed as a weighted sum of several channels $g_i(\vec{\Phi})$, each adapted to a
certain event topology:
\begin{equation}
g(\vec{\Phi})  = \sum_i \alpha_i \cdot g_i(\vec{\Phi}). 
\end{equation}
\vspace{-0.2cm}
The values of relative weights $\alpha_i$ are determined from  multi-channel
self-optimisation procedure  in order to minimise the variance of the 
weights $w(\vec{\Phi})$ \cite{Kleiss1994}. The phase space points are than sampled from
the function $g(\vec{\Phi})$, first by randomly choosing a channel $i$ according to the
relative frequencies $\alpha_i$ and then deriving the required four momenta from the
chosen $g_i(\vec{\Phi})$ using unitary\footnote{Unitary in this context meaning that 
there is no event rejection in the algorithm.} algorithms \cite{fermisv}. 

In order to have a closer look at the event generation steps, one first has to write down
the generic differential cross-section formula:
\begin{equation}
d\sigma = \sum_{a,b} f_a(x_1,Q^2) f_b(x_2,Q^2) \frac{|{\mathcal{M}}|^2}{(2 \pi)^8 (2\hat{s})} dx_1 dx_2
d\vec{\Phi},
\end{equation}
where $f_{a,b}(x,Q^2)$ represent the gluon or (anti)quark parton density functions,
$|{\mathcal{M}}|^2$ the squared matrix element divided by the flux factor $
2\hat{s}$ and $d\vec{\Phi}$ denotes the phase space differential. The quantity
$\hat{s} = x_1\, x_2\, s$ is the effective centre-of-mass energy, and the sum
$\sum_{a,b}$ runs in case of quark-antiquark incident partons over all possible
quark-antiquark combinations ($a,b = u,d,s,c,\bar{u},\bar{d},\bar{s},\bar{c}$). 
In case of $g g$ initial state the sum has only one term with $a=b=g$.

Alternatively, in order to use the {\tt PDFLIB} built-in structure
functions $x f(x,Q^2)$, it is convenient to rewrite the differential cross-section to the
form:
\begin{equation}
d\sigma = \sum_{a,b} x_1 f_a(x_1,Q^2) \; x_2 f_b(x_2,Q^2) \frac{|{\mathcal{M}_{ab}}|^2}{(2 \pi)^8 (2 s)} dy
\frac{d\tau}{\tau^2} d\vec{\Phi},
\label{e:dsig}
\end{equation}
with the two new variables given by $\tau = x_1 \cdot x_2$ and $y = 0.5
\log(x_1/x_2)$.  The matrix element used in the calculation depends explicitly
on the four-momenta of the incoming and outgoing partons:
\begin{equation}
{\mathcal{M}_{ab}} = {\mathcal{M}_{ab}}(q_1,q_2,p_1,p_2,p_3,p_4),
\end{equation}
where the $p_i$ represent the four-momenta of the final state particles and
$q_1, q_2$ the four-momenta of the incident partons. Consequently, all the 
four-momenta have to be explicitly generated.

While generating events with (anti-)quarks in the initial state, an additional
step is required to pick the flavours of the incoming pair. The selection is
again done using importance sampling, using as sampling weights:
\begin{equation}
\lambda_i = x_1 f_{q_{i}}(x_1,Q^2)\cdot  x_2 f_{\bar{q}_{i}}(x_2,Q^2) \cdot {|\mathcal{M}_{i}|^2},
\end{equation}
where the index $i$ runs over all possible quark-antiquark combinations.  When
the matrix element of the hard scattering process is for a given processes
independent of the incoming quark flavour, it is excluded from the above weights
for the sake of simplicity (note that the incoming quarks are treated as
massless so only flavour/(weak-)isospin dependent couplings can introduce the
flavour dependence of the $ {|\mathcal{M}_{i}|^2}$ term).

Only four flavours ($u,d,s,c$) of the incoming quarks are considered at the
moment, the contributions of the incoming $b$ quarks are excluded from
calculation due to the very high suppression induced by either the structure
functions and/or ${\rm CKM}$ matrix suppression.

In the event generation procedure, a generation channel is thus chosen by
weighted sampling using a (pre-determined) set of $\alpha_i$. Next, the values
of $\tau$ are sampled from a distribution:
\begin{equation}
\frac{1}{(\tau)^\mu} ~~~~~~~~ \mu \sim 1,
\end{equation}
and $y$ from the distribution
\begin{equation}
\frac{1}{\cosh(\nu\;y)} ~~~~~~~~ \nu \sim 1,
\end{equation}
as also used in {\tt PYTHIA} standard phase-space algorithm \cite{Pythia57}. 
From the two values $x_1$ and $x_2$, the
momenta $q_1,q_2$ of incoming particles and the effective centre-of-mass energy
$\sqrt{\hat{s}}$ are derived. In the following step, the four momenta of the final state
particles $p_i$ are sampled by re-parametrising the general four-body phase space:
\begin{equation}
d\vec{\Phi}=\left( \prod_{i=1}^4 d^4 p_i \delta(p_i^2 - m_i^2) \theta(p^0_i) \right) 
\delta^4(q_1+q_2 - \sum_{i=1}^4 p_i).
\end{equation}
in terms of kinematic variables that are expected to exhibit the strongest peaking
behaviour for a specific channel topology and consequently introducing appropriate
sampling functions for these variables. 

The modelling of kinematic channels relies heavily on the procedures developed
in  {\tt NEXTCALIBUR} program \cite{nextcalibur}; nevertheless, many additions
and improvements were made. Two examples of the extended/added procedures used
in {\bf AcerMC} are given below.

The detailed description of the implementations of four-momenta sampling in all
existing kinematic channels is omitted for the sake of brevity; the two examples
below should serve as a representative illustration.

Subsequently, the four-momenta constructed from the obtained set of kinematic
variables are used in the matrix element calculation. Each event is further
weighted by the appropriate phase space weight corresponding to the importance
sampling procedure and calculated using the unitary algorithms; for further
details on the applied method and unitary algorithms the reader is referred to
the original papers (e.g. \cite{fermisv,nextcalibur}).  \newline

\subsubsection{Breit-Wigner Function with s-dependent Width}

In some topologies of the processes involving $W^\pm$ or $Z^0$ bosons, a bias of the matrix
element towards large values in the high $s^*_{W/Z}$ region is evident, which in turn
means that a more accurate description of the tails of $s^*_{W/Z}$ distribution is
needed. Consequently, the Breit-Wigner sampling function was replaced by\footnote{To our
knowledge this implementation is original and done for the first time in
{\bf AcerMC}.}:
\begin{equation}
BW_s(s^*_W)=\frac{s^*_W}{(s^*_W - M_W^2)^2 + M_W^2 \Gamma_W^2},
\label{e:bwnew}
\end{equation}
which is proportional to the (more accurate) Breit-Wigner function with an $s^*_W$ dependent
width (W in the above formula denotes either a $W^\pm$ or a $Z^0$ boson).

In order to implement a unitary algorithm (an algorithm that produces a result for every
trial, i.e. there is no rejection) of value sampling on the above function one first has
to calculate the normalisation integral (cumulant) and then its inverse
function. Introducing a new variable $\eta = (s^*_W-M_W^2)/(M_W\;\Gamma_W)$ the integral
of the above function can be expressed as:
\begin{equation}
\int BW_s(s^*_W) \; ds^*_W  = \int \biggl\{ \frac{M_W^2}{M_W \Gamma_W} \cdot
\frac{1}{1+\eta^2} + \frac{\eta}{1+\eta^2} \biggr\} \; d\eta,
\label{e:bws}
\end{equation}
where the upper integral limit is left as a free parameter. The integral thus
gives a function:
\begin{equation}
F(\eta)=  \biggl\{ \frac{M_W^2}{M_W \Gamma_W} \cdot {\rm atan}(\eta) \biggr\}  + 
\biggl\{ \frac{1}{2} \cdot \log(\eta^2+1) \biggr\}, =
F_1(\eta) + F_2(\eta)
\label{e:bwint}
\end{equation}
with $F(\eta_{\rm max})-F(\eta_{\rm min})$ defining the normalisation. One of
the undesirable features is that the function $F(\eta)$ does not have a (simple)
analytical inverse, which is a prerequisite for unitary sampling. Taking a
closer look at the two above expressions one can quickly spot another
undesirable feature, namely that the second term  in the Equation
\ref{e:bws} is an odd function of $\eta$, which after the integration gives an
even term $F_2(\eta)$ in $\eta$ in Equation \ref{e:bwint}. In other words the
second term alone is neither a non-negative function nor does it have an unique
inverse - one has to deal with a {\it negative probability}. 
A reasonably elegant solution to this problem has been developed and implemented here:
\begin{itemize}
\item One samples values of $\eta$ by using only the first term of the above
expressions (the {\it usual} Breit-Wigner function). 
\item One then re-samples the obtained value of $\eta$ using the full expression of
Equation \ref{e:bws}: If $\eta$ is less than zero the value is mapped to $-\eta$
with the probability given by Equation \ref{e:bws}.
\end{itemize}

\begin{figure}[ht]
\vspace{-0.6cm}
\begin{center}
\mbox{
     \epsfxsize=7cm
     \epsffile{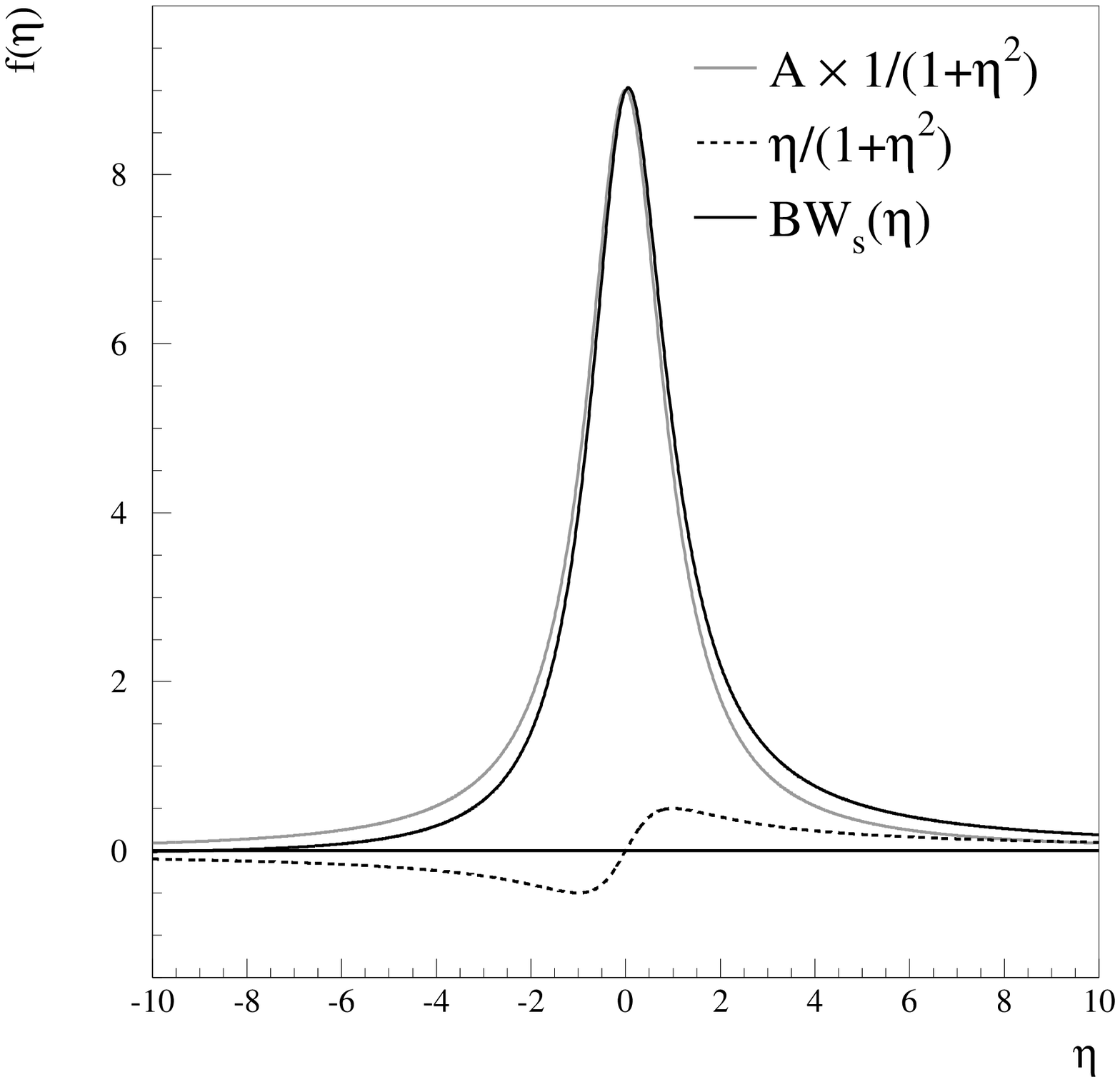} 
     \epsfxsize=7cm
     \epsffile{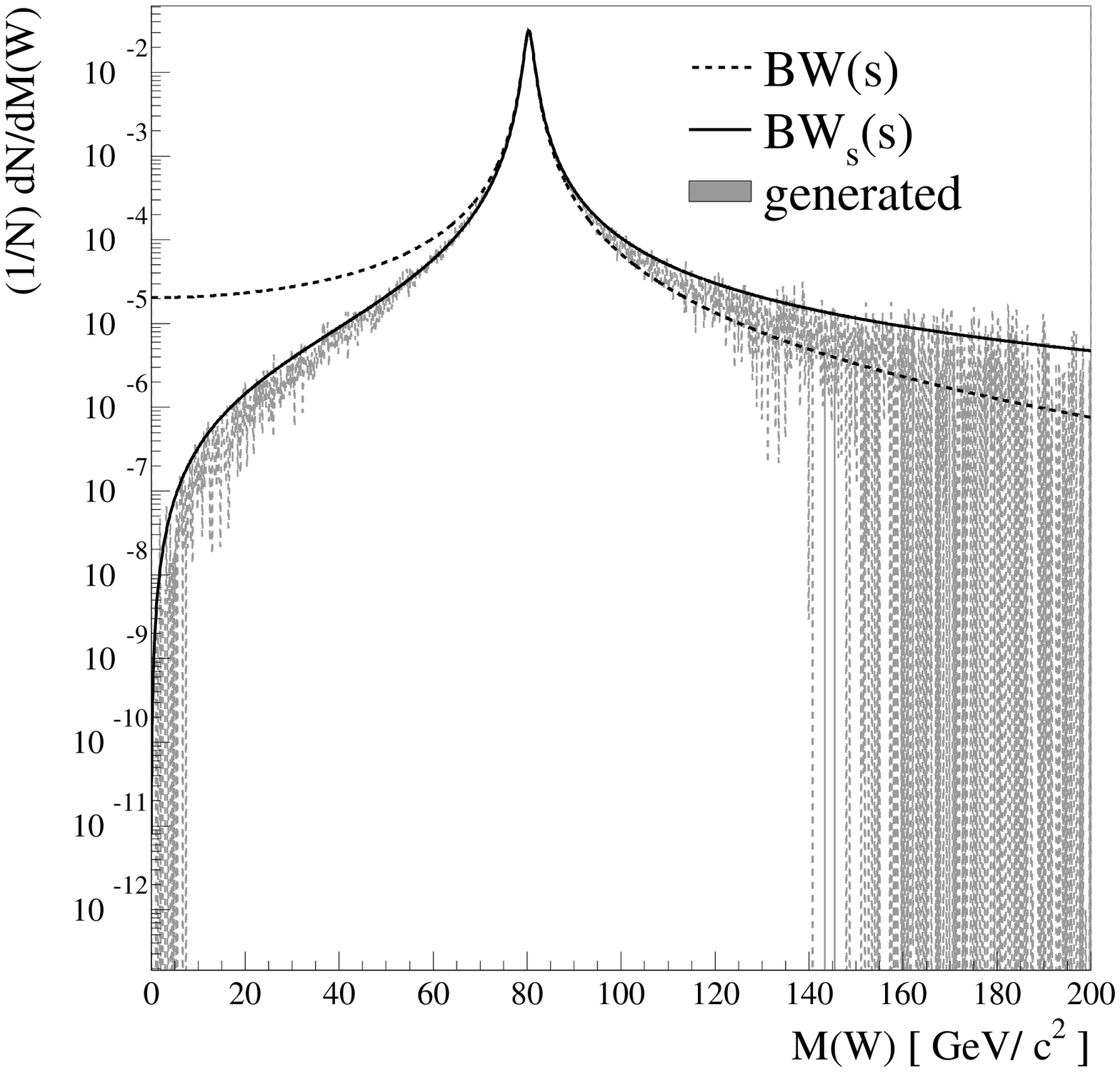} 
}
\end{center}
\vspace{-0.6cm}
\caption{\em {\bf Left} Comparisons of the two functional terms of Eq. \ref{e:bws} to 
${\rm BW}_s(\eta)$ given by Equation \ref{e:bwnew}. Note that the scaling factor $A$ is
chosen in view of making the contributions more transparent; it is much too small compared
to the real case of $W^\pm/Z^0$ bosons. \newline {\bf Right} Comparison of the (normalised)
distributions of differential cross-section for the process $q \bar q \to W b \bar b$
(dashed) and sampling functions (solid line) with respect to the variables obtained by
importance sampling, as described in the text.
\label{f:bwcomp}}
\vspace{0.3cm}
\end{figure}

Why this works can quickly be deduced by looking at the Figure \ref{f:bwcomp}:
At negative values of $\eta$ the second term of Equation \ref{e:bws} gives a
{\it negative probability} in the region $\eta<0$, i.e. using a simple
Breit-Wigner (Cauchy) probability function too many events are generated in this
region. Correspondingly, since the  second term of Eq. \ref{e:bws} is an odd
function, exactly the same fraction (distribution) of events is {\it missing} in the
region $\eta>0$. By mapping events with $\eta<0$ over the $\eta=0$ axis one thus
solves both problems at the same time. Using the above re-sampling procedure the
whole approach remains unitary, i.e. no events are rejected when there are no
limits set on the value of $\eta$ or they are symmetric $|\eta_{\rm min}|=\eta_{\rm
max}$. In the contrary case, a small fraction of sampling values is rejected.

After some calculation the whole unitary procedure can thus be listed as follows:
\begin{itemize}
\item Calculate the kinematic limits $\eta_{\rm min}$ and $\eta_{\rm max}$.
\item Calculate the {\it normalisation} factors $\Delta_1=F_1(\eta_{\rm
max})-F_1(\eta_{\rm min})$, $\Delta_2=F_2(\eta_{\rm max})-F_2(\eta_{\rm min})$
and $\Delta_s = \Delta_1 + \Delta_2$; the term $\Delta_2$ can actually be
negative and thus does not represent proper normalisation.

\item Obtain a (pseudo-)random number $\rho_1$. 
\item If $\rho_1 \leq \Delta_2/\Delta_s$ then:
\begin{itemize}
\item Obtain a (pseudo-)random number $\rho_2$;
\item Construct $\eta$ as:
\begin{eqnarray*}
X & = & \Delta_2 \cdot \rho_2 + F_2(\eta_{\rm min}),\\
\eta & = & \sqrt(e^{2X}-1),
\end{eqnarray*}
which is the inverse of the (normalised) cumulant $(F_2(\eta)-F_2(\eta_{\rm
min}))/\Delta_2$.
\item Note that the condition $\rho_1 \leq \Delta_2/\Delta_s$ can be fulfilled
only if $\Delta_2 \geq 0$, which means that $\eta_{\rm max}$ is positive and
greater than $\eta_{\rm min}$.
\end{itemize}
\item Conversely, if $\rho_1 > \Delta_2/\Delta_s$ then:
\begin{itemize}
\item Obtain a (pseudo-)random number $\rho_2$;
\item Construct $\eta$ as:
\begin{eqnarray*}
X & = & \Delta_1 \cdot \rho_2 + F_1(\eta_{\rm min}),\\
\eta & = & \tan(\frac{M_W \Gamma_W}{M_W^2}\cdot X)
\end{eqnarray*}
which is the inverse of the (normalised) cumulant $(F_1(\eta)-F_1(\eta_{\rm
min}))/\Delta_1$.
\item If the obtained $\eta$ is less than zero then calculate the normalised
probability densities:
\begin{eqnarray*}
P_1 & = & \frac{1}{\Delta_1} \cdot \{\frac{M_W^2}{M_W \Gamma_W} \cdot
\frac{1}{1+\eta^2} \} \\
P_s & = &  \frac{1}{\Delta_s} \cdot \{\frac{M_W^2}{M_W \Gamma_W} \cdot
\frac{1}{1+\eta^2} + \frac{\eta}{1+\eta^2} \}
\end{eqnarray*}
\item Obtain a (pseudo-)random number $\rho_3$;
\item If $\rho_3 > P_s/P_1$ map $\eta \to -\eta$.
\item If the new $\eta$ falls outside the kinematic limits $[\eta_{\rm
min},\eta_{\rm max}]$ the event is rejected.
\item Note also that the last mapping can only occur if the original $\eta$ was
negative, since $P_s < P_1$ only in the region $\eta < 0$. 
\end{itemize}
\item Calculate the value of $s^*_W$ using the inverse of $\eta$ definition:
\begin{equation}
s^*_W = (M_W\;\Gamma_W) \cdot \eta + M_W^2
\end{equation}
The weight corresponding to the sampled value $\eta$ is exactly:
\begin{equation}
\Delta_s \cdot \frac{(s^*_W - M_W^2)^2 + M_W^2 \Gamma_W^2}{s^*_W},
\end{equation}
which is the (normalised) inverse of Equation \ref{e:bwnew} as requested.
\end{itemize}
As it turns out in subsequent generator level studies, this generation procedure
gives much better agreement with the  differential distributions than the
{\it usual} (width independent) Breit-Wigner; an example obtained for the  $q \bar q \to W b
\bar b$ process is shown in Figure \ref{f:bwcomp}. The evident consequence is
that the unweighting efficiency is substantially improved due to the reduction
of the event weights in the high $s^*_W$ region.

\subsubsection{Unitary Generation of Multipheral Topology}  

The procedure implemented in {\bf AcerMC} for generating events with
multipheral topology is an extension of the three-body phase-space
sampling, developed in  \cite{nextcalibur} for the phase space 
generation in {\tt NEXTCALIBUR} program (kinematic
channel {\tt MULTI1}, Appendix A). In order to describe the features
implemented in the extension of that procedure, let us recapitulate
first its basic principles: The procedure used in {\tt MULTI1} starts with
splitting the four-body phase space into a three-body times a two-body
decay:
\begin{equation}
\int d\Phi_4 = \int ds_{34}\; I_3(p_1,p_2,p_{34}) \; I_2(p_3,p_4),
\label{e:phi4}
\end{equation}
where the integral $I_3$ represents the three-body and $I_2$ the two-body phase
space, the labels $p_i, \; i=1,4$ represent the four-momenta of the produced
particles with $p_{34} = p_3 + p_4$ and $s_{34} = p_{34}^2$. The key issue
thus becomes parametrisation of the (more difficult) three body phase-space
integral $I_3(P_1,P_2,P_3)$:
\begin{equation}
I_3(P_1,P_2,P_3) = \int \prod_{i=1}^3 d^4 P_i \delta(P_i^2 - M_i^2)\theta(E_i). 
\label{e:i3}
\end{equation}
It should be stressed that  the masses $M_i$ in the above on-shell constraint do
not necessarily correspond to physical masses of the elementary particles,
can as well be values derived by a pre-sampling procedure (e.g. $M_3^2 = s_{34}$ in
the above example, $M_3$ representing the off-shell mass of a gluon
decaying into $p_3$ and $p_4$). Defining the reduced energies $x$ and $y$ as:
\begin{equation}
E_1=\frac{\sqrt{s}}{2}\; x, ~~~~~~~~~~~~~ E_2=\frac{\sqrt{s}}{2}\; y,
\end{equation}
where the $s$ is the square of centre-of-mass energy of the system, and setting 
$\mu_i = M_i^2/s$ one can re-parametrise the integral $I_3(P_1,P_2,P_3)$ as:
\begin{equation}
I_3(P_1,P_2,P_3) = \frac{s}{32} \int_{2\sqrt{\mu_1}}^{x_{-}} dx \int dy\, d\Omega_1\,
d\Omega_2\, \delta[F(x,y)].
\end{equation}
The integral is now parametrised with the energies and solid angles of
two particles with a constraint in form of a $\delta[F(x,y)]$ due to the presence
of the third particle. The integration limits on $x$ can be easily derived from
the Dalitz constraints, with $x_{-}$ given by:
\begin{equation}
x_{-}=(1+\mu_1) - (\sqrt{\mu_2} + \sqrt{\mu_3})^2.
\end{equation}
The delta-function $\delta[F(x,y)]$ is actually a constraint on the value of the
cosine of the angle between the particles $1$ and $2$:
\begin{equation}
F(x,y)=C(x,y)-c_{12},
\label{e:c12}
\end{equation}
the full expression for $C(x,y)$ is given in \cite{nextcalibur}. A short calculation
shows that, given values $x \in [2\sqrt{\mu_1},x_{-}]$ and $c_{12} \in [-1,1]$,
 the constraint of Equation \ref{e:c12} gives either two, one or no solutions
for $y$, i.e. the procedure of event generation can involve some event rejection and is
thus not necessarily unitary.

\begin{figure}[hb]
\begin{center}
\mbox{
     \epsfxsize=5.cm
     \epsffile{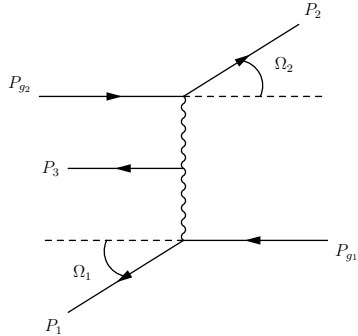}
}
\end{center}
\vspace{-0.2cm}
\caption{\em
A diagrammatic representation of a three-body decay as modelled by the procedure described in 
the text. 
\label{f:3fdec}}
\vspace{0.5cm}
\end{figure}

In the original {\tt MULTI1} channel implementation, the event generation
uses the sampling sequence:
\begin{itemize}
\item  Sample the value of $x$.
\item Calculate the value of $\tilde{y}$ assuming $c_{12}=-1$ (which is favoured by
multipheral singularity).
\item Use the $(x,\tilde{y})$ to sample the four values $\cos(\theta_{1,2}),\phi_{1,2}$.
\item From the angles calculate $c_{12} = \cos(\theta_{12})$ and hence determine the
possible values of $y$ using Equation \ref{e:c12}. If there is no solution for
$y$, the event is rejected. 
\end{itemize}

As it turns out, the rejection rate, very low in cases of processed where it was
applied  originally  \cite{nextcalibur},   increases  dramatically  for the QCD
process where  heavy quarks are involved.   Therefore, the approach was extended
to an unitary one (albeit at the cost of computation time).

The developed extension is based on a short calculation, which shows the
existence of a turning point (maximum) at $y_T=y_T(x)$ in the $c_{12}(y) =
C(x,y)$ function, i.e. there is a value of $x$ where $c_{12}(y_T)$ reaches its
maximal value. At given values of $x$ and $c_{12}$, the number of solutions for
$y$ thus depends on the location and value of the extremum
$c_T=c_{12}(y_T)$. Basing on these deductions, a new approach was developed:
\begin{itemize}
\item  Sample the value of $x$ from a distribution:
\begin{equation}
\frac{1}{(1 - x)^\nu} ~~~~~~~~ \nu \sim 1,
\label{e:xe}
\end{equation}
\item Sample the angles $\phi_1 \in [0,2\pi]$ from a flat distribution and $\theta_1$ from:
\begin{equation}
\frac{1}{a_1 - \cos(\theta_1)} ~~~~~~~~ a_1=\frac{2 E_1
E_{g_1}-s_1-s_{g_1}+s_3}{2|\vec{P_1}| |\vec{P_{g_1}}|},
\label{e:cth1} 
\end{equation}
where the index ${g_1}$ represents the incoming particle that splits into
the t-channel exchange and the $s_i$ terms correspond to $s_i=M_i^2$ from Equation \ref{e:i3} 
(remember that $E_1=\sqrt{s}/2 \cdot x$).
\item Calculate the values of kinematic limits on $[y_{\rm min}(x),y_{\rm max}(x)]$
and the values $y_T(x)$. If the $y_T$ is within kinematic limits $[y_{\rm
min}(x),y_{\rm max}(x)]$, calculate the maximum limit $c_T$ and hence the limits on
$\phi_2$ and $\theta_2$. 
\item Calculate the value of $\tilde{y}$ assuming $c_{12}=-1$ (which is always allowed).
\item Sample the value of $\theta_2$ within the calculated limits from a
distribution analogous to Equation \ref{e:cth1} using $\tilde{y}$.
\item Get the value of $\phi_2$ from a flat distribution within the allowed
limits. 
\item From the generated angles calculate $c_{12} = \cos(\theta_{12})$ and
 hence determine the
possible values of $y$ using Equation \ref{e:c12}. If the calculations were done
correctly, there should always be at least one solution.
\end{itemize}
The modified procedure is thus an unitary one, giving the four-vectors
$P_1,P_2,P_3$.
\vspace{0.3cm}

Going back to the initial issue of the four body phase space (Eq. \ref{e:phi4}),
the whole sampling procedure for such kinematic channels is as follows:
\begin{itemize}
\item The squared invariant mass $s_{34}$ is sampled from:
\begin{equation}
\frac{1}{(s_{34})^\mu}.
\end{equation}
The value of $\mu$ depends on the channel implementation (virtual gluon, a
t-channel object, masses of the particles $3$ and $4$). 
\item The described above three-body decay procedure is implemented.
\item The virtual $P_{34}$ object is decayed according to the actual channel (a
t-channel angular dependence, isotropic decay of a virtual gluon).
\end{itemize}

\subsubsection{Modified {\tt VEGAS} Algorithm \label{s:ac-veg}}

Using the described multi-channel approach, the total generation (unweighting)
efficiency amounts to about $3-10\%$ depending on the complexity of the chosen process.
 In order to further improve the efficiency, a set of modified {\tt VEGAS}
\cite{vegas} routines was used as a (pseudo-)random number generator for
sampling the peaking quantities in each kinematic channel. The conversion into a
{\it (pseudo-)random number generator} consisted of re-writing the calling routines
so that instead of passing the analysed function to {\tt VEGAS} for sampling and
integration, {\tt VEGAS} calls produce only (weighted) random numbers in the region
$[0,1]$ and the corresponding sampling weight, while the {\tt VEGAS} grid training is
done using a separate set of calls.

\begin{figure}[hb]
\begin{center}
\mbox{
     \epsfxsize=7cm
     \epsffile{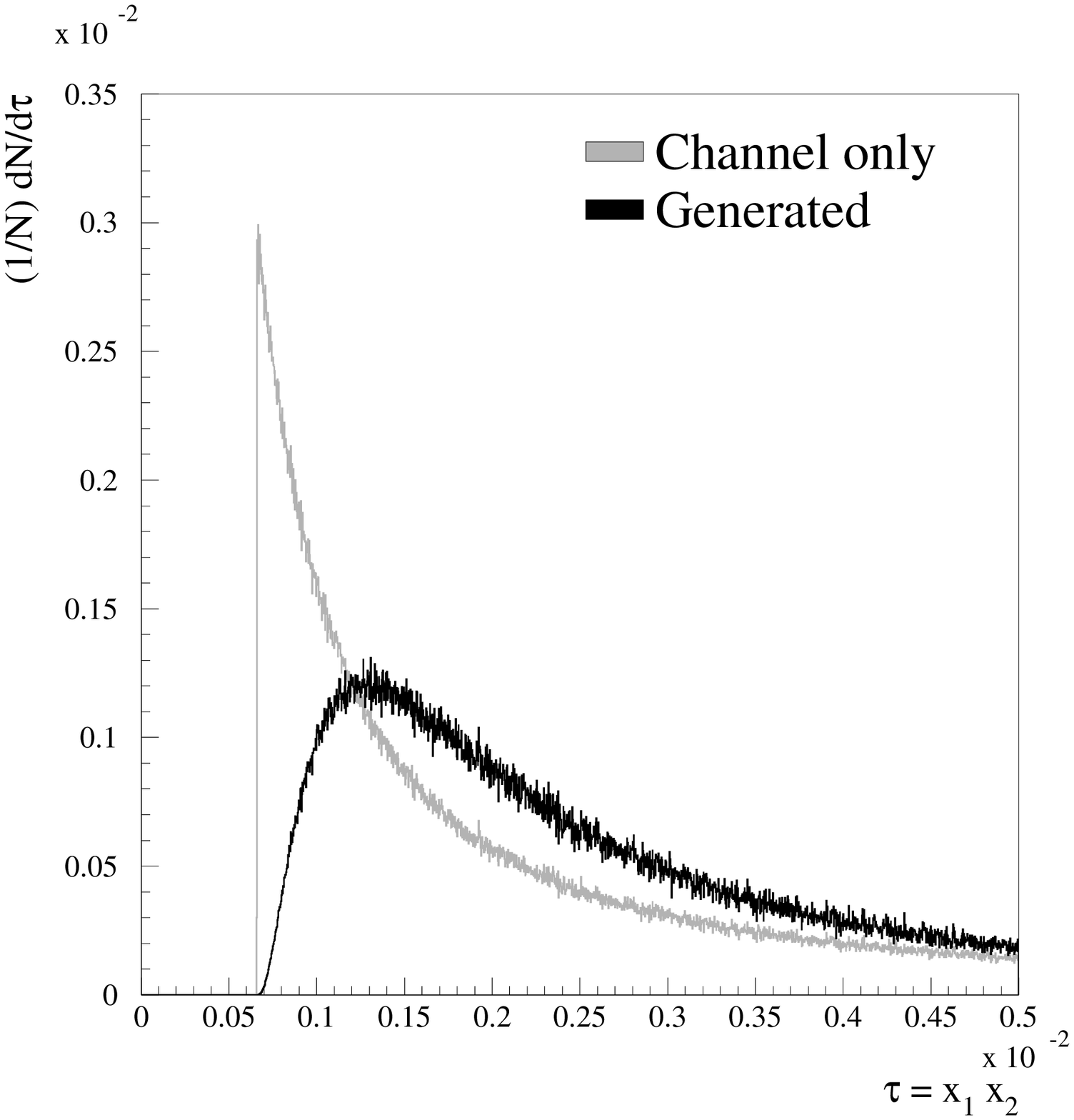}
     \epsfxsize=7cm
     \epsffile{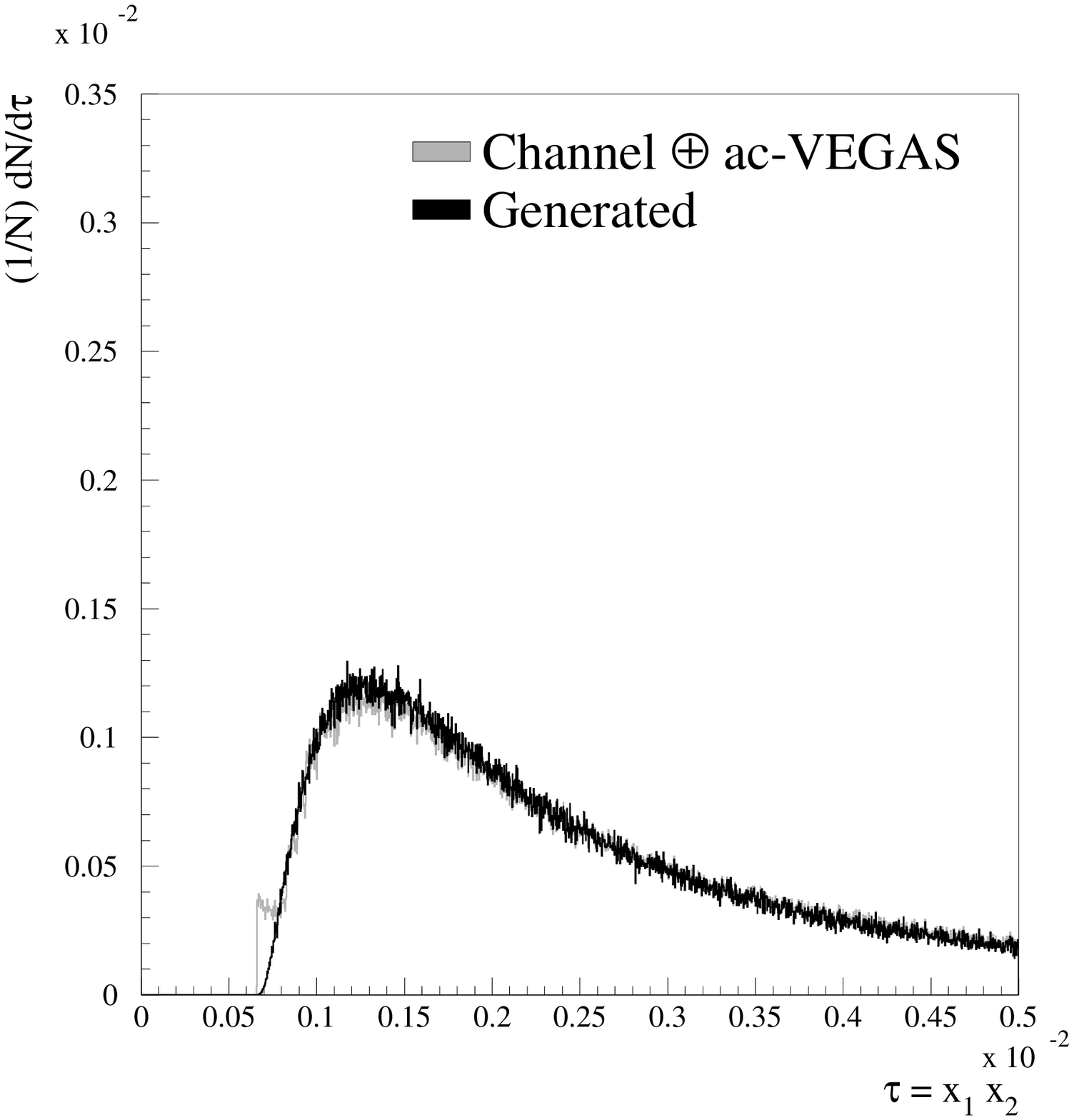}
}
\end{center}
\vspace{-0.4cm}
\caption{\em
Comparison between the sampling distribution for the $\tau = \hat{s}/s
\in [\tau_{\rm min},1]$ variable in $g g -> t \bar{t} b \bar{b}$
process before and after the application of modified {\tt ac-VEGAS}
\cite{vegas} smoothing procedure (light gray histogram). The generated (normalised) 
differential cross-section is also drawn (black histogram, labelled {\sf Generated}).
\label{f:tauveg}}
\end{figure}

After training all the sampling grids (of dimensions 4-7, depending on the
kinematic channel), the generation efficiency increased to the order of
$6-14\%$. The motivation for this approach was that in unitary algorithms
only a very finite set of simple sampling functions is available, since the
functions have to have simple analytic integrals for which an inverse function
also exists. Consequently, the non-trivial kinematic distributions can not be
adequately described by simple functions at hand in the whole sampling domain
(e.g. the $\tau$ distribution, c.f. Figure \ref{f:tauveg}) and some additional
smoothing might be welcome. In addition, the random numbers distributions
should, due to the applied importance sampling, have a reasonably flat behaviour
to be approached by an adaptive algorithm such as {\tt VEGAS}\footnote{At this
point also a disadvantage of using the adaptive algorithms of the {\tt VEGAS}
type should be stressed, namely that these are burdened with the need of
training them on usually very large samples of events before committing them to
event generation.}. 

The further modification of {\tt VEGAS}, beside adapting it to function as a
(pseudo-)random number generator instead of the usual {\it integrator}, was based on
the discussions \cite{ohl,foam} that in case of event generation,
i.e. unweighting of events to the weight one, reducing the maximal value of
event weights is in principle of higher importance than achieving the minimal
weight variance. Since the {\tt VEGAS} algorithm was developed with the latter
scope, some modification of the algorithm was necessary. As it turned out, the
modification was fairly easy to implement: Instead of the usual cumulants:
\begin{equation}
<I>_{\rm cell} ~= \sum_{\rm cell} {\rm wt}_i,
\end{equation}
according to the size of which {\tt VEGAS} decides to split its cells, the values:
\begin{equation}
<F>_{\rm cell} ~= \Delta_{\rm cell} \cdot {\rm wt}^{\rm max}_{\rm cell} \;- 
\sum_{\rm cell} {\rm wt}_i,
\end{equation}
were collected and used as the splitting criterion. The above value
(called {\it loss integral} in \cite{foam}) is basically a measure of the deviation
between the maximal weight sampled in the given cell ${\rm wt}^{\rm max}_{\rm
cell}$ and the average weight in the cell $<\rm wt_{\rm cell}> = (\sum_{\rm cell} {\rm
wt}_i)/\Delta_{\rm cell}$ (the quantity $\Delta_{\rm cell}$ denoting the cell
width, i.e. the integration range). Re-writing the above expression as:
\begin{equation}
<F>_{\rm cell} ~= \bigl( \Delta_{\rm cell} \cdot {\rm wt}^{\rm max}_{\rm cell}
\bigr)
\cdot \biggl\{ 1 - \frac{<\rm wt_{\rm cell}>}{{\rm wt}^{\rm max}_{\rm cell}} \biggr\}
\end{equation}
clearly indicates that the value $<F>_{\rm cell}$ is actually a measure of the
generation {\it inefficiency} in the cell, since the term in the curly brackets is
equivalent to one minus the generation efficiency $<\rm wt_{\rm cell}>/{\rm
wt}^{\rm max}_{\rm cell}$. In addition, the inefficiency is weighted with the
{\it crude}/maximal estimation of the function integral over the cell $\Delta_{\rm
cell} \cdot {\rm wt}^{\rm max}_{\rm cell}$ and cells with the highest $ <F>_{\rm
cell}$ are split. 

This method is of relevance because the {\tt VEGAS} cells are actually
projections of the whole phase space on the (chosen) side axes, i.e. {\tt VEGAS}
cannot isolate a maximal weight in a certain point in phase-space and build a
cell around it, which in principle would be an ideal solution.
An implementation with this scope in view has been made in {\tt FOAM}
\cite{foam}, nevertheless we have not found it competitive with respect to
the modified {\tt VEGAS} for the given application.                                                          

The thus modified {\tt ac-VEGAS} algorithm further increased the unweighting
efficiency for almost a factor of two.

One of the sampling distributions is shown in Fig.~\ref{f:tauveg} as a gray
histogram (marked {\it channel}) and the actual ({\it generated})
differential cross-section dependence is drawn in black. In the first figure,
the random variable used for sampling values from $1/\tau^\mu$ distribution was
drawn from a flat probability in the interval $[0,1]$; in the second plot the
{\tt ac-VEGAS} algorithm was used to give an optimal grid for sampling the random
variables needed for parameter generation (the grid is trained for each
kinematic channel separately, the sum of all channels is shown in the plot). The
improvement is evident; one has to stress that the use of {\tt ac-VEGAS} algorithm
to generate the values of $\tau$ directly would be much less efficient since
{\tt VEGAS} gives a grid of 50 bins/dimension, which would give a very crude
description of the $\tau$ distribution compared to the one at hand.
\begin{figure}[ht]
\begin{center}
\mbox{
     \epsfxsize=8cm
     \epsffile{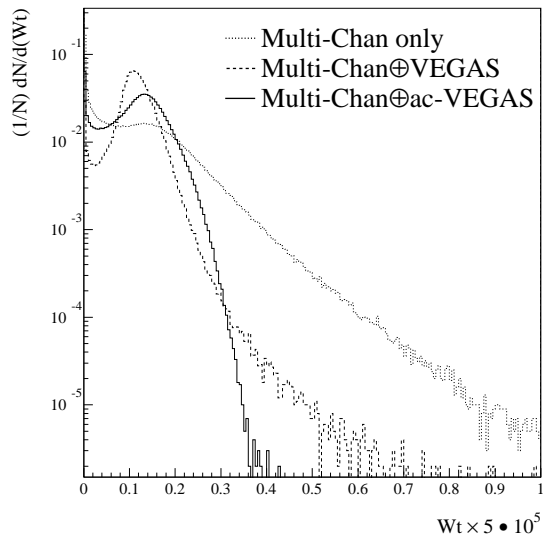}
}
\end{center}
\vspace{-0.5cm}
\caption{\em
The distribution of event weights using only the Multi-Channel approach (dotted histogram)
 and after application of {\tt VEGAS} (dashed histogram) and {\tt ac-VEGAS} (full histogram) 
algorithms in the $ g g \to (Z^0 \to) l \bar{l} b \bar{b}$ process. 
\label{f:weights}}
\end{figure}

Observing the distributions of the event weights before and after the inclusion
of the modified {\tt ac-VEGAS} algorithm (Fig.~\ref{f:weights}) it is evident
that {\tt ac-VEGAS} quite efficiently clusters the weights at lower values.
Note that the principal effect of original {\tt VEGAS} is indeed to cluster event
weights in a narrow region, nevertheless a tail towards the high-weight region
remains. On the other hand, the {\tt ac-VEGAS} efficiently reduces the tail in
the high weight region; only a few of the event weights still retain their large
values, thus reducing the generation efficiency.  Given the difference in
distributions, the observed increase of the generation efficiency seems
relatively modest. To better understand this result one should consider that the
formula for the MC generation efficiency is given by:
\begin{equation}
\epsilon = \frac{<{\rm wt}>}{\rm wt_{\rm max}},
\label{e:cleff}
\end{equation}
where $<{\rm wt}>$ is the average weight of the sample and equals the total
event cross-section, while $\rm wt_{\rm max}$ represents the maximum event weight in
the applied generation procedure and is determined through a pre-sampling run
with a high statistic.  Since the average weight $<{\rm wt}>$ equals the total
cross-section of the process, it remains (necessarily) unchanged after the
application of the {\tt VEGAS} refining; consequently the change of efficiency
results in the reduction of the maximum weight $\rm wt_{\rm max}$ by approximately a
factor two, which is from technical point of view quite an achievement.

A further step to profit from the clustering of weights induced by {\tt ac-VEGAS}
is to adopt a re-definition of the MC generation efficiency as proposed by
\cite{bhlumi,foam}. In this approach, the alternative definition of $\rm wt_{\rm
max}$ is: For a given precision level $\alpha << 1$, the $\rm wt_{\rm max}$ is
determined from the total weight distribution in such a way that the
contribution of the events exceeding this value to the total weight sum
(i.e. cross-section integral) equals $\alpha$. Such a quantity is referred to as
$\rm wt^\alpha_{\rm max}$ and the efficiency expression becomes:
\begin{equation}
\epsilon = \frac{<{\rm wt}>}{\rm wt^\alpha_{\rm max}}.
\label{e:neweff}
\end{equation}

The argument presented in \cite{bhlumi,foam} seems to be quite reasonable since
the {\it true} event weight is in any case only estimated from a finite sample of
events and the new definition simply takes into account a certain level of
accuracy in the maximum weight determination. In addition, certain very weak
singularities that might exist in the simulated process and might occasionally
result in a very high event weight are automatically taken into account.  The
use of new $\rm wt^\alpha_{\rm max}$ consequently results in a generation efficiency
of about $\epsilon \geq 20\%$ for all the implemented processes, which is a
significant improvement in terms of time needed for MC generation.

\boldmath
\subsubsection{\label{s:cols}Colour Flow Information}
\unboldmath

Before the generated events are passed to {\tt PYTHIA/HERWIG} to complete the
event generation, additional information on the colour flow/connection of the
event has to be defined.  Below we discuss the implemented method of the colour
flow determination on the example of two processes, $ g g \to t \bar{t} b \bar{b}$ 
and $ q \bar{q} \to t \bar{t} b \bar{b}$.

\begin{figure}[ht]
\vspace{-0.3cm}
\begin{center}
\mbox{
     \epsfxsize=10cm
     \epsffile{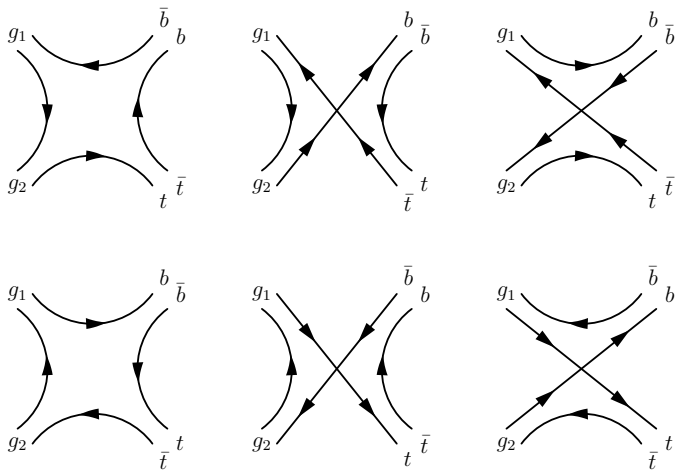}
}
\end{center}
\vspace{-0.3cm}
\caption{\em 
A diagrammatic representation of the six colour flow configurations in the process
$g g \to t \bar{t} b \bar{b}$. Certain colour combinations, leading for example to 
colourless (intermediate) gluons, are not allowed.
\label{f:cols}}
\vspace{-0.5cm}
\end{figure}

For the process $ g g \to t \bar{t} b \bar{b}$ six colour flow configurations are
possible, as shown in Figure \ref{f:cols}. With 36 Feynman diagrams contributing to the
process and at least half of them participating in two or more colour flow configurations,
calculations by hand would prove to be very tedious. Consequently, a slightly modified
colour matrix summation procedure from {\tt MADGRAPH} \cite{Madgraph} was used to
determine the colour flow combinations of the diagrams and the corresponding colour
factors. The thus derived squared matrix elements for separate colour flow combinations
$|{\mathcal{M}_{\rm flow}}|^2$ were used as sampling weights on an event-by-event basis to
decide on a colour flow configuration of the event before passing it on to {\tt
PYTHIA/HERWIG} for showering and fragmentation. The procedure was verified to give
identical results regarding the colour flow combinations and corresponding colour factors
when applied to the processes published in \cite{cflows}. As one can see this approach
neglects the interference terms between the distinct colour-ordered amplitudes and is
indeed exact only in the $N_C \to \infty \;$ limit\footnote{The matrix elements used in
the cross-section calculation and event generation are of course complete and do not
employ any approximation.} \cite{mangano_cflow,alpha}. Since there is no {\it a priori}
rule of how to split the interference terms between the colour ordered amplitudes this
approach is generally deemed to be the best one can do; recent developments in this field
\cite{odagiri} however suggest additional improvements to the method that indeed might be
incorporated into later versions of {\bf AcerMC}.

\begin{figure}[ht]
\vspace{-0.4cm}
\begin{center}
\mbox{
     \epsfxsize=8cm
     \epsffile{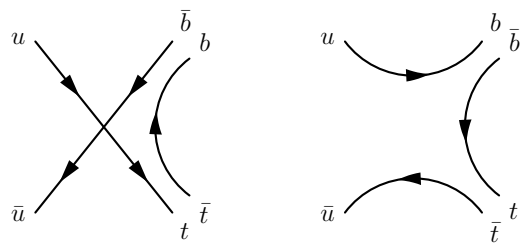}
}
\end{center}
\vspace{-0.3cm}
\caption{\em 
A diagrammatic representation of the two colour flow configurations in the process
$q \bar{q} \to t \bar{t} b \bar{b}$.
\label{f:colsqq}}
\end{figure}

The colour flow configuration in the $ q \bar{q} \to t \bar{t} b \bar{b}$ channel is much
simpler since only two colour flow topologies exist (Fig. \ref{f:colsqq}); the choice
between the two has been solved in a manner identical to the one for the $ g g \to t
\bar{t} b \bar{b}$ process, as described above.

Some specifications of the implemented matrix-element-based processes: number of
Feynman diagrams, channels used in the phase-space generation and colour flow
configurations are collected in Table~\ref{t:acmeps}.

\begin{table}[h]
\vspace{0.25cm}
\newcommand{\lstrut}{{$\strut\atop\strut$}}
  \caption {\em Some details on matrix-element-based process implementation in {\bf
AcerMC} library. In case of $ q \bar {q}$ initial state the number of Feynman diagrams
corresponds to one flavour combination. The $f=e,\mu,\tau,b$. 
\label{t:acmeps}}
\vspace{2mm} 
\begin{center}
\begin{tabular}{|c|c|c|c|c|} \hline \hline
Process id & Process specification & Feyn. diagrams & Channels & Colour flows \\ 
\hline\hline
 1      &  $gg  \to   t \bar t b \bar b $ & 36 & 11 & 6 \\
\hline
 2      &  $q \bar q  \to   t \bar t b \bar b $ & 7 & 4 & 2 \\
\hline
 3      &  $q \bar q  \to W(\to \ell \nu)  b \bar b $ & 2 & 2 & 1 \\
\hline
 4      &  $q \bar q  \to W(\to \ell \nu)  t \bar t $ & 2 & 2 & 1 \\
\hline
 5      &  $gg \to Z/\gamma^*(\to \ell \ell)  b \bar b $ & 16 & 6 & 2 \\
\hline
 6      &  $q\bar q  \to Z/\gamma^*(\to \ell \ell)  b \bar b $ & 8 & 5 & 1 \\
\hline
 7      &  $gg \to Z/\gamma^*(\to f f, \nu \nu)  t \bar t $ & 16 & 6 & 2 \\
\hline
 8      &  $q \bar q  \to Z/\gamma^*(\to f f, \nu \nu)  t \bar t $ & 8 & 5 & 1 \\
\hline
 9      &  $gg  \to (Z/W/\gamma^* \to) t \bar t b \bar b $ & 72 & 21 & 12 \\
\hline \hline
\end{tabular}
\end{center}
\end{table}

\boldmath
\subsection{The $\alpha_{\rm QED}$ and $\alpha_{s}$ calculations \label{s:alphas}} 
\unboldmath

Native functions of running $\alpha_{\rm QED}(Q^2)$ and $\alpha_{s}(Q^2)$ have been
implemented inside {\bf AcerMC} with the main objective of providing a means to keep the
(total) cross-sections of the processes unchanged when interfacing with the two
supervising generators, since the implementations of the two functions in {\tt
PYTHIA} and {\tt HERWIG} differ to some extent. Especially the  $\alpha_{s}(Q^2)$ is
subject to  experimental and theoretical uncertainties, however obtaining a
different cross-sections for the same {\bf AcerMC} process due to different interface,
could be regarded (at least to some extent) as an inconsistency\footnote{The values  will 
still differ by a small amount in processes containing W bosons (processes 3,4) 
due to different values of the CKM matrix in the two supervising generators.}. 
\begin{figure}[ht]
\begin{center}
\mbox{
     \hspace{-0.3cm}
     \epsfxsize=7.4cm
     \epsffile{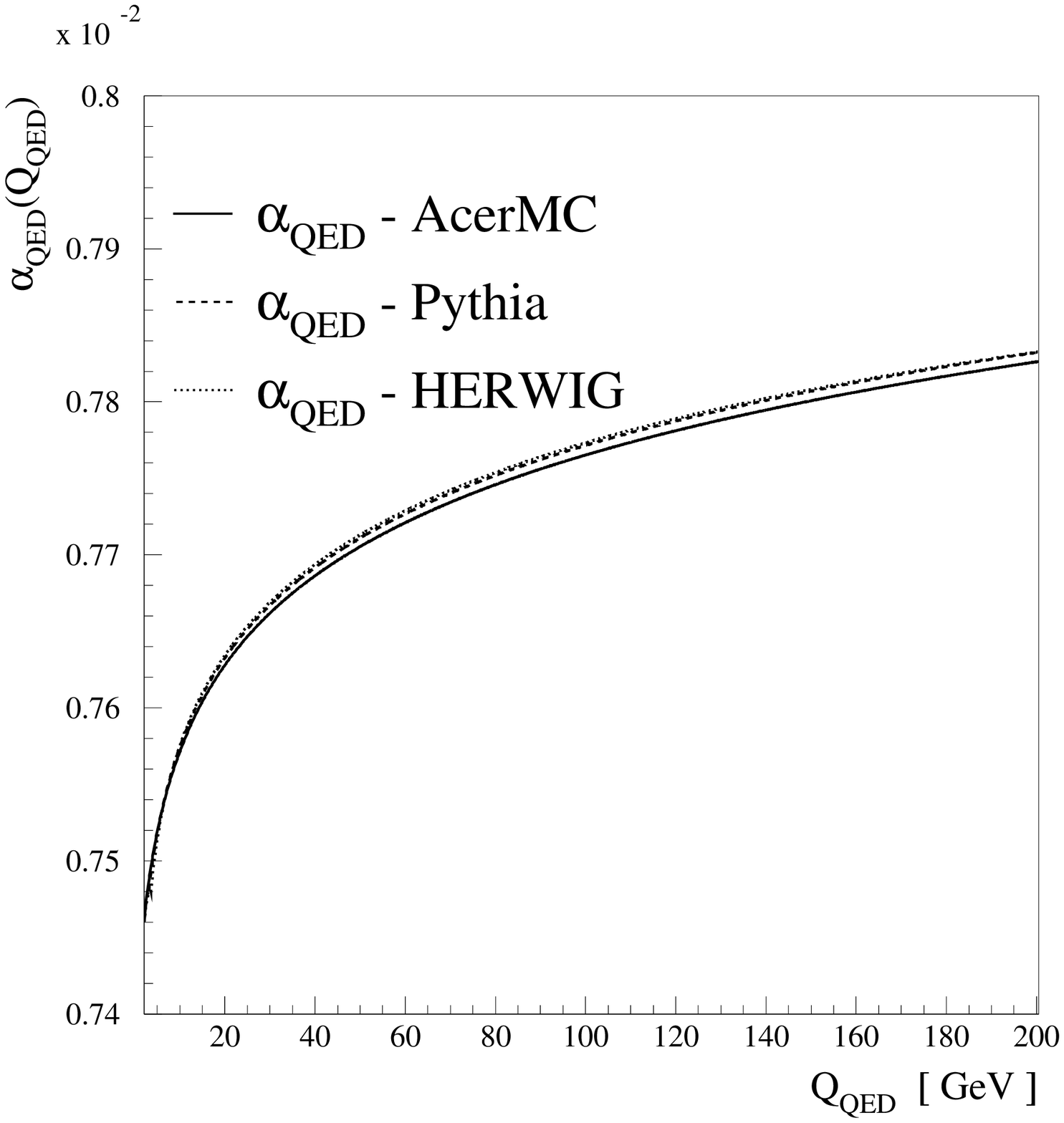}
     \hspace{-0.7cm}
     \epsfxsize=7.4cm
     \epsffile{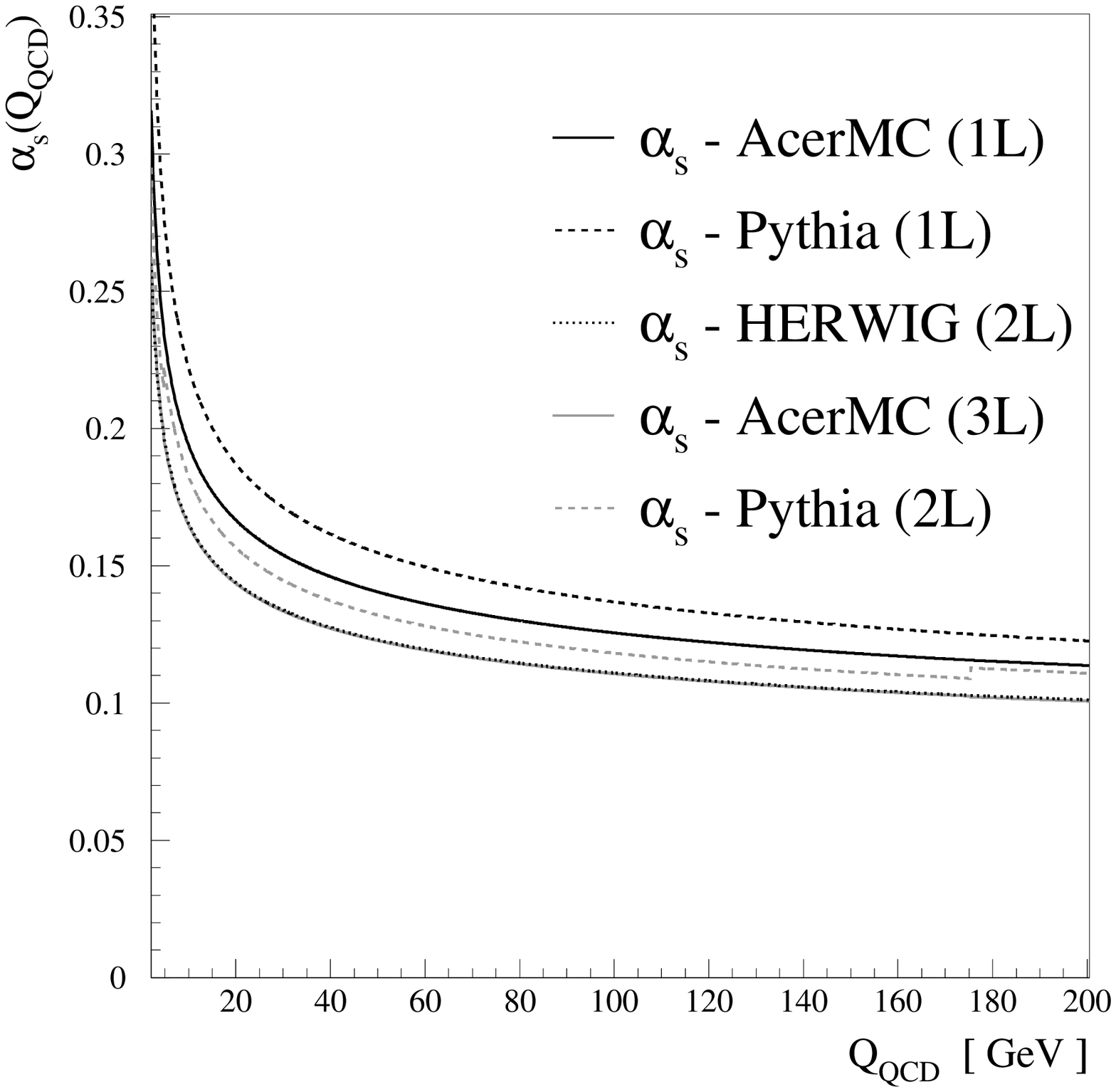}
}
\end{center}
\caption{\em
Comparison between the $\alpha_{\rm QED}(Q^2)$ ({\bf Left}) and $\alpha_{s}(Q^2)$ ({\bf
Right}) implementations in {\bf AcerMC}, {\tt PYTHIA} and {\tt HERWIG}. For
$\alpha_{s}(Q^2)$ calculations with different loop orders ({\sf L}) are given where
applicable.
\label{f:alps}}
\end{figure}

\begin{itemize}
\item {\boldmath $\alpha_{\rm QED}$} is implemented in {\bf AcerMC} using the formulae given
in \cite{field} and is in complete accordance with the implementations in {\tt
PYTHIA} and {\tt HERWIG} apart from the updated hadronic component published recently by
Burkhardt {\it et. al.} \cite{burk2001}. As one can see in Figure \ref{f:alps}, the latter 
minimally lowers the $\alpha_{\rm QED}$ values.
\item{\boldmath $\alpha_{s}$} has one and three loop implementations in {\bf AcerMC}
following the calculations of W. J. Marciano \cite{marciano} and using
$\Lambda^{(nf)}_{\bar{MS}}$ transformations for flavour threshold matches. The
three loop version gives good agreement with the {\tt HERWIG} implementation
(both functions have been set to the same $\Lambda^{(nf=5)}_{\bar{MS}}$ value)
as one can see in Figure \ref{f:alps}. The {\tt PYTHIA} two loop
implementation deviates somewhat from the latter two; the kinks observed in the plot 
are due to approximate $\Lambda^{(nf)}_{\bar{MS}}$ transformations at flavour thresholds, 
which are exact to one loop only.
\end{itemize}

Although the {\bf AcerMC} and {\tt PYTHIA} one loop implementations are
identical in form the resulting values differ by a small amount because the
default {\tt PYTHIA} implementation reads the $\Lambda^{(nf=4)}_{\bar{MS}}$
value from {\tt PDFLIB} instead of the $\Lambda^{(nf=5)}_{\bar{MS}}$ one used by
{\bf AcerMC} and {\tt HERWIG}; the difference thus occurs due to
$\Lambda^{(nf)}_{\bar{MS}}$ propagation at flavour thresholds.


\boldmath
\section{Structure of the package} 
\unboldmath

The {\bf AcerMC} package consist of a library of the matrix-element-based generators
for selected processes, interfaces to the {\tt PYTHIA 6.2} and {\tt HERWIG 6.3}
generators, sets of data files and two main programs: 
{\tt demo\_hw.f} and {\tt demo\_py.f}. Provided makefiles
allow to build the executables with either of these generators as the 
{\it supervising generator}:
{\tt demo\_hw.exe} and {\tt demo\_py.exe}.
\enlargethispage{2cm}

\boldmath
\subsection{Main event loop and  interface to {\tt  PYTHIA/HERWIG}} 
\unboldmath

\begin{figure}[ht]
\begin{center}
\mbox{
     \epsfxsize=\linewidth
     \epsffile{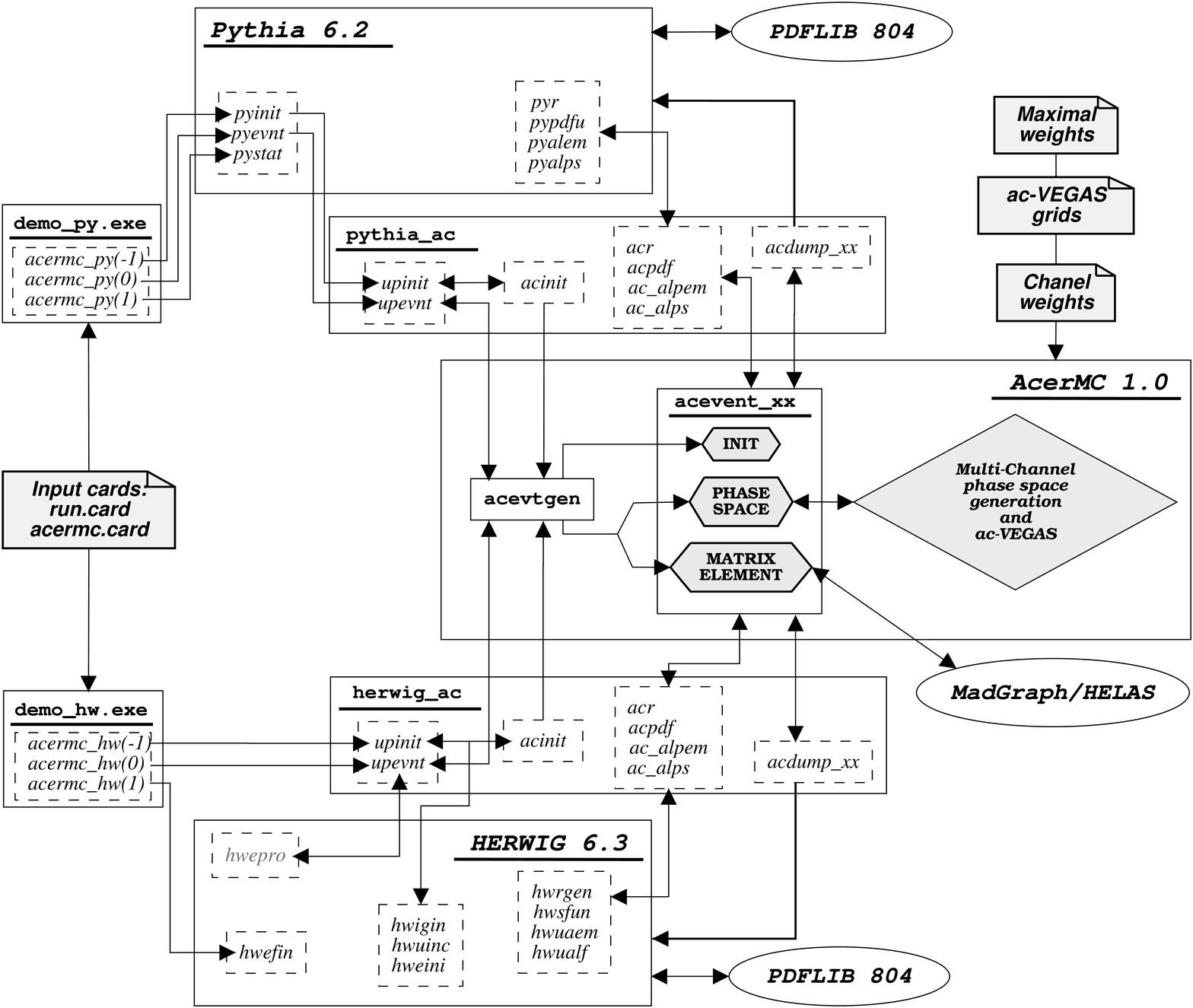}
}
\end{center}
\caption{\em
The calling sequence of the main event generation routine {\tt acevent\_xx}.
The routine is called either through {\tt demo\_py $\to$ acermc\_py} sequence
when interfacing the {\tt PYTHIA~6.2} generator or {\tt demo\_hw $\to$
acermc\_hw} sequence when the {\tt HERWIG~6.3} is linked. The structure of the
interface subroutines and relations with the corresponding ones from supervising
generators and/or external libraries is also evident. 
\label{f:callstruc}}
\vspace{0.4cm}
\end{figure}

The main event loop is coded in the {\tt demo\_hw.f} or {\tt demo\_py.f} files,
where the opening/closing of the input/output files, reading of the data-cards
and event-loop execution is performed.  Main event loop consists only
of calls to the {\tt acermc\_py} or {\tt acermc\_hw} subroutines, with parameter
{\tt MODE = -1,~0,~1} respectively set for initialisation, generation and finalisation of
the event loop.  The call to {\tt acermc\_xx} activates respective procedures of
the supervising generator, which in  turn activates the {\tt acevtgen}
procedure steering the native {\bf AcerMC} generation of the matrix element
event. Fig.~\ref{f:callstruc} illustrates this calling sequence in some details.

As one can deduce from the diagram in Fig.~\ref{f:callstruc}, certain functions
called by {\bf AcerMC}, as e.g. pseudo-random number generator {\tt acr} are
re-routed through the interfaces to the linked supervising generator, depending
on the choice at compilation time (e.g. {\tt acr} function giving
(pseudo-)random numbers is linked to either {\tt pyr} or {\tt hwrgen} as shown
in the plot), providing the internal consistency of the package. The generated
event is rewritten to the format required by the supervising generator by means
of the {\tt acdump\_xx} routines.

The {\tt pythia\_ac.f} and  {\tt herwig\_ac.f} files contain sets of re-routing/interface
functions, specialised for the respective supervising generator. The main library 
of {\bf AcerMC} is well screened from dependencies on the supervising generator,
all dependencies are hidden in {\tt herwig\_ac.f} and {\tt pythia\_ac.f} respectively.

The {\tt PYTHIA} and {\tt HERWIG} libraries remain essentially untouched\footnote{For specification of exceptions see Section 6.4 and 6.5.},
without introducing any dependencies on the {\bf AcerMC} code.
 The input cards are common for both interfaces.

\newpage

\boldmath
\subsection{Structure of the {\bf AcerMC} matrix-element and phase-space code} 
\unboldmath

\begin{figure}[ht]
\begin{center}
\mbox{
     \epsfxsize=\linewidth
      \epsffile{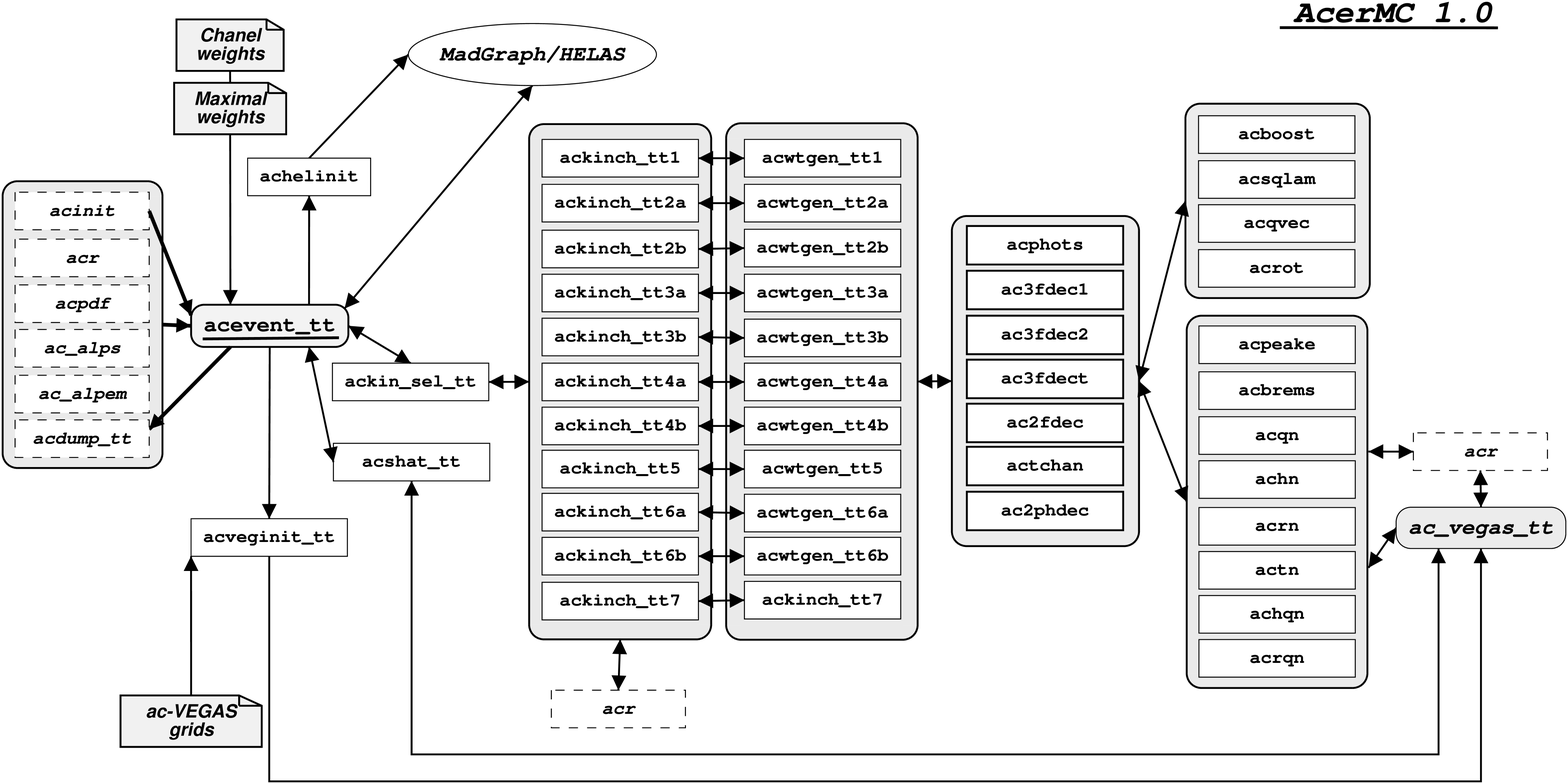}
}
\end{center}
\caption{\em
The event generation sequence controlled by {\tt acevent\_tt} subroutine. Phase
space generation is sequenced by calling first the {\tt acshat\_tt} routine to
obtain the incoming gluon momenta and next the phase space generation via the
{\tt ackin\_sel\_tt} routine which re-routes the generation to a certain kinematic
channel {\tt ackin\_ch}. The latter routine handles the possible momenta
permutations and calls the explicit four-momenta generation (and PS weight
calculation) in the routines {\tt acwtgen\_tt}. These (channel-specific) routines 
are constructed from common building blocks listed in the next two columns. 
The {\tt acevent\_tt} routine also initialises {\tt MADGRAPH/HELAS} package and retrieves
the matrix element values. All the generated four-momenta, as well as the event
weight are finally passed back to the supervising generator via the {\tt
acdump\_tt} call.
\label{f:ttbbflow}}

\vspace{0.25cm}
\end{figure}

The {\bf AcerMC} core code performs the generation of a matrix-element-based event.
Fig.~\ref{f:ttbbflow} illustrates an example of the calling sequence for
generating $gg \to t \bar t b \bar b$ event. The steering subroutine is called
{\tt acevent\_xx}, where {\tt xx} denotes an unique label corresponding to the
process at hand\footnote{The label does actually not have any relation with the
process at hand, it is just an unique two character choice.} (in case of $gg \to
t \bar t b \bar b$ we have  {\tt xx=tt}).  To stress again, this subroutine calls only a
sequence of the native {\bf AcerMC} subroutines, any call to the supervising
generator goes via the respective interface function/subroutine. A more detailed
representation of calling sequence is shown in the Figure \ref{f:ttbbflow}.

\begin{figure}[ht]
\begin{center}
\mbox{
     \epsfxsize=0.6\linewidth
      \epsffile{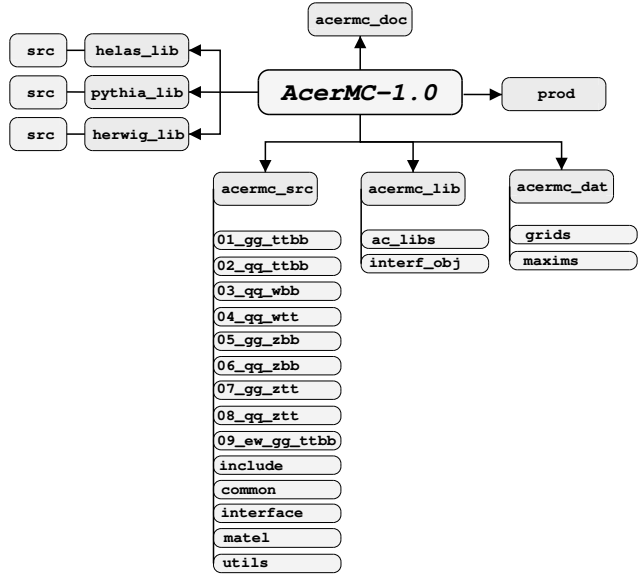}
}
\end{center}
\caption{\em The structure of the {\bf AcerMC} directories.
\label{f:acstruct}}
\end{figure}

Code for the phase space generation is grouped together in the subdirectories,
one per subprocess, e. g. code for generating $gg \to t \bar t b \bar b$ event is
in subdirectory {\tt 01\_gg\_ttbb}. Code for matrix element calculations is
grouped together for all processes in subdirectory {\tt matel}. Code with
different utility subroutines, e.g. kinematic transformations used by all
subprocesses, is in the subdirectory {\tt common}.  Subdirectory {\tt interface}
contains code with interfaces to supervising generators, finally
subdirectory {\tt include} contains all include files. The overall view on 
the structure of the {\bf AcerMC} directories is shown in Fig.~\ref{f:acstruct}.

\vspace{0.25cm}

The core code builds one library  {\tt libacermc.a}. 

\boldmath
\subsection{Data files for the phase-space optimisation} 
\unboldmath

The {\bf AcerMC}  matrix-element-based generators are very highly optimised, using
multi-channel optimisation and additional improvement with the {\tt ac-VEGAS} grid. 
The generation modules require three kinds of the input data
to perform the generation of unweighted events:
\begin{itemize}
\item A file containing the list of the values of relative channel weights obtained by
the multi-channel optimisation, defaults being stored in {\tt acermc\_src/include}.
\item A file containing the pre-trained {\tt ac-VEGAS} grid, the pre-trained
(default) ones located in {\tt acermc\_dat/grids}.
\item A file containing the maximum weight $\rm wt_{\rm max}$,  $\alpha$-cutoff
maximum weight $\rm wt_{\rm max}^\alpha$ and the 100 events with the highest
weights, the default ones being provided in {\tt acermc\_dat/maxims}.
\end{itemize}
In case of changing the default running conditions, like parton density functions or
centre-of-mass energy, the user should repeat the process of preparation of the listed
data files containing the inputs for the phase-space generator modules in order to
preserve the initial event generation efficiency.

\begin{figure}[hb]
\vspace{0.2cm}
\begin{center}
\mbox{
     \epsfxsize=0.6\linewidth
      \epsffile{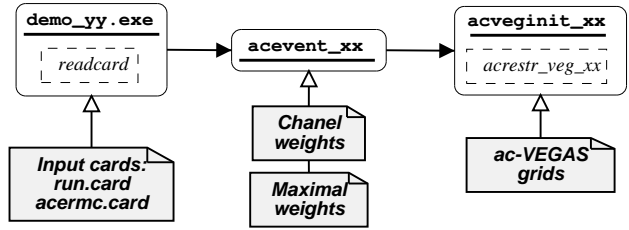}
}
\end{center}
\vspace{-0.2cm}
\caption{\em The reading sequence of the input files and the performing subroutines 
in the {\bf AcerMC} code.
\label{f:acread}}
\end{figure}

The reading sequence of data files inside {\bf AcerMC} is shown in Figure \ref{f:acread}.

Pre-trained data sets are obtained using $\sqrt{s}=14\;$ TeV, {\tt PYTHIA} default
$\alpha_s(Q^2)$ and $\alpha_{\rm QED}(Q^2)$ and CTEQ5L (parametrised) parton density function
set and are provided for each implemented process\footnote{These can also be used for a series of 
other settings, see Section \ref{s:training} for details}.  For these, the relative channel
weights are stored in the {\tt INCLUDE} files in {\tt acermc\_src/include/chanwt\_xx.inc}
where {\tt xx} denotes the process id (c.f. Table \ref{T3:1}); the
default/pre-trained {\tt ac-VEGAS} grids are listed in the directory {\tt
acermc\_dat/grids/vscalA\_xxYYY.veg}, where {\tt A} denotes the scale choice of the
process {\tt xx} and {\tt YYY} denotes the cutoff value of the $m_{Z^0/\gamma^*}$ for the
{\bf AcerMC} processes {\tt xx = 05 $\to$ 08}. The files containing the maximal weights
$\rm wt_{\rm max}$ and $\rm wt_{\rm max}^\alpha$ as well as the 100 events with the highest
weights are stored in the directory {\tt acermc\_dat/grids/vtmaxA\_xxYYY.dat}, following
the same labelling convention. Both the trained {\tt ac-VEGAS} grids and the weight files
were obtained from test runs with at least $2 \cdot 10^6$ weighted events being generated.

The number of required input files might at first look seem large, considering that many
event generators do not require any input files for operation; the difference is not in so
much in the complexity of the phase space generation as in the fact that many event
generators require a {\it warming run} instead, i.e. before the generation of unweighted
events is performed a certain number of weighted events (typically of the order of $10^4$)
is generated in order to obtain the relative multi-channel weights (in case multi-channel
phase space generation is used) and/or the optimised {\tt VEGAS} grid and/or an estimate
of the maximal weight. Such an approach can have an advantage when event generation is
very fast and the phase space regions with the highest weights are well known (as done for
the $2 \to 2$ processes in {\tt PYTHIA}); on the other hand, when the phase space topology
of the process is more complex and the event generation is comparatively slow, generating
a relatively small number of e.g. $10^4$ weighted events {\it every time} a generator is
started can become CPU wasteful and/or inaccurate in terms of maximum weight estimation.

Reasonably accurate estimation of the latter is namely crucial for correct event
unweighting; event generators using {\it warming-up} method for maximal weight search often
find still higher weights during the production run and reset the maximal weight
accordingly. In this case however, statistically correct approach would be to reject all
events generated beforehand and start the event generation anew, which is almost never
implemented due to the CPU consumption and the possibility of hitting a weak singularity
(the same argument leads to the definition of the $\rm wt_{\rm max}^\alpha$, c.f. Section
\ref{s:ac-veg}). With a small pre-sampled set the generator can however badly under-estimate the
maximum weight and a large number of events can be accepted with a too-high
probability. The only hope of obtaining correct results is in such cases that the weight
{\it plateau} will be hit sufficiently early in the event generation process. Consequently,
such approach can be very dangerous when generating small numbers of
events\footnote{{\it Small} being a somewhat relative quantifier, since the size of an
representative sample should depend on the phase space dimension, i.e. the number of
particles in the final state; with e.g. 4 particles in the final state, $10^5$ events can
still be considered a relatively small statistics.}.

In contrast to the {\it warming-up} approach the {\bf AcerMC} we decided that
using separate {\it training} runs with large numbers of weighted events to obtain the
optimised grids and maximum weight estimates are preferable, in case user wants
to produce data sets for non-defaults setting, this can easily be done
by configuring the switches in the {\tt acermc.card} (see Section 6.2).


\boldmath
\section{How to use the package} 
\unboldmath

There are two steering input files: {\bf run.card} and {\bf acermc.card} which share a
common format for both executables.  The {\bf run.card} (see App. \ref{app:run}) provides
switches for modifying: generated process, number of events, parton density functions,
predefined option for hadronisation/fragmentation in the {\it supervising generator},
random number, etc..  The {\bf acermc.card} (see App. \ref{app:acmc}) provides switches for
modifying more specialised settings for the {\bf AcerMC} library itself. Once the user
decides on a setup for the generated process, only {\bf run.card} is very likely to be
modified for the job submission. Both input files are read by AcerMC executables through
the {\tt CERNLIB FFREAD} routines, some commands given in the input files (e.g. LIST
entry, see Appendix B) are internal {\tt FFREAD} commands which should be disregarded by
the user. 

The same executables can also be used for running standard {\tt PYTHIA 6.2} and
{\tt HERWIG 6.3} processes. The example how to require such process is provided as
well, in {\tt demo\_hw.f} and {\tt demo\_py.f} respectively.  If the user
requires that the {\bf AcerMC} library is not used, the $ttH$ production will be
generated with {\tt demo\_py.exe} and {\tt HERWIG 6.3} implementation of the
$Zbb$ production will be generated with {\tt demo\_hw.exe}. In this case only
the {\tt run.card} file will be read, so in case the user requires different
processes and/or settings of the supervising generators the user has to
implement her/his steering there or create another xxx.card file, together with
the corresponding code added to the {\tt demo\_xx.f}.
  
\boldmath
\subsection{Steering switches of the overall run} 
\unboldmath

The overall run is controlled by the switches read from the {\tt
run.card} file (see also App. \ref{app:run}).
Some of these general switches are  also passed to the {\bf AcerMC} library.

\begin{itemize}
\item{\tt CMS }\ : \ Sets the centre-of-mass energy in GeV. 
\item{\tt ACER}\ : \ Specifies if the internal {\bf AcerMC} process will be used\\
{\tt ACER=0} - use process from {\tt PYTHIA/HERWIG} \\
{\tt ACER=1} - use internal {\bf AcerMC} process
\item{\tt PROCESS}\ : \ Sets process id
\item{\tt HAD}\ : \ Sets predefined option for QCD ISR/FSR and hadronisation\\
{\tt HAD=0} - only hard process \\
{\tt HAD=1} - only ISR (works for {\tt PYTHIA} interface only)\\
{\tt HAD=2} - only ISR and FSR\\
{\tt HAD=3} - full treatment\\
{\tt HAD=4} - only FSR \\
{\tt HAD=5} - only FSR and hadronisation
\item{\tt PDFLIB804} {\tt NGROUP}\\
Sets the value of the {\tt PDFLIB804} {\tt NGROUP} parton density function choice.
\item{\tt PDFLIB804} {\tt NSET}\\
Sets the value of the {\tt PDFLIB804} {\tt NSET} parton density function choice.
\item{\tt RSEED}\ : \ Choose the random seed for (pseudo-)random generator initialisation
\item{\tt NEVENT}\ : \ Required number of generated events 
\end{itemize}

\boldmath  
\subsection{Steering switches of the  {\bf AcerMC} processes} 
\unboldmath

The {\bf AcerMC} processes are controlled by values set in a simple arrays specified
in {\tt acermc\_src/include/AcerMC.inc}:
\begin{scriptsize}
{\tt 
\\
C CROSS-TALK PARAMETERS \\
      DOUBLE PRECISION ACSET \\
      INTEGER IACPROC\\
      COMMON/ACPAR1/ACSET(200),IACPROC(200)\\
C PARTICLE PROPERTIES\\
      DOUBLE PRECISION ACCHG,ACMAS,ACCKM \\
      COMMON/ACPAR2/ACCHG(50,4),ACMAS(50,4),ACCKM(4,4) \\ 
C ROUTINE I/O\\	
      INTEGER LACSTD,LACIO \\
      COMMON/ACPAR3/LACSTD,LACIO 
}
\end{scriptsize}

The {\tt IACPROC} array activates the process {\tt IPROC=PROCESS}
(read from {\tt run.card} file) by setting {\tt IACPROC(IPROC)=1}.

The list of currently implemented processes in {\bf AcerMC}
can be found in Table \ref{T3:1}. When running in the generation mode
with {\tt ACER=0} full list of processes implemented in either {\tt PYTHIA 6.2}
or {\tt HERWIG 6.3} can be activated, however the mechanism for passing information 
about process id to either of these generators has to be coded by user individually
in {\tt demo\_xx.f}.

The main control switches reside in the array {\tt ACSET}.
The {\tt COMMON} block {\tt ACPAR2} contains the particle charges, masses and
decay widths as well as the CKM matrix using the {\tt PYTHIA} convention. The
values are filled by the interface routines to be equal to the {\tt PYTHIA/HERWIG}
internal values in order to preserve consistency within the generation
stream. In case the user wants to change some of the particle properties this
should be done through the native {\tt PYTHIA/HERWIG} switches; {\bf AcerMC} will copy
them and use the new values.

The {\tt COMMON} block {\tt ACPAR3} contains the two logical I/O unit numbers
used by {\bf AcerMC}. The {\tt LACSTD} value determines the output unit of the {\bf AcerMC}
messages and the {\tt LACIO} unit is used for reading/writing the {\bf AcerMC} data files. 
\newpage

The main control switches which reside in the array {\tt ACSET} (see also App. \ref{app:acmc}):
\begin{itemize}
\item{\tt ACSET(1)}\ : \ Sets the centre-of-mass energy in GeV. 
\item{\tt ACSET(2)}\ : \ Scale of the hard process\\
Choose the $Q^2$ scale for the active {\bf AcerMC} process.
The implemented values differ for various processes, the currently implemented
settings are specified in Section~6.6.
\item{\tt ACSET(3)}\ : \ Fermion code\\
The flavour of the final state fermions produced in $W^{\pm}, Z^0/\gamma^* \to f \bar f$ decays of {\bf AcerMC} processes $3 \to 8$. The {\tt PYTHIA/PDG} naming convention is used:\\
{\tt ACSET(3)=11} - $W \to e \nu_e$; $Z^0/\gamma^* \to e^+ e^-$ \\
{\tt ACSET(3)=12} - $Z^0/\gamma^* \to \nu_e \nu_e$,\\\
{\tt ACSET(3)=13} - $W \to \mu \nu_{\mu}$; $Z^0/\gamma^* \to \mu^+ \mu^-$  \\
{\tt ACSET(3)=14} - $Z^0/\gamma^* \to \nu_{\mu} \nu_{\mu}$\\
{\tt ACSET(3)=15} - $W \to \tau \nu_{\tau}$; $Z^0/\gamma^* \to \tau^+ \tau^-$  \\
{\tt ACSET(3)=16} - $Z^0/\gamma^* \to \nu_{\tau} \nu_{\tau}$  \\
{\tt ACSET(3)=5} -  $Z^0/\gamma^* \to b \bar b$ \\
At present the {\tt ACSET(3)=5} and {\tt ACSET(3)=12,14,16} are implemented only for 
processes $7 \to 8$.
\item{\tt ACSET(4)}\ : \ $Z^0/\gamma^*$ propagator\\
Use full $Z^0/\gamma^*$ propagator instead of the pure $Z^0$ propagator in
matrix element calculation for the {\bf AcerMC} processes $5 \to 8$.
 The switch is provided
since in some of the analyses the $\gamma^*$ contribution is of relevance in the
selected mass windows; for the analyses selecting the mass window around the
$Z^0$ peak this contribution can safely be neglected.\\
{\tt ACSET(4)=0} - only $Z^0$ propagator. \\
{\tt ACSET(4)=1} - full $Z^0/\gamma^*$ propagator. 
\item{\tt ACSET(5)}\ : \ $m_{Z^0/\gamma^*}$ mass cut\\
Cutoff value on the invariant mass $m_{Z^0/\gamma^*}$ in GeV when {\tt ACSET(4)=1}. 
Note that the provided data files exist only for values of
{\tt ACSET(5)}=2,5,10,15,30,60, 120,270,300 and 500 GeV which should satisfy most 
user requirements for the analyses foreseen at LHC.
 In case a different value is set the user has also to provide 
the user data files for the run.
\item{\tt ACSET(6)}\ : \ Sets the value of the {\tt PDFLIB804} {\tt NGROUP} 
parton density function choice.
\item{\tt ACSET(7)}\ : \ Sets the value of the {\tt PDFLIB804} {\tt NSET}
 parton density function choice.
\item{\tt ACSET(8)}: The implementation of $\alpha_s(Q^2)$\\
Selects the implementation of $\alpha_s(Q^2)$ to be used in the matrix element
calculation:\\
{\tt ACSET(8)=0} - Use the $\alpha_s(Q^2)$ as provided by the supervising generator/\\
{\tt ACSET(8)=1} - Use the  $\alpha_s(Q^2)$ (one loop) provided by the {\bf AcerMC}; this
option gives $\alpha_s(Q^2)$ values equal to the default {\tt PYTHIA} implementation.\\
{\tt ACSET(8)=2} - Use the  $\alpha_s(Q^2)$ (three loop) provided by the {\bf AcerMC}.
\newpage

\item{\tt ACSET(9)}: $\Lambda^{(nf=5)}_{\bar{MS}}$ value\\
Sets the $\Lambda^{(nf=5)}_{\bar{MS}}$ value to be used in the $\alpha_s(Q^2)$
calculations in case the {\bf AcerMC} native implementation ({\tt ACSET(8)=1}) is
used.\\
{\tt ACSET(9)=-1} - The $\Lambda^{(nf=5)}_{\bar{MS}}$ value is taken from the
{\tt PDFLIB804} for the selected parton density function set.\\
{\tt ACSET(9)$>$0} - The provided value is taken.
\item{\tt ACSET(10)}: The implementation of $\alpha_{\rm QED}(Q^2)$\\
Selects the implementation of $\alpha_{\rm QED}(Q^2)$ to be used in the matrix
element calculation:\\ 
{\tt ACSET(10)=0} - Use the $\alpha_{\rm QED}(Q^2)$ as provided by
the  supervising  generator.\\ 
{\tt ACSET(10)=1} - Use the $\alpha_{QED}(Q^2)$ implemented in the {\bf AcerMC}.
\item{\tt ACSET(11)}: $\alpha_{\rm QED}(0)$ value\\
Specifies the value of $\alpha_{\rm QED}(0)$ for {\bf AcerMC} $\alpha_{\rm QED}(Q^2)$
calculation.\\
{\tt ACSET(11)=-1} - The $\alpha_{\rm QED}(0)$ value is set to 
$\alpha_{\rm QED}(0) = 0.0072993$. \\
{\tt ACSET(11)$>$0} - The provided value is taken.
\item{\tt ACSET(12)}: Decay mode of the produced $t \bar t$ pair\\
Sets the decay mode of the W boson pair from the $t \bar t$ final
state in the {\bf AcerMC} processes 1,2,4,7,8 and 9. 
For {\tt ACSET(12)$>$0} the combinatoric  value of the $\sigma \times BR$ is
recalculated and printed in the output. This switch was implemented
since the supervising generators 
({\tt PYTHIA/HERWIG}) do not allow for forcing specific decays of the
top quark pairs generated by external processes. This switch imposes a  modification 
of the decay tables of the supervising generators on an event by event
basis.\\ 
{\tt ACSET(12)=0} - both W bosons decay according to {\tt
PYTHIA/HERWIG} switches.\\ 
{\tt ACSET(12)=1} - $W_1 \to e \nu_e$ and $W_2 \to q \bar q$.\\ 
{\tt ACSET(12)=2} - $W_1 \to \mu \nu_{\mu}$ and $W_2 \to q \bar q$.\\ 
{\tt ACSET(12)=3} - $W_1 \to \tau \nu_{\tau}$ and $W_2 \to q \bar q$.\\ 
{\tt ACSET(12)=4} - $W_1 \to e \nu_e, \mu \nu_{\mu}$ and $W_2 \to q \bar q$.\\ 
{\tt ACSET(12)=5} - $W_1 \to e \nu_e, \mu \nu_{\mu}$ and $W_2 \to q \bar q$. \\ 
The setting {\tt ACSET(12)=5} works for {\tt PROCESS=4} only and implies
leptonic decay of the $W$-boson with the same charge as the one of the primary 
$W$ boson produced in the hard process. Folowing configurations are posible:
\[
 q \bar q \to W^+ t \bar t 
\to (W^+ \to) L^+ \nu_L ~~~ (W^+_1 \to) l^+ \nu_l\; b ~~~ (W^-_2 \to) q' \bar{q}' \; \bar b,
\]
or:
\[
 q \bar q \to W^- t \bar t 
\to  (W^- \to) L^- \bar{\nu}_L ~~~ (W^+_1 \to) q' \bar{q}' \;  b  ~~~ (W^-_2 \to) l^- \bar{\nu}_l \; \bar b ,
\]
where $L^\pm$ is the lepton from the primary $W$ decay (controlled by {\tt ACSET(3)} switch) 
and $l^\pm$ is either an $e^\pm$  or $\mu^\pm$ as for {\tt ACSET(12)=4}. Since the charge of the 
semi-leptonic decaying $W$ is correlated
with the charge of the primary W boson, the $\sigma \times BR$ is consequently a factor 
two smaller than the one for {\tt ACSET(12)=4}.
\newpage

\item{\tt ACSET(50)}: {\bf AcerMC} training mode\\
The switch controls the mode in which {\bf AcerMC} is run:\\
{\tt ACSET(50)=0} - production run, generate unweighted events.\\
{\tt ACSET(50)=1} - perform multi-channel optimisation and output the user file
with channel weights.\\
{\tt ACSET(50)=2} - perform {\tt ac-VEGAS} grid training and output the user file
with trained {\tt ac-VEGAS} grid.\\
{\tt ACSET(50)=3} - perform {\tt ac-VEGAS} grid training as in {\tt ACSET(50)=2}
but do this by updating a provided grid.
\item{\tt ACSET(51)}: Required number of generated events {\tt NEVENT}\\
In case the switch {\tt ACSET(50)} is set to the non-zero value (i.e. in one of
the training modes) the  {\tt ACSET(51)} entry is used and defines the number of
(weighted) events that will be generated; this information is necessary for the
learning algorithms to decide on steps in the learning sequence. 
\item{\tt ACSET(52)}: User data files\\
Use the data files provided by user:\\
{\tt ACSET(52)=0} - no, use native (default) {\bf AcerMC} data files.\\
{\tt ACSET(52)=1} - use the user's multi-channel optimisation and {\tt VEGAS} grid
files.\\ 
{\tt ACSET(52)=2} - use the default multi-channel optimisation
 and user's {\tt VEGAS} grid files.\\
{\tt ACSET(52)=3} - use the default multi-channel optimisation and {\tt VEGAS}
                    grid files; read the user maximal weight file.
\item{\tt ACSET(53)}: Maximum weight search\\
Mode for the maximum weight search needed for unweighting procedure:\\
{\tt ACSET(53)=0} - no, use the provided files containing maximal weights.\\
{\tt ACSET(53)=1} - use the provided files for max. weights and re-calculate 
the max. weights using the stored 100 events with the highest weight.\\
{\tt ACSET(53)=2} - perform the search and give the new {\tt wtmax\_xx\_new.dat}
file; the switch is  equivalent to generation of weighted events.
\item{\tt ACSET(54)}: Maximum weight choice\\
Use the $\alpha$-cutoff maximal weight $\rm wt_{\rm max}^\alpha$ or the overall maximal
weight $\rm wt_{\rm max}$ found in training (see Section \ref{s:ac-veg} for the explanation on
these two options).\\
{\tt ACSET(54)=0} - use the $\rm wt_{\rm max}^\alpha$ weight.\\
{\tt ACSET(54)=1} - use the $\rm wt_{\rm max}$ weight.
\end{itemize}

\boldmath 
\subsection{How to prepare data-files for the non-default setup \label{s:training}} 
\unboldmath

The following actions are possible, to recover better efficiency of the generator
modules with the non-default settings:

\begin{itemize}
\item{\it The user wants to generate events using different parton density function
sets and/or different coupling values (e.g. {\bf AcerMC} third order $\alpha_s(Q^2)$
instead of the first order one)}:\\
 
It should suffice to set the the switch {\tt ACSET(53)=1}, which signals {\bf AcerMC} to
re-calculate the $\rm wt_{\rm max}$ and $\rm wt_{\rm max}^\alpha$ using the 100 events stored in
the file \\ {\tt acermc\_dat/grids/vtmaxA\_xxYYY.dat}. The coupling and
parton density functions values should not change significantly the process topology but affect
foremost the overall scale of the event weights; thus, the stored hundred events should
still remain the ones with the highest weights and the re-calculated approximate estimates
of the highest weight should be accurate enough.

In case the user is not confident in the obtained result, the new maximal weight estimation
can be initiated by setting the switch {\tt ACSET(53)=2}, which will result in generation
of weighted events. The number of generated events is determined by the usual {\tt NEVENT}
in {\tt run.card}.  At the end of the run {\bf AcerMC} will produce a file called {\tt
wtmax\_xx\_new.dat}, with {\tt xx} specifying the process number. The user should then
start the generation of unweighted events with the setting {\tt ACSET(52)=3} and
linking(renaming) the new file to {\tt wtmax\_xx\_usr.dat}, with {\tt xx} denoting the
process number (e.g. {\tt wtmax\_01\_usr.dat}).\\

\item{\it The user wants to generate events using different values of particle/boson
masses or other significant changes of the parameters apart from the
centre-of-mass energy and/or $m_{Z^0/\gamma^*}$ cutoff value for processes 5-8}:\\

In this case the user should re-train the {\tt VEGAS} grid since the process
topology is assumed to undergo minor changes. This is done by setting the switch
{\tt ACSET(50)=2} or {\tt ACSET(50)=3}; in the first case {\bf AcerMC} starts with an
untrained grid and in the second one it starts modifying the existing grid
provided for the process at the selected hard process scale. In general the
second option should be preferable since the topology should still be close to
the pre-trained one. {\bf AcerMC} again produces weighted events and at the end of the
run outputs a file {\tt grid\_xx\_new.veg}. The number of generated events is determined
by the usual {\tt NEVENT} in {\tt run.card}.  As
in the previous case, the user should re-name the file to {\tt grid\_xx\_usr.veg},
re-set the switch to {\tt ACSET(50)=0} and repeat the maximal weight search
procedure described above, by setting the switch {\tt ACSET(53)=2} etc..
When the maximum weight search is completed the user switch {\tt ACSET(52)=2},
which will cause {\bf AcerMC} to read the {\tt wtmax\_xx\_usr.dat} as well as {\tt
grid\_xx\_usr.veg} files and produce unweighted events with the new setup.\\ 

\item{\it The user wants to generate events at a different $m_{Z^0/\gamma^*}$
cutoff value and/or different centre-of-mass energy $\sqrt{s}$}:\\

When the user changes at least one of these two parameters the event topology
 is significantly changed
as well as the contributions from different kinematic channels. The user should thus start
with a new multi-channel optimisation by setting the mode switch {\tt ACSET(50)=1} and
start an {\bf AcerMC} run.  The number of generated events is determined by the usual {\tt
NEVENT} in {\tt run.card}.  At the end of the run {\bf AcerMC} will produce a file {\tt
chanwt\_xx\_new.dat} which should be renamed/linked to {\tt chanwt\_xx\_usr.dat}. The
user should then set the switch {\tt ACSET(52)=1} and first put {\tt ACSET(50)=2} and
and repeat the {\tt VEGAS} grid training as described above and consequently {\tt
ACSET(50)=0} and {\tt ACSET(53)=2} to perform the maximum weight search. After obtaining
all three user files the {\tt ACSET(53)} should again be put back to {\tt ACSET(53)=0} and
a normal run should be started; the switch {\tt ACSET(52)=1} will in this case
force {\bf AcerMC} to read all three user files and produce unweighted events.
\end{itemize}
At the first look procedure for listed action scenaria might seem a bit complex
but should after a few trials and errors become a straightforward routine; it is
expected that the vast majority of users would have to deal with at most the
first scenario.

\newpage
\boldmath 
\subsection{Details on the interface to {\tt PYTHIA 6.2}} 
\unboldmath

The {\bf AcerMC} interface to {\tt Pythia 6.2} is implemented close to the new
standard specified at the Les Houches workshop 2001 \cite{Boo01}. The full
description of the standard can be found in the {\tt PYTHIA 6.2} manual
(\cite{Pythia62}). In addition to the {\tt UPINIT} and {\tt UPEVNT} routines the file
{\tt acermc\_src/interface/pythia\_ac.f} provides links between a list of {\bf AcerMC}
routines and the corresponding {\tt PYTHIA} ones, as e.g. the (pseudo-)random
number generator, $\alpha_s$ and $\alpha_{\rm QED}$ calculations as well as a
series of routines that re-write the {\bf AcerMC} event output to the required {\tt
PYTHIA} format. Using this strategy, the native {\bf AcerMC} code is completely
de-coupled from the linked hadronisation library (at the moment {\tt
\tt PYTHIA/HERWIG}) and new interfaces can thus easily be added. 
The special {\bf AcerMC}
requirement is the call to the {\tt ACFINAL} subroutine at the end of the run
which signals the {\bf AcerMC} to close the various I/O files and produce the final
output. An example of the implementation of the {\tt PYTHIA}/{\bf AcerMC} interface
can be found in the provided {\tt demo\_py.f}. The {\tt PYTHIA} code is
unmodified apart from making a small modification in {\tt PYINIT} routine:

\noindent {\tt CALL UPINIT(1)}\\
\noindent {\sf ..parameter initalisation..}\\
\noindent {\tt CALL UPINIT(2)}\\
\noindent {\sf ..process initialisation..}

since the user-supplied processes in this new interface are not
allowed to (re-)estimate
maximal weights (as e.g. the native {\tt PYTHIA} processes do). 
In the original code
the call to {\tt UPINIT} is set before the {\tt PYTHIA}  parameters and functions 
(e.g. {\tt PYALPS}
for $\alpha_s(Q^2)$ calculation) are initialised with the user settings\footnote{This
was however possible in the old {\tt PYTHIA} interface}. 

The user can add the most recent {\tt PYTHIA} library without other
 modifications but for
the two lines of code in {\tt PYINIT} routine as described above (the dummy routines
{\tt UPINIT, UPEVNT, STRUCTM, STRUCTP} and {\tt PDFSET} however 
have to be removed from the code
for an external process to work and to activate the {\tt PDFLIB} interface).

By setting {\tt  ACER=1} user decides to generate hard process from
 {\bf AcerMC} library.
Modeling of ISR/FSR shower, hadronisation and decays are generated by {\tt PYTHIA} 
generator. All steering parameters, relevant for these steps of full event generation
remain the same as in standard {\tt PYTHIA} execution. 

By setting {\tt  ACER=0} user decides to generate standard {\tt PYTHIA}
 process. The simple
example how to generate $ttH$ production process within {\bf AcerMC} framework
  is provided in {\tt demo\_py.f}.
\newpage

\boldmath
\subsection{Details on the interface to {\tt HERWIG 6.3}} 
\unboldmath

Interfacing the {\bf AcerMC} to {\tt HERWIG 6.3} has proved to be more
of a problem\footnote{This interface is temporary and we plan to move 
to HERWIG 6.4 version in the next future},
since the interfaced version does not comply with the Les Houches standard  \cite{Boo01}. The
interface routines are nevertheless written in the way to mimic the Les Houches
description; in the {\tt HERWIG 6.3} interface the {\tt UPINIT} routine however has
 to be called in two steps in order to enable the user to add the changes
of the {\tt HERWIG 6.3} internal settings:\\
 
\noindent {\tt CALL UPINIT(1)}\\
\noindent {\sf ..user values..}\\
\noindent {\tt CALL UPINIT(2)}\\

The {\tt UPEVNT} subroutine in the {\tt HERWIG} case replaces the
native {\tt HWEPRO} subroutine as an iterface to {\bf AcerMC}
processes; in addition, it enables a user to activate more than one
process in the same run, which is absent from the native {\tt HWEPRO}
implementation. As in the {\tt PYTHIA} implementation all the
interface subroutines needed for communication between {\tt HERWIG}
and {\bf AcerMC} are stored in {\tt
acermc\_src/interface/herwig\_ac.f}.  Some minor modifications of the
original {\tt HERWIG 6.3} code were albeit necessary:
\begin{itemize}
\item {\tt IMPLICIT NONE} was commented out in the {\tt herwig6301.inc} file; this
was needed since the {\bf AcerMC} code is written with the implicit {\tt IMPLICIT
DOUBLE PRECISION(A-H,O-Z)}.
\item The {\tt HERWIG} interface to {\tt PDFLIB} was changed in order to link it to
the latest {\tt PFFLIB804} implementation which has a different {\tt PDFSET}
syntax.
\end{itemize}
In principle the implemented changes should be very easy and transparent for the transfer
into new {\tt HERWIG} releases; it is also anticipated that a future {\tt HERWIG} version
will already have the interface to {\tt PDFLIB} updated accordingly, and interface to Les
Houches standard implemented.  An example of the use of {\bf AcerMC}/{\tt HERWIG}
interface is provided in the file {\tt demo\_hw.f}.

By setting {\tt  ACER=1} the user decides to generate a hard process from {\bf AcerMC} library.
Modeling of ISR/FSR shower, hadronisation and decays are generated by {\tt HERWIG} 
generator. All steering parameters, relevant for these steps of full event generation
remain valid as for the standard {\tt HERWIG} execution. 

By setting {\tt  ACER=0} the user decides to generate standard {\tt HERWIG} processes. The simple
example how to generate $Zb \bar b$ production process within the {\bf AcerMC} framework
 is provided in {\tt demo\_hw.f}.

The output logs of the run are produced in the directory {\tt prod}, the {\tt acermc.out}
file containing the {\bf AcerMC} specific information and the outputs {\tt pythia.out}
and/or {\tt herwig.out} listing the outputs of the respective supervising generators. The
information about the input values of the steering files is stored in {\tt run.out} in
order to facilitate the event generation 'bookkeeping'. The sample outputs are given in
Appendix C.

\boldmath
\subsection{Definition of the energy scale \label{s:scdef}} 
\unboldmath

A few different values of scale $Q^2$ used in the evolution of parton density functions 
as well as the running couplings $\alpha_s(Q^2)$ and $\alpha_{\rm QED}(Q^2)$ can be set by
the switch {\tt ACSET(2)} (remember that the factorisation and renormalisation scales are
assumed to be equal in {\bf AcerMC}). Note that the {\it correct} value of the scale to be used for
certain processes is in principle not known; what was implemented in {\bf AcerMC} are the
most probable/usual choices on the market; in measurements the {\it best}
value will have to be determined by data analysis.
\begin{itemize}
\item Processes $1,2,9$:\\
{\sf 
  ACSET(2): (D=1)\\
\hspace{1cm} 1 - $Q^2 = \hat{s}$ \\       
\hspace{1cm} 2 - $Q^2 = \sum{({p^i_T}^2 + m_i^2)}/4~=\;<m_T^2>$\\
\hspace{1cm} 3 - $Q^2 = \sum{({p^i_T}^2)}/4 ~=\;<p_T^2>$\\
\hspace{1cm} 4 - $Q^2 = (m_t + m_H/2)^2,~~~m_H = 120$~GeV/$c^2$ 
}
\item Processes $3 \to 4$:\\
{\sf 
  ACSET(2): (D=1)\\
\hspace{1cm} 1 - $Q^2 = M_W^2$ \\       
\hspace{1cm} 2 - $Q^2 = s^*_{q \bar q}$, where $q=b,t$ \\       
\hspace{1cm} 3 - $Q^2 = M_W^2+pT_W^2$ \\       
\hspace{1cm} 4 - $Q^2 = 0.5 \cdot (s^*_W + s^*_{q \bar q}) +(p_T^W)^2$, where $q=b,t$        
}
\item Processes $5 \to 8$:\\
{\sf 
  ACSET(2): (D=1)\\
\hspace{1cm} 1 - $Q^2 = M_Z^2$      
}
is the only setting implemented at the moment.
\end{itemize}

\boldmath
\subsection{Installation procedure} 
\unboldmath

The installation requires availability of the {\tt CERNLIB} fortran library.
\begin{itemize}
\item
Ungzip and untar distribution file.
\item
Go to the directory {\tt ./pythia\_lib/src\_pythia}; type {\tt make install}.
It will compile the sources and install the created library {\tt libpythia.a} to 
directory {\tt pythia\_lib}.
\item
Go to the directory {\tt ./herwig\_lib/src\_herwig}; type {\tt make install}.
It will compile the sources and install the created library {\tt libherwig.a} to
directory  {\tt herwig\_lib}.
\item
Go to the directory {\tt ./helas\_lib/src\_helas}; type {\tt make install}.
It will compile the sources and install the created library {\tt libhelas.a} to
directory  {\tt helas\_lib}.
\item
Go to the directory {\tt ./acermc\_src}; type {\tt make install}.
It will the compile sources and install the necessary object files and the 
created library {\tt libacermc.a} to the directory {\tt ./acermc\_lib}.
\item
In the main directory type {\tt make demo\_py} or {\tt make demo\_hw}.
It will compile {\tt demo\_py.f} or {\tt demo\_hw.f} and produce 
the executables {\tt demo\_py.exe} or {\tt demo\_hw.exe} depending on the selected option.
\item
To execute the programs type {\tt run demo\_py} or {\tt run demo\_hw}.  The
scritps will change directory to {\tt prod} and create respective links to data
directories there. The execution will also be performed there. All input files
should be accessible/routed from directory {\tt prod}, the output files will
also be produced in that directory.
\end{itemize}

\boldmath
\section{Outlook and conclusions} 
\unboldmath

In this paper we presented the {\bf AcerMC} Monte Carlo Event Generator, based on the library
of the matrix-element-based generators and interfaces to the universal event generators
{\tt PYTHIA~6.2} and {\tt HERWIG 6.3}. The interfaces are based on the standard proposed
in \cite{Boo01}. 

The presented library fulfills the following goals:
\begin{itemize}
\item
It gives a possibility to generate the few Standard Model background processes which
were recognised as very dangerous for the searches for the {\it New Physics} at LHC,
and generation of which was either unavailable or not straightforward so far.
\item
Although the hard process event is generated with matrix-element-based generator, the
provided interface allows to complete event generation with initial and final
state radiation, multiple interaction, hadronisation, fragmentation and decays,
using implementation of either {\tt PYTHIA~6.2} or {\tt HERWIG 6.3}.
\item
These interfaces can be also used for studying systematic differences between 
 {\tt PYTHIA~6.2} or {\tt HERWIG~6.3} predictions for the underlying QCD processes.
\end{itemize}

The complete list of the native {\bf AcerMC} processes implemented so far is:
$gg, q \bar q \to t \bar t b \bar b$, $q \bar q \to W(\to \ell \nu) b \bar b$,
$gg, q \bar q \to Z/\gamma^*(\to \ell \ell) b \bar b$, $q \bar q \to W(\to \ell
\nu) t \bar t$, $gg, q \bar q \to Z/\gamma^*(\to \ell \ell, \nu \nu, b \bar b) t
\bar t$ and the complete EW $g g \to (Z/W/\gamma^* \to)t \bar t b \bar b$.  We
plan to extend this not too exhaustive, but very much demanded list of
processes, in the near future.

Several improvements of the existing Monte Carlo algorithms/programs have been developed
in the process of this work. Let us make short list of the most interesting
ones: (1)The colour flow information has been obtained after some modification of
{\tt MADGRAPH} package; (2) The power of the multi-channel optimisation was
enhanced by using the modified {\tt ac-VEGAS} package.  We believe that the
modification in the {\tt VEGAS} code represents a very powerful extension of
this package; (3) The additions and extensions to the available (multi-channel)
phase space algorithms (e.g. Breit-Wigner function with s-dependent width) lead
to substantial improvement of the unweighting efficiency;
Figs.~\ref{f:bwcomp},~\ref{f:tauveg} and~\ref{f:weights} illustrate the
improvements achieved in the generation efficiency.

Having all these different production processes implemented in the consistent
framework, which can be also directly used for generating standard processes
available in either {\tt PYTHIA~6.2} or {\tt HERWIG 6.3} Monte Carlo, represents
very convenient environment for several phenomenological studies dedicated to
the LHC physics.  Such frame was not available to our knowledge so far. We hope
that it can serve as an interesting example or even a framework.  This way some
tools for discussing the ambiguities due to QCD effects are collected, however
the necessary discussion for the appropriate uncertainties is still not
exhausted.  Nevertheless some discussions using this tool can be already found
in \cite{ATLCOMP013}, \cite{ATLCOMP014}, \cite{ATLCOMP025},
\cite{ATLCOMP032}.

\begin{ack}

We would like to thank Daniel Froidevaux, Torbjorn Sjostrand and 
Alessandro Ballestrero for several very valuable discussions. We would like
also to thank all our colleagues from ATLAS Collaboration who were the first 
and very enthusiastic users of the preliminary versions of this package.
We both very warmly acknowledge the support from the CERN EP division.
\end{ack}


\appendix


\boldmath
\section{Feynman Diagrams}
\unboldmath

The 38+7 Feynman diagrams contributing to the 
$g g,q \bar{q} \to t \bar t  b \bar b$ production. Only four flavours are included 
for incoming quarks. Contribution of the incoming b-quarks could be excluded 
from the calculations thanks to very high suppression induced by either 
the parton density functions and/or CKM matrix elements.

\begin{figure}[hb]
\begin{center}
\hspace{-2cm}
\mbox{
     \epsfxsize=6.5cm
     \epsffile{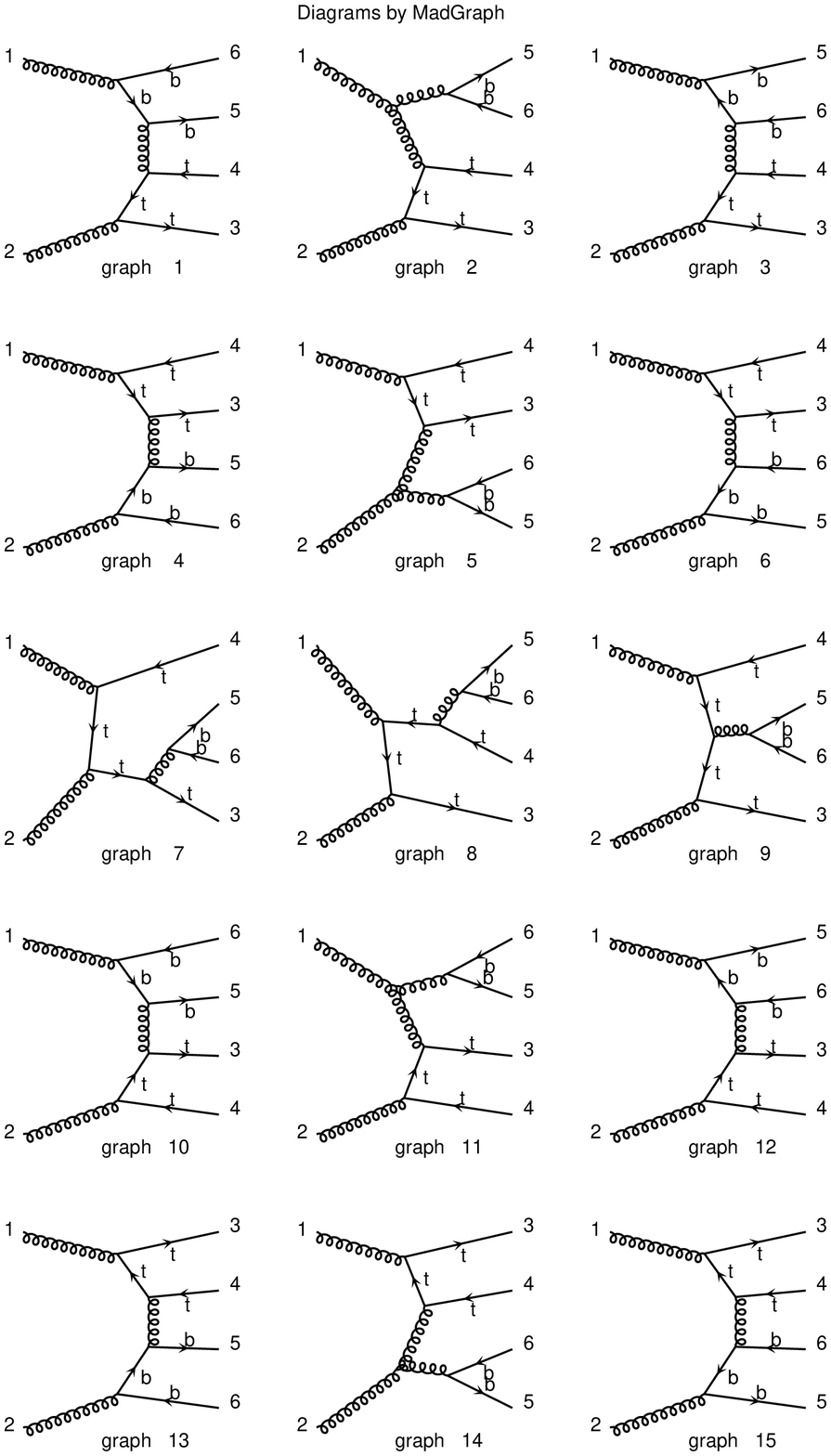}
}
\mbox{
     \epsfxsize=6.5cm
     \epsffile{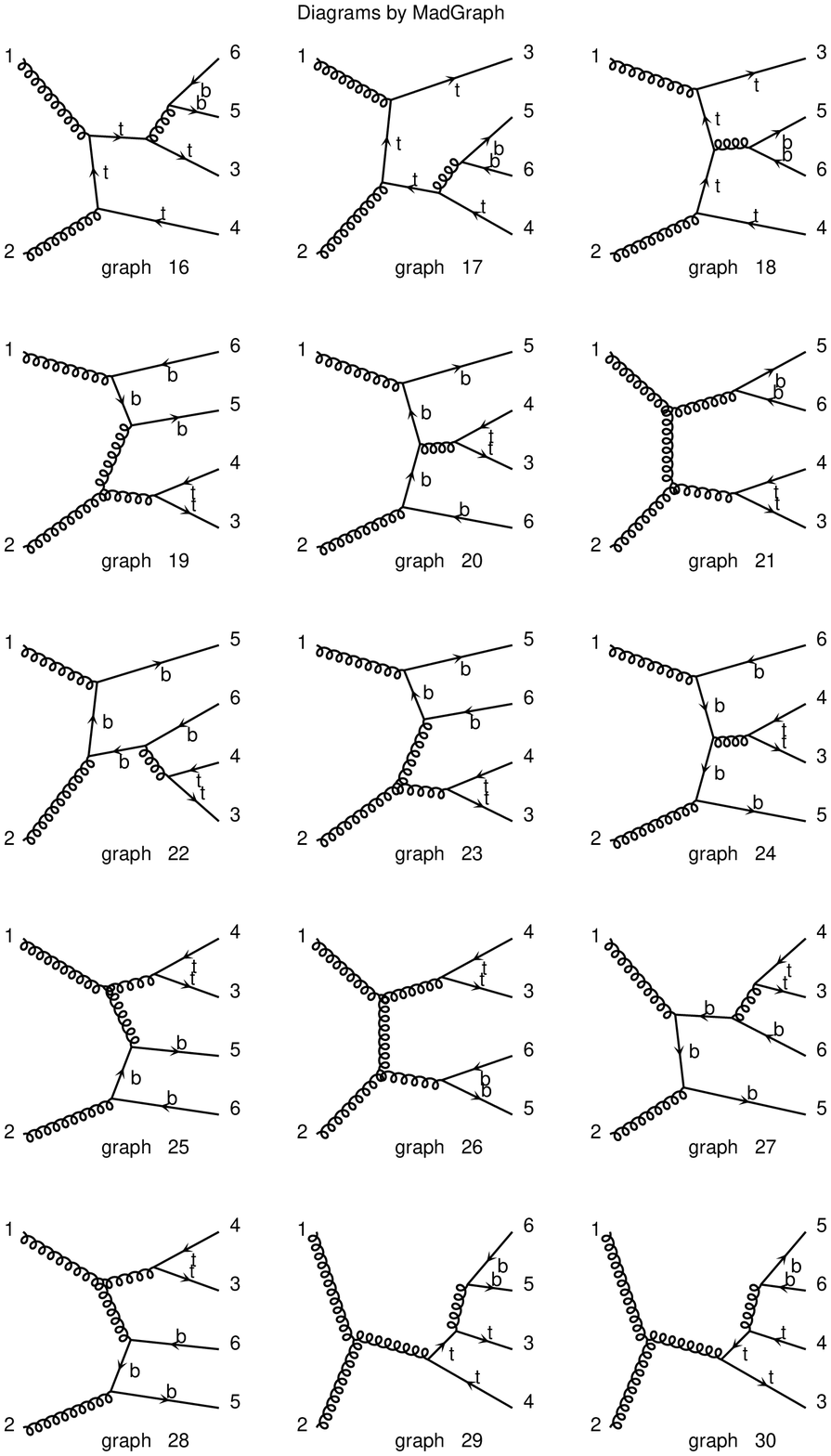}
}
\end{center}
\begin{center}
\hspace{-2cm}
\mbox{
     \epsfxsize=6.5cm
     \epsffile{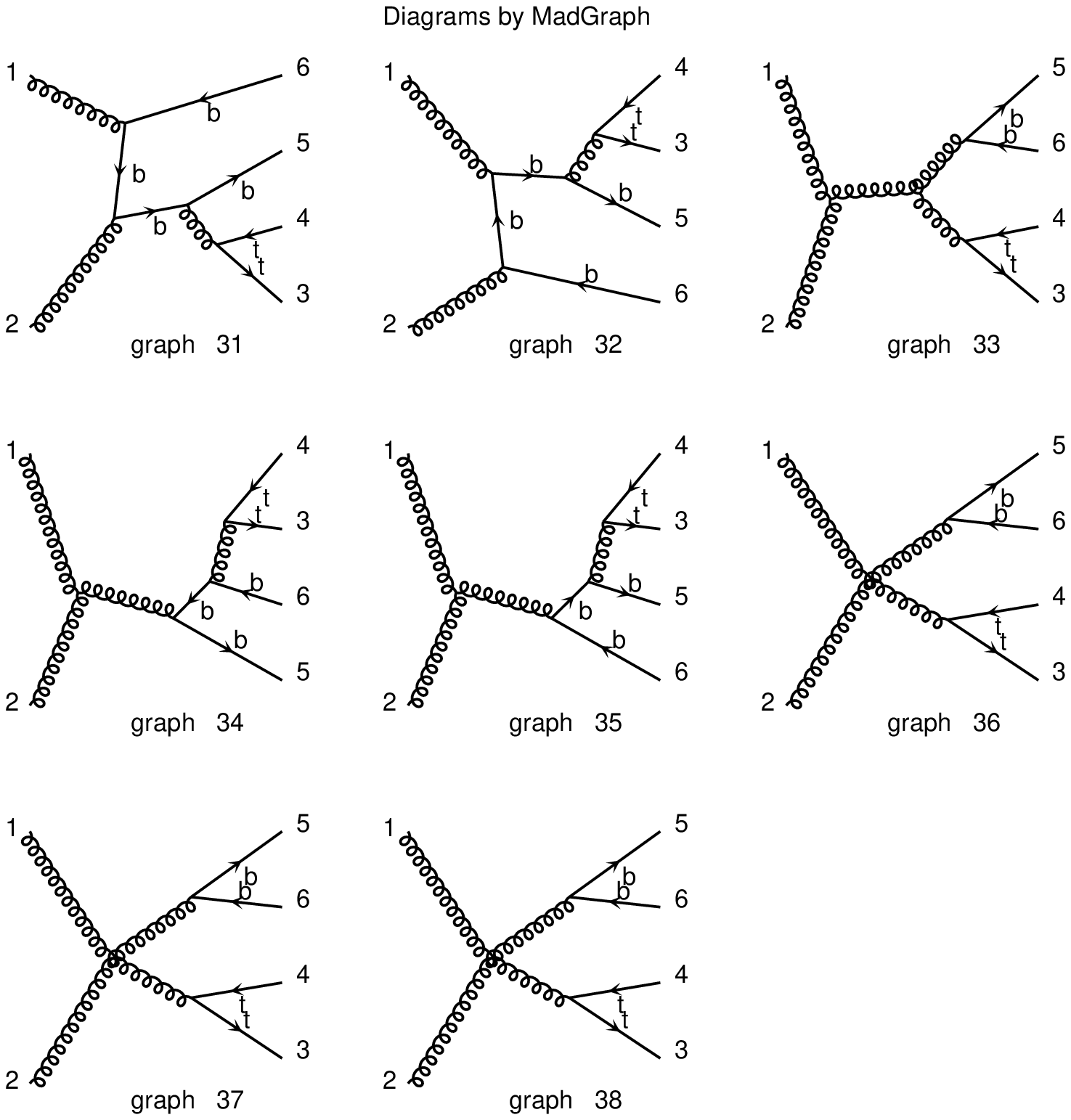}
}
\mbox{
     \epsfxsize=6.5cm
     \epsffile{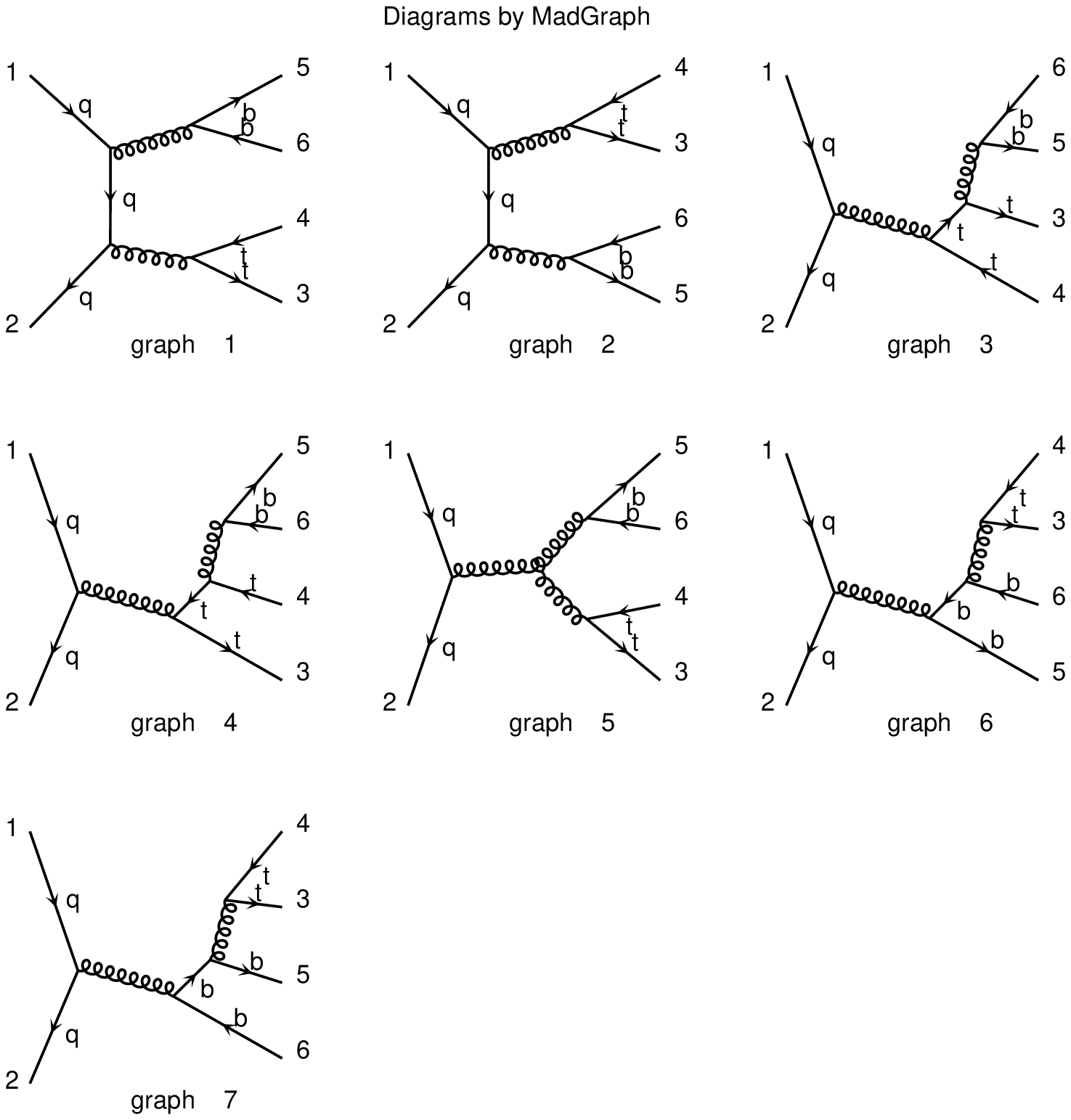}
}
\end{center}

\vglue -0.5cm
\caption{\em
The Feynman diagrams for the processes
$gg,q \bar{q} \to t \bar t b \bar b$.
\label{f:ttbbdia}}
\end{figure}

\begin{figure}[hb]
\vspace{1cm}
\begin{center}
\mbox{
     \epsfxsize=12.5cm
     \epsffile{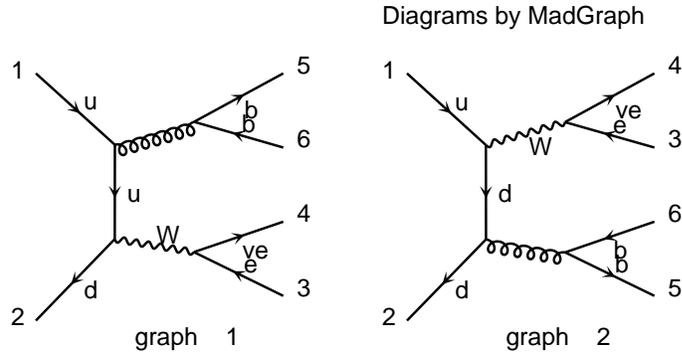}
}
\end{center}
\caption{\em
The Feynman diagrams for the process
 $q \bar q \to W b \bar b \to e^+ \nu_e\; b \bar b $. The same set is used for
$q \bar q \to W b \bar b \to e^+ \nu_e\; t \bar t $ process, with b-quarks 
replaced by top-quarks. 
\label{PS:1}}
\end{figure}

\newpage
The Feynman diagrams contributing to the $q \bar q \to W (\to \ell \nu) b \bar b$ and
 $q \bar q \to W (\to \ell \nu) t \bar t$ matrix
element are just two t-channel diagrams with fermion exchange and double conversion into
an off-shell W boson and a virtual gluon; the W boson subsequently decays leptonically
into $\ell \nu$ and the gluon splits into a $b \bar b$ pair or $t \bar t$ pair
 respectively.
(c.f. Figure \ref{PS:1}).

\begin{figure}[hb]
\vspace{1cm}
\begin{center}
\hspace{-2cm}
\mbox{
     \epsfxsize=6.5cm
     \epsffile{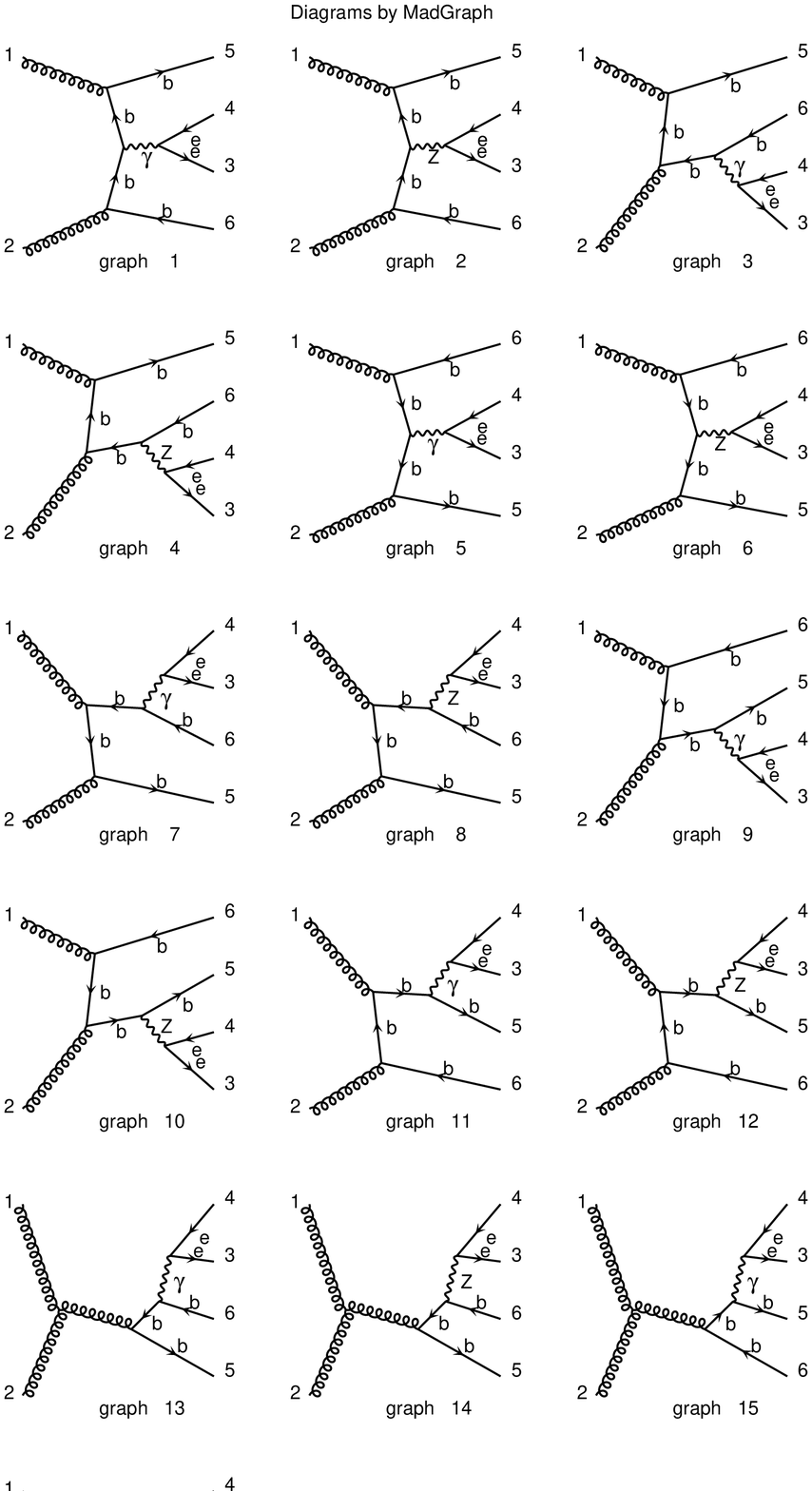}
}
\mbox{
     \epsfxsize=6.5cm
     \epsffile{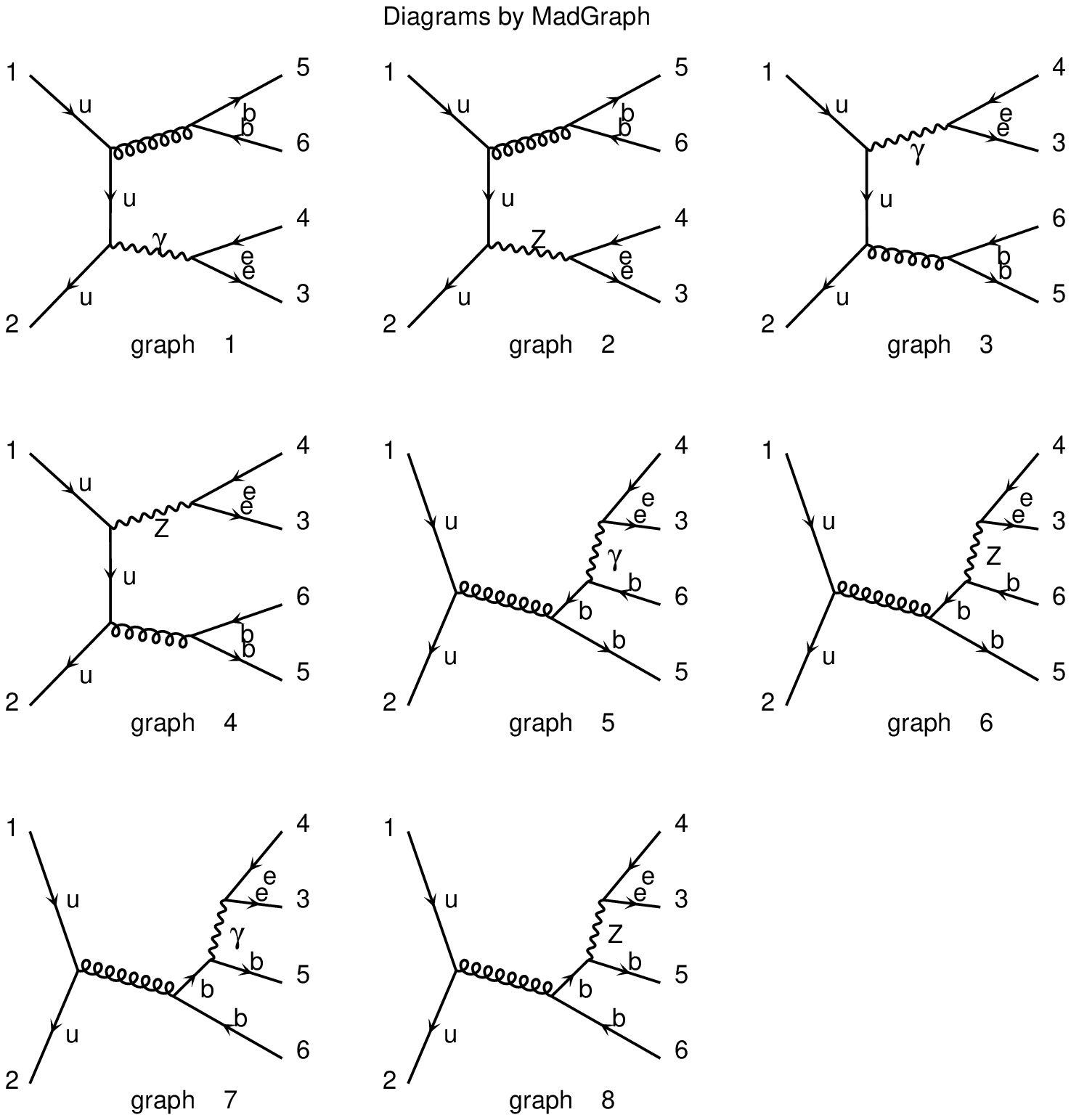}
}
\end{center}
\vspace{1cm}
\caption{\em
The Feynman diagrams for the processes
 $gg, q \bar q \to Z/\gamma^* b \bar b \to e e b \bar b $. 
\label{FA3:1}}
\end{figure}

\newpage
The Feynman diagrams contributing to the
 $g g, q \bar q \to Z/\gamma^*(f \bar f, \nu \nu) b \bar b$
production are shown in Figure ~\ref{FA3:1}.  The dominant contribution comes from the (2)
and (6) configurations for the processes with $gg$ initial state and the double conversion
configuration (2),(4) for the ones with $q \bar q$ initial state.
The same set of Feynman diagrams is used for the 
 $g g, q \bar q \to Z/\gamma^*(f \bar f, \nu \nu ) t \bar t$ process, with b-quarks being replaced by the top-quarks. 
If the  $Z/\gamma^*(\to b \bar b)$ decay mode is simulated, 
it represents only subset of the
EW production of $t \bar t b \bar b$ final state.

\begin{figure}[hb]
\hspace{-1cm}
\mbox{
     \epsfxsize=6.45cm
     \epsffile{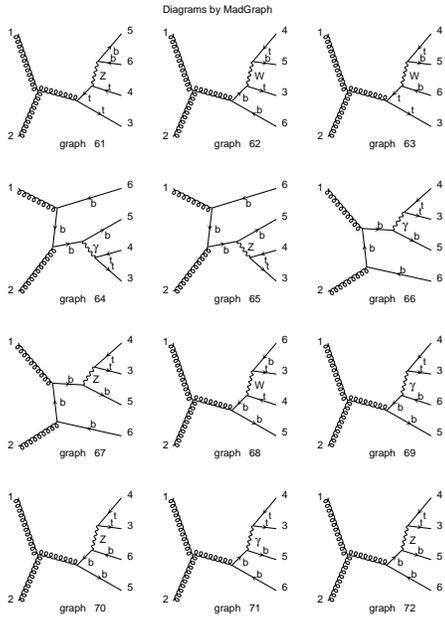}
}
\vspace{-2cm}
\caption{\em
The Feynman diagrams for the processes
 $gg \to (Z/W/\gamma^* \to) b \bar b t \bar t $. 
\label{FA3:2}}
\end{figure}

\newpage
The complete set of th the Feynman diagrams contributing to the full electro-weak
 $gg \to (Z/W/\gamma^* \to) b \bar b t \bar t$ production mediated by exchange of the
$Z/W/\gamma^*$ bosons is shown in Fig.~\ref{FA3:2}.
\begin{figure*}[hb]
\vspace{-0.2cm}
\begin{center}
\hspace{-2cm}
\mbox{
     \epsfxsize=6.45cm
     \epsffile{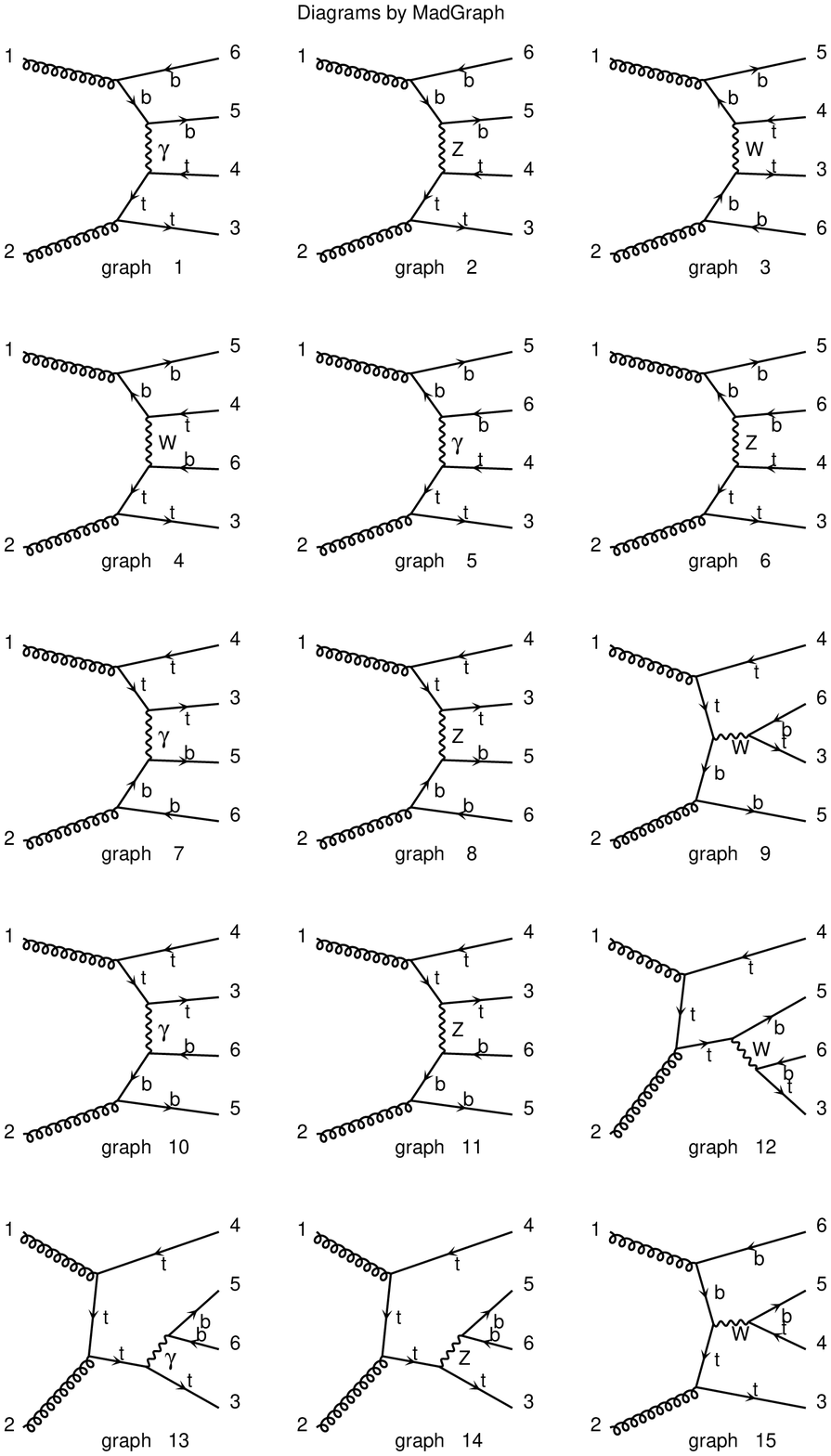}
}
\mbox{
     \epsfxsize=6.45cm
     \epsffile{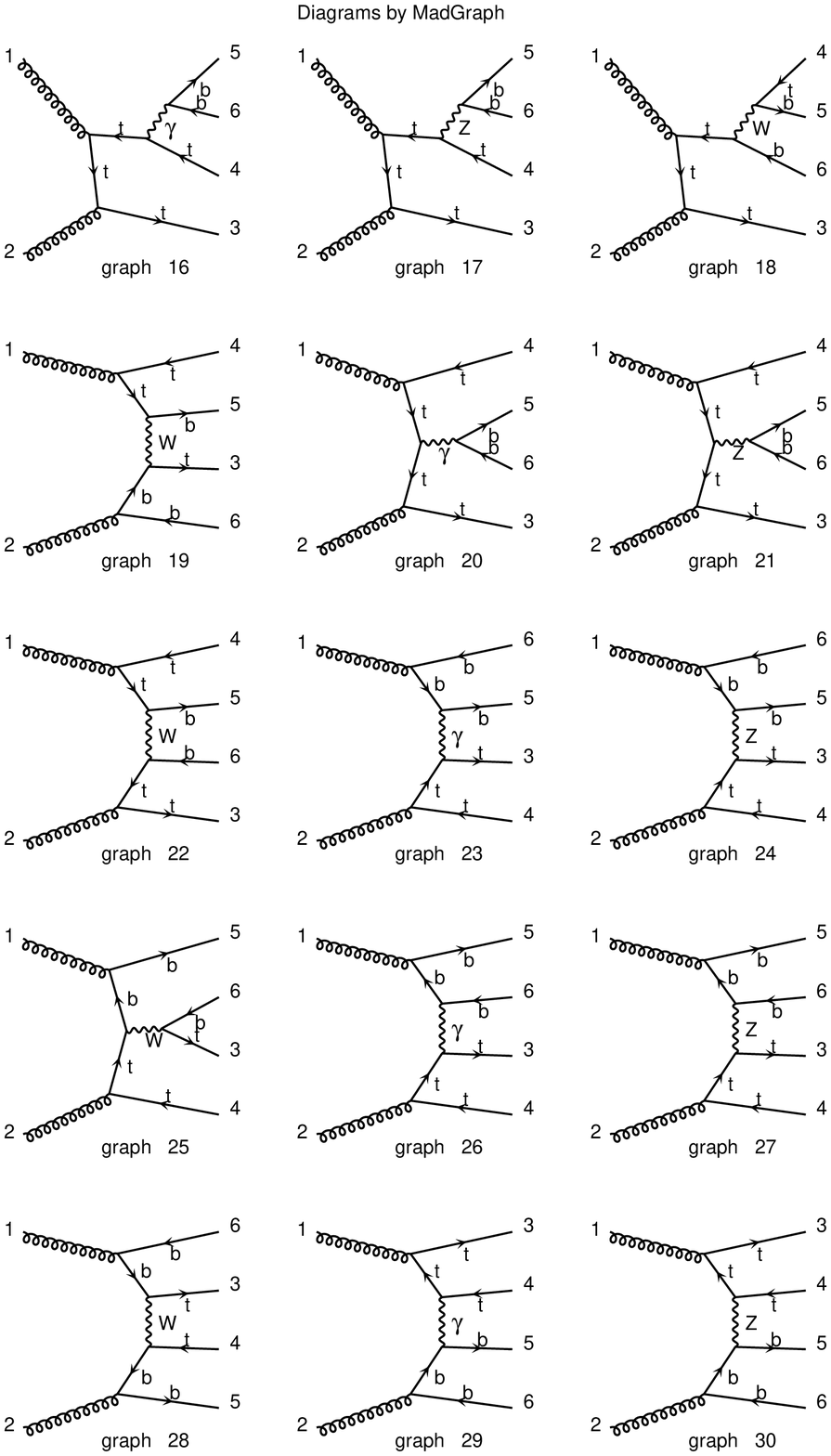}
}
\end{center}
\vspace{-0.6cm}
\begin{center}
\hspace{-2cm}
\mbox{
     \epsfxsize=6.45cm
     \epsffile{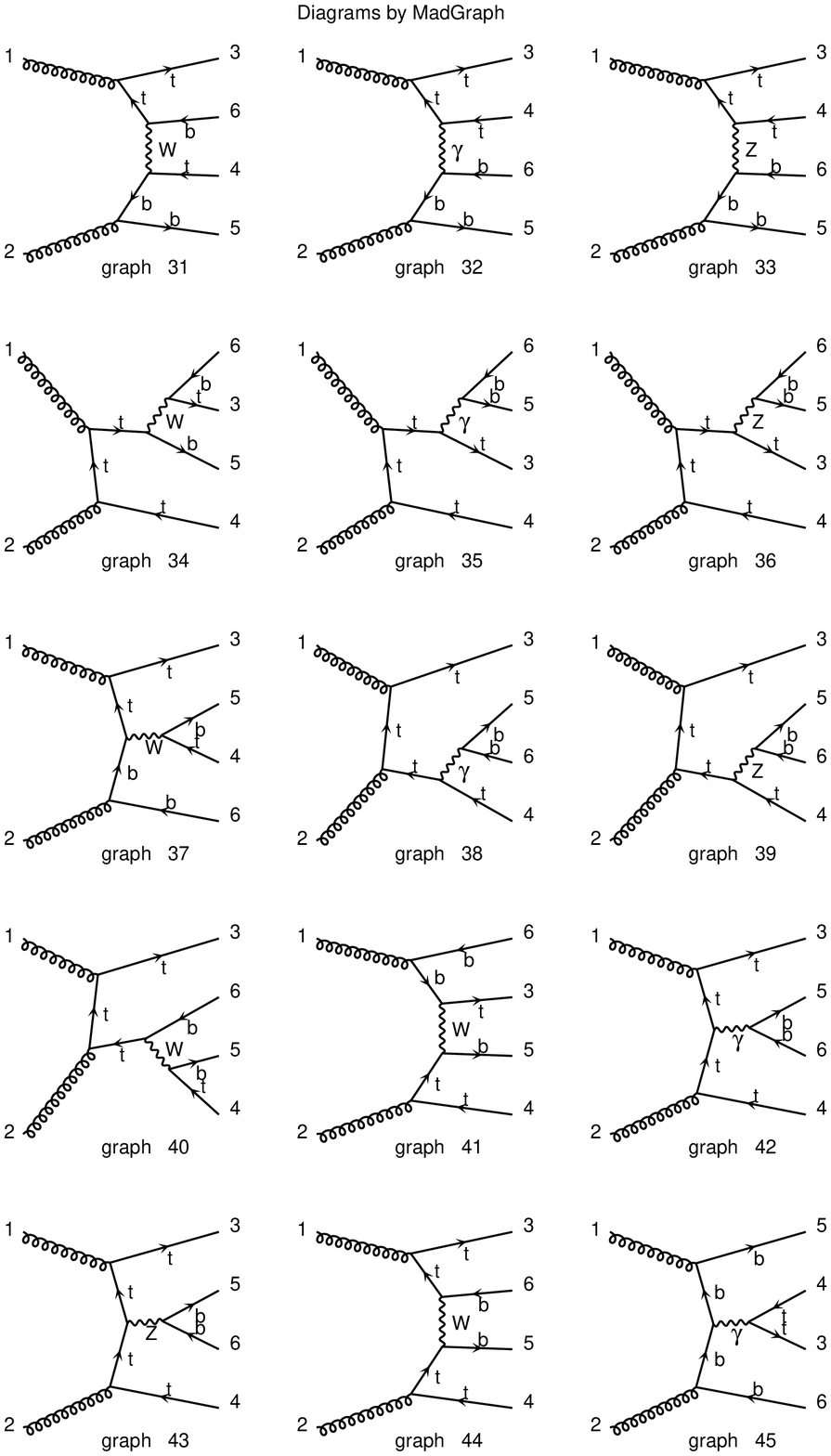}
}
\mbox{
     \epsfxsize=6.45cm
     \epsffile{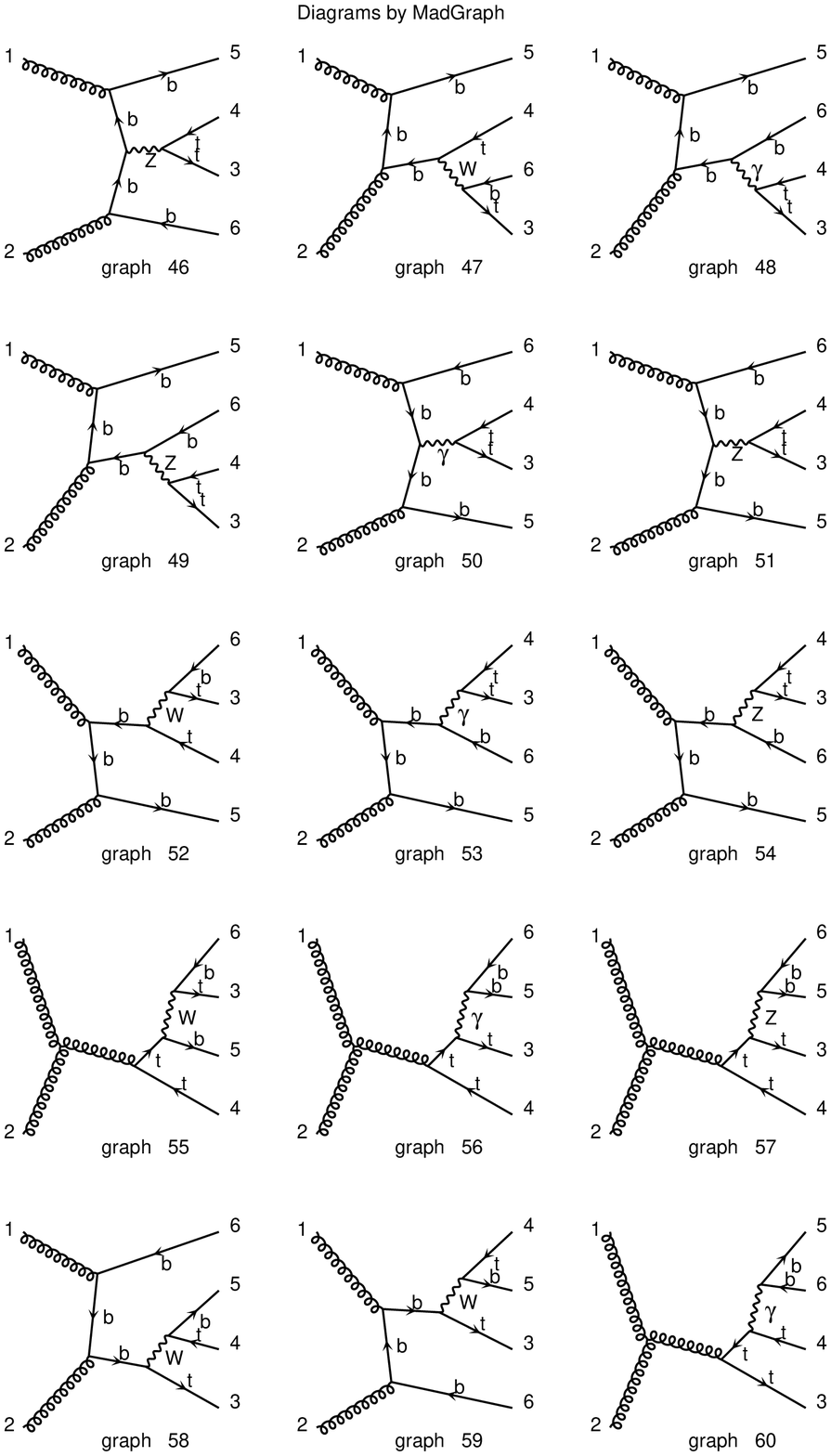}
}
\end{center}
\end{figure*}

\boldmath
\section{Example input files}
\unboldmath

\boldmath
\subsection{File run.card \label{app:run}}
\unboldmath

{\scriptsize
\begin{verbatim}

C------------------------------------------------------------------------------
C                 STEERING FILE FOR ACERMC (1.0BETA) - BASIC SETTINGS
C------------------------------------------------------------------------------

C==== TURN ON FFKEY STEERING FILE (DEBUG)
LIST 

C==== CMS/ACSET(1)
C Specify the centre-of-mass energy in GeV 

CMS 14000.0
C====

C==== ACER
C Use AcerMC code
C ACER=1 - yes
C ACER=0 - no

ACER 1
C====

C==== PROCESS/IACSET ARRAY
C Specify the process to generate. The available AcerMC processes are:
C
C  1)  g + g  ->  t t~ b b~ (MG)    
C  2)  q + q~ ->  t t~ b b~ (MG)    
C  3)  q + q~ ->  (W->) l nu_l b b~ (MG)
C  4)  q + q~ ->  (W->) l nu_l t t~ (MG)
C  5)  g + g  ->  (Z0->) l l~ b b~ (MG) 
C  6)  q + q~ ->  (Z0->) l l~ b b~ (MG) 
C  7)  g + g  ->  (Z0->) f f~ t t~ (MG) 
C  8)  q + q~ ->  (Z0->) f f~ t t~ (MG) 
C  9)  g + g  ->  (Z0/W/gamma->)t t~ b b~ (MG)    
C
C In case ACER=0 the native Pythia/Herwig conventions should be used 

PROCESS 9
C====

C==== HAD
C Control of hadronization/fragmentation/ISR/FSR switches:
C HAD=0 - switch off radiation in initial and final state, multinteraction and
C        hadronization
C HAD=1 - switch off radiation in final state and hadronization
C HAD=2 - switch off hadronization
C HAD=3 - full treatment 
C HAD=4 - switch off radiation in initial state, multiinteraction and hadronization 
C HAD=5 - switch off radiation in initial state and multiinteraction

HAD 3
C====

C==== PDF-GROUP/ACSET(6)
C Choose a PDF group according to PDFLIB804 naming scheme

PDFGROUP 4
C====

C==== PDF-SET/ACSET(7)
C Choose a PDF set according to PDFLIB804 naming scheme
C PDFSET=55 in not (yet) in the manual but represents CTEQ5L parametrised set

PDFSET 55
C====

C==== RSEED
C Choose the random seed for random generator initialisation

RSEED 611123
C====

C==== NEVENT/ACSET(51)
C Specify the number of events to generate

NEVENT 100 
C====

C ------------------------------------------------------------------------------
END
\end{verbatim}      

\boldmath
\subsection{File acermc.card \label{app:acmc}}
\unboldmath

{\scriptsize
\begin{verbatim}      
C------------------------------------------------------------------------------
C                 STEERING FILE FOR ACERMC (1.0BETA) - ACERMC SETTINGS
C------------------------------------------------------------------------------

C==== TURN ON FFKEY STEERING FILE (DEBUG)
LIST 

C THE AcerMC EVENT SETTINGS ---------------------------------------------------

C==== SCALE/ACSET(2)
C Choose the Q^2 scale for the active AcerMC process.
C The implemented values differ for various processes, please look into the manual
C for details

ACSET2 4
C====

C==== FERMION/ACSET(3)
C The flavour of the final state leptons produced in W or Z decays of AcerMC
C processes 3-6.The Pythia/PDG naming convention is used:
C FERMION=11 - electron
C FERMION=13 - muon
C FERMION=15 - tau
C FERMION=12,14,16 - neutrinos, a single flavour is generated and the cross-section is
C                   calculated accordingly. For a neutrino final state the user should
C                   generate all three flavours!
C FERMION=5 - b-quark final state
C The last two settings  ACSET(3)=12,14,16,5 work only for processes 7 and 8!

ACSET3 5
C====

C==== Z/GAMMA/ACSET(4)
C Use the full Z/gamma* propagator in AcerMC processes 5-6. 
C ZGAMMA=0 - only Z propagator
C ZGAMMA=1 - full Z/gamma* propagator

ACSET4 1
C====

C==== Z/G CUT/ACSET(5)
C Cutoff value on the invariant mass m_Z/gamma* in GeV when ZGAMMA=1. 
C Note that the provided data files exist only for values of
C ZGCUT=2,5,10,15,30,60,120,270,300,500 GeV which should satisfy most 
C users. In case a different value is set the user has also to provide 
C the user data files for the run.

ACSET5 10.0
C====

C THE AcerMC ADVANCED SWITCHES ------------------------------------------------

C==== ALPHA_S/ACSET(8)
C Use the alpha_s provided by the linked generator (Pythia/Herwig) or the
C one provided by AcerMC
C ALPHAS=0 - use the linked generator's alpha_s
C ALPHAS=1 - use the AcerMC's alpha_s (one loop calculation)
C ALPHAS=2 - use the AcerMC's alpha_s (three loop calculation)
ACSET8 0
C====

C==== LAMBDA_S/ACSET(9)
C Specify the value of lambda^(nf=5)_MSbar for AcerMC alpha_s calculation
C LAMBDAS=-1 - the value is taken from PDFLIB804 for the corresponding 
C parton density function set
C LAMBDAS>0 - the provided value is taken
ACSET9 -1.
C====

C==== ALPHA_EM/ACSET(10)
C Use the alpha_QED provided by the linked generator (Pythia/Herwig) or the
C one provided by AcerMC
C ALPHAEM=0 - use the linked generator's alpha_QED
C ALPHAEM=1 - use the AcerMC's alpha_QED
ACSET10 0
C====

C==== ALPHA_EM(0)/ACSET(11)
C Specify the value of alpha_QED(0) for AcerMC alpha_QED calculation
C ALPHAEM0=-1 - the default AcerMC value is used 
C ALPHAEM0>0 - the provided value is taken

ACSET11 -1.
C====

C==== TOP S-L/ACSET(12)
C Specify the decay mode of WW pair produced by top decays in AcerMC
C processes 1,2,4,7,8 and 9:
C TOPDEC=0 - both W bosons decay according to Pythia/Herwig switches
C TOPDEC=1 - one W decays into electron + nu and the other one hadronically
C TOPDEC=2 - one W decays into muon + nu and the other one hadronically
C TOPDEC=3 - one W decays into tau + nu and the other one hadronically
C TOPDEC=4 - one W decays into el or mu + nu and the other one hadronically
C TOPDEC=5 - one W decays into el or mu + nu and the other one hadronically, the 
C            W decaying leptonically has the same charge as the primary W; the decay
C            mode makes sense only for AcerMC processes 4!
C When TOPDEC>0 the output cross-section is ALREADY MULTIPLIED by the corresponding 
C branching ratio! (Courtesy of AcerMC authors) 

ACSET12 0
C====

C THE AcerMC TRAINING SETUP AND UNWEIGHTING TREATMENT -------------------------

C==== MODE/ACSET(50)
C Specify the AcerMC training mode:
C MODE=0 - normal run, generate unweighted events
C MODE=1 - perform multi-channel optimisation.
C MODE=2 - perform VEGAS grid training.
C MODE=3 - perform VEGAS grid training as MODE=2 but does this by updating a provided grid

ACSET50 0
C====

C==== USER/ACSET(52)
C Use the data files provided by user
C USER=0 - no, use internal files
C USER=1 - use the user's multi-channel optimisation and VEGAS grid files 
C USER=2 - use the default multi-channel optimisation and user's VEGAS grid files 
C USER=3 - use the default multi-channel optimisation and VEGAS grid files, read the user
C           maximal weight file. 

ACSET52 0
C====

C==== MAXFIND/ACSET(53)
C Search for the maximum weight needed for event unweighting
C MAXFIND=0 - no, use the provided file for max. weights
C MAXFIND=1 - use the provided file for max. weights, re-calculate the max. weights using
C             the stored 100 highest events
C MAXFIND=2 - perform the search and give the wtmax file, equivalent to generation of
C             weighted events

ACSET53 1
C====

C==== EPSILON/ACSET(54)
C Use the epsilon maximal weight or the overall maximal weight found in training (see the 
C manual for the difference)
C EPSILON=0 - use the epsilon max. weight
C EPSILON=1 - use the overall maximal weight

ACSET54 0
C====

C ------------------------------------------------------------------------------
END
\end{verbatim}

\newpage
\boldmath
\section{Example output files}
\unboldmath

\boldmath
\subsection{File acermc.out}
\unboldmath

{\scriptsize
\begin{verbatim}      
      
  -----------------------------------------------------------------------------
                                   ._                           
                                 .j%3]:,                        
                               ~!%%%%%%% ,._.                   
                            _|xx%xxxx%%%%%+`                    
                             :~]%xxxx]xx%x_,+_x_%`              
                      -__||x||xx]+]]]+]]]]x|xxx]`               
                       -+%%xxxx]]]]+]]++]+]x]|>- .;..;.:_/`     
                          -+x]]]]|+]+]+]++]++]]+|+|]+|]]+-      
              ,.   .  ., |x]+]+||=++]+]=++++++=|++]=+]|~-       
              -|%x]]];x]]||=++++++++|];|++++++++++=+]=];   ..   ..  :..,; 
                -/]|]+]|]++++++++||||==|:;:=|==++==+;|,   :;;. :;=;;===|` 
                 _|]|+>+++]+]]|+|||=|=;|;;:|;=:===|==;::.;;:;,;:;;;;;;:-  
                     -++]+++:+|x+::||=||:=:::::;;::::::::;:;:;:;:;;;=::   
                   .|x+|+|]++:,-..:::|=;=:-.:.:-::.:.:.::-::::::::;:-     
         .,  .:||_   --||;:|:::.-.:.-:|;||::.:...-.-.-....::::-:::;::.    
     __._;++;;;|=|;. -:::::::---:.--;;|==|:;:.:-:-:.-.-.:::.--:::;:;--    
     -+++=+=======;==:::::-:.::...:.|+=;;===:-..:.:.:::::-...--:  -       
     :|:-:|===;:;:;::::::--::.:-::::.|||===:::.:.:::::.-......--:: .      
        -;|===;;:;;;;::;::.:-:: -:=;;|+||=;==|=::::::::...-.:..-.::..     
         ---|;:,::::.::::=;=:=::  -++===|======:--:   ...-...-:-:.-       
             ---;:::::;:--:===;;;:|-|+=+|+|==;:`    ...-....-..           
             .::;;:::;:;;:--- -- --.|=||=+|=:=,      ..... .              
            .;;=;;:-:::;:::.        -+;,|]| ;;=`                          
              - -:- --:;;:- -         - :|; -                             
                       -                 :`                               
                                          :                               
                                          :                               
                                          .                               
                                                                  
   40000L,                                           |0000i   j000&  .a00000L#0
   --?##0L        .aaaa aa     .aaaa;     aaaa, _aaa, -000A  _0001- _d0!`  -400
     d0 40,     _W0#V9N0#&    d0#V*N#0,   0##0LW0@4#@' 00j#; J0|01  d0'      40
    J0l -#W     #0'    ?#W   ##~    -#0;    j##9       00 4#|01|01  00     
   _00yyyW0L   :0f      ^-  :000###00001    j#1        00 ?#0@`|01  #0      
   ##!!!!!#0;  -0A       _  -0A             j#1        00  HH< |01  j0L       _
 ad0La,  aj0Aa  4#Aaa_aj#0`  ?0Laa_aaa0L  aaJ0Laaa,  _a00aa  _aj0La  *0Aaa_aad
 HHHRHl  HHHHH   `9##009!     `9NW00@!!`  HHHHRHHHl  :HHHHH  ?HHHRH   ?!##00P!`
      
  -----------------------------------------------------------------------------
      
        AcerMC 1.0 (February 2002),  B. P. Kersevan, E. Richter-Was
      
  -----------------------------------------------------------------------------
      
  ---------------------------< ACTIVATED PROCESSES >---------------------------
      
                      1)  g + g  ->  t t~ b b~ (MG)               OFF 
                      2)  q + q~ ->  t t~ b b~ (MG)               OFF 
                      3)  q + q~ ->  (W->) l nu_l b b~ (MG)       OFF 
                      4)  q + q~ ->  (W->) l nu_l t t~ (MG)       OFF 
                      5)  g + g  ->  (Z0->) l l~ b b~ (MG)        OFF 
                      6)  q + q~ ->  (Z0->) l l~ b b~ (MG)        OFF 
                      7)  g + g  ->  (Z0->) f f~ t t~ (MG)        OFF 
                      8)  q + q~ ->  (Z0->) f f~ t t~ (MG)        OFF 
                      9)  g + g  -> (Z0/W/gamma->) t t~ b b~ (MG) ON  
      
  -----------------------------< ACERMC SETTINGS >-----------------------------
      
  -----------------------------------------------------------------------------
                      C.M.S ENERGY  =       14000.00       [ACSET(1)]
                      SCALE CHOICE  =              4       [ACSET(2)]
                      ACERMC ALPHA_QCD =           0       [ACSET(8)]
                      LAMBDA(5)_MS     =   -1.000000       [ACSET(9)]
                      ACERMC ALPHA_QED =           0       [ACSET(10)]
                      ALPHA_QED(0)     =   -1.000000       [ACSET(11)]
                      TOP->W S-L DECAY =           0       [ACSET(12)]
       
                      OPTIMIZATION   =             0       [ACSET(50)]
                      OPTIM. STEPS   =           100       [ACSET(51)]
                      USER FILES     =             0       [ACSET(52)]
                      MAX. SEARCH    =             1       [ACSET(53)]
                      EPSILON CUTOFF =             0       [ACSET(54)]
       
  
  
      ------< APPROXIMATE MAXIMUM WEIGHT ESTIMATION >------
  
         NEW MAXIMUM WEIGHT(MB)  =   0.991294E-08
         NEW EPSILON WEIGHT(MB)  =   0.865417E-08

      ------> APPROXIMATE MAXIMUM WEIGHT ESTIMATION <------
  
  
    ---------< FINALIZATION FOR PROCESS: 9 >---------
      
      --------------< WEIGHT SURVEY >--------------
      
      ------------< TOTAL STATISTICS >-------------
  
       CROSS-SECTION ESTIMATE =   0.877415E+00 PB
                            +/-   0.378414E-01 PB
            VARIANCE ESTIMATE =   0.143197E-02 PB^2
                            +/-   0.107049E-03 PB^2
  
       MAXIMUM WEIGHT  =   0.656660E-08
       NO.WEIGHTS NE 0 =           1009
       NO.WEIGHTS EQ 0 =              0
       NO.WEIGHTS LT 0 =              0
       EFFICIENCY FOR ALL WEIGHTS     =  13.362 %
       EFFICIENCY FOR NONZERO WEIGHTS =  13.362 %

       NO.WEIGHTS ABOVE EPSILON-CUT  =           0

      --------------> WEIGHT SURVEY <--------------
\end{verbatim}

\newpage

\boldmath
\subsection{File pythia.out}
\unboldmath

{\scriptsize
\begin{verbatim}
 ******************************************************************************
 ******************************************************************************
 **                                                                          **
 **                                                                          **
 **              *......*                  Welcome to the Lund Monte Carlo!  **
 **         *:::!!:::::::::::*                                               **
 **      *::::::!!::::::::::::::*          PPP  Y   Y TTTTT H   H III   A    **
 **    *::::::::!!::::::::::::::::*        P  P  Y Y    T   H   H  I   A A   **
 **   *:::::::::!!:::::::::::::::::*       PPP    Y     T   HHHHH  I  AAAAA  **
 **   *:::::::::!!:::::::::::::::::*       P      Y     T   H   H  I  A   A  **
 **    *::::::::!!::::::::::::::::*!       P      Y     T   H   H III A   A  **
 **      *::::::!!::::::::::::::* !!                                         **
 **      !! *:::!!:::::::::::*    !!       This is PYTHIA version 6.203      **
 **      !!     !* -><- *         !!       Last date of change: 13 Nov 2001  **
 **      !!     !!                !!                                         **
 **      !!     !!                !!       Now is  0 Jan 2000 at  0:00:00    **
 **      !!                       !!                                         **
 **      !!        lh             !!       Disclaimer: this program comes    **
 **      !!                       !!       without any guarantees. Beware    **
 **      !!                 hh    !!       of errors and use common sense    **
 **      !!    ll                 !!       when interpreting results.        **
 **      !!                       !!                                         **
 **      !!                                Copyright T. Sjostrand (2001)     **
 **                                                                          **
 ** An archive of program versions and documentation is found on the web:    **
 ** http://www.thep.lu.se/~torbjorn/Pythia.html                              **
 **                                                                          **
 ** When you cite this program, currently the official reference is          **
 ** T. Sjostrand, P. Eden, C. Friberg, L. Lonnblad, G. Miu, S. Mrenna and    **
 ** E. Norrbin, Computer Physics Commun. 135 (2001) 238.                     **
 ** The large manual is                                                      **
 ** T. Sjostrand, L. Lonnblad and S. Mrenna, LU TP 01-21 [hep-ph/0108264].   **
 ** Also remember that the program, to a large extent, represents original   **
 ** physics research. Other publications of special relevance to your        **
 ** studies may therefore deserve separate mention.                          **
 **                                                                          **
 ** Main author: Torbjorn Sjostrand; Department of Theoretical Physics 2,    **
 **   Lund University, Solvegatan 14A, S-223 62 Lund, Sweden;                **
 **   phone: + 46 - 46 - 222 48 16; e-mail: torbjorn@thep.lu.se              **
 ** SUSY author: Stephen Mrenna, Physics Department, UC Davis,               **
 **   One Shields Avenue, Davis, CA 95616, USA;                              **
 **   phone: + 1 - 530 - 752 - 2661; e-mail: mrenna@physics.ucdavis.edu      **
 **                                                                          **
 **                                                                          **
 ******************************************************************************
 ******************************************************************************
1****************** PYINIT: initialization of PYTHIA routines *****************

 ==============================================================================
 I                                                                            I
 I         PYTHIA will be initialized for p+ on p+ user configuration         I
 I            with   7000.000 GeV on   7000.000 GeV beam energies             I
 I                                                                            I
 I           corresponding to  14000.000 GeV center-of-mass energy            I
 I                                                                            I
 ==============================================================================

 ******** PYMAXI: summary of differential cross-section maximum search ********

           ==========================================================
           I                                      I                 I
           I  ISUB  Subprocess name               I  Maximum value  I
           I                                      I                 I
           ==========================================================
           I                                      I                 I
           I    4   User process 669              I    8.6542E-09   I
           I                                      I                 I
           ==========================================================

 ********************** PYINIT: initialization completed **********************



                            Event listing (summary)

    I particle/jet KS     KF  orig    p_x      p_y      p_z       E        m

    1 !p+!         21    2212    0    0.000    0.000 7000.000 7000.000    0.938
    2 !p+!         21    2212    0    0.000    0.000-7000.000 7000.000    0.938
 ==============================================================================
    3 !g!          21      21    1    1.144    0.067 1425.499 1425.499    0.000
    4 !g!          21      21    2    1.111    0.247-1179.739 1179.739    0.000
    5 !g!          21      21    3    1.144    0.067 1425.499 1425.499    0.000
    6 !g!          21      21    4    1.111    0.247-1179.739 1179.739    0.000
    7 !t!          21       6    0  -24.631  -28.056   43.153  184.068  175.000
    8 !tbar!       21      -6    0   80.035  -23.611 1084.153 1101.351  175.000
    9 !b!          21       5    0  -30.160   24.308  217.096  220.578    4.800
   10 !bbar!       21      -5    0  -22.988   27.673-1098.642 1099.241    4.800
 ==============================================================================
   11 t         A   2       6    7  -24.631  -28.056   43.153  184.068  175.000
   12 uu_1      V   1    2203    2   -0.215   -0.195-5351.371 5351.371    0.771
   13 tbar      A   2      -6    8   80.035  -23.611 1084.153 1101.351  175.000
   14 u         V   1       2    1   -0.494   -0.216 1493.138 1493.138    0.330
   15 b         A   2       5    9  -30.160   24.308  217.096  220.578    4.800
   16 ud_0      V   1    2101    1   -0.651    0.149 4081.363 4081.363    0.579
   17 bbar      A   2      -5   10  -22.988   27.673-1098.642 1099.241    4.800
   18 d         V   1       1    2   -0.896   -0.052 -468.889  468.890    0.330
 ==============================================================================
                   sum:  2.00          0.00     0.00     0.00 14000.00 14000.00
1********* PYSTAT:  Statistics on Number of Events and Cross-sections *********

 ==============================================================================
 I                                  I                            I            I
 I            Subprocess            I      Number of points      I    Sigma   I
 I                                  I                            I            I
 I----------------------------------I----------------------------I    (mb)    I
 I                                  I                            I            I
 I N:o Type                         I    Generated         Tried I            I
 I                                  I                            I            I
 ==============================================================================
 I                                  I                            I            I
 I   0 All included subprocesses    I          100          1009 I  8.774E-10 I
 I   4 User process 669             I          100          1009 I  8.774E-10 I
 I                                  I                            I            I
 ==============================================================================

 ********* Fraction of events that fail fragmentation cuts =  0.00000 *********

\end{verbatim}
}

\newpage

\boldmath
\subsection{File herwig.out}
\unboldmath

{\scriptsize
\begin{verbatim}



          HERWIG 6.301    9 July 2001 

          Please reference:  G. Marchesini, B.R. Webber,
          G.Abbiendi, I.G.Knowles, M.H.Seymour & L.Stanco
          Computer Physics Communications 67 (1992) 465
                             and
          G.Corcella, I.G.Knowles, G.Marchesini, S.Moretti,
          K.Odagiri, P.Richardson, M.H.Seymour & B.R.Webber,
          JHEP 0101 (2001) 010

          INPUT CONDITIONS FOR THIS RUN

          BEAM 1 (P       ) MOM. =   7000.00
          BEAM 2 (P       ) MOM. =   7000.00
          PROCESS CODE (IPROC)   =   90660
          NUMBER OF FLAVOURS     =    6
          STRUCTURE FUNCTION SET =    8
          AZIM SPIN CORRELATIONS =    T
          AZIM SOFT CORRELATIONS =    T
          QCD LAMBDA (GEV)       =    0.1460
          DOWN     QUARK  MASS   =    0.3200
          UP       QUARK  MASS   =    0.3200
          STRANGE  QUARK  MASS   =    0.5000
          CHARMED  QUARK  MASS   =    1.5500
          BOTTOM   QUARK  MASS   =    4.8000
          TOP      QUARK  MASS   =  175.0000
          GLUON EFFECTIVE MASS   =    0.7500
          EXTRA SHOWER CUTOFF (Q)=    0.4800
          EXTRA SHOWER CUTOFF (G)=    0.1000
          PHOTON SHOWER CUTOFF   =    0.4000
          CLUSTER MASS PARAMETER =    3.3500
          SPACELIKE EVOLN CUTOFF =    2.5000
          INTRINSIC P-TRAN (RMS) =    0.0000
          MIN MTM FRAC FOR ISR   =1.0000E-04
          1-MAX MTM FRAC FOR ISR =1.0000E-06

          NO EVENTS WILL BE WRITTEN TO DISK

          B_d: Delt-M/Gam =0.7000 Delt-Gam/2*Gam =0.0000
          B_s: Delt-M/Gam = 10.00 Delt-Gam/2*Gam =0.2000

          PDFLIB USED FOR BEAM 1: SET455 OF AcerMC              
          PDFLIB USED FOR BEAM 2: SET455 OF AcerMC              


          Checking consistency of particle properties


          Checking consistency of decay tables


          INPUT EVT WEIGHT   =  6.9480E-03
          INPUT MAX WEIGHT   =  6.9480E-03

          SUBROUTINE TIMEL CALLED BUT NOT LINKED.
          DUMMY TIMEL WILL BE USED. DELETE DUMMY
          AND LINK CERNLIB FOR CPU TIME REMAINING.



 EVENT       1: 7000.00 GEV/C P        ON  7000.00 GEV/C P        PROCESS: 90669
 SEEDS:       17673 &       63565   STATUS:   10 ERROR:   0  WEIGHT:  8.2681E-04


                                                   ---INITIAL STATE---    

 IHEP    ID      IDPDG IST MO1 MO2 DA1 DA2   P-X     P-Y     P-Z  ENERGY    MASS     V-X   ..
    1 P           2212 101   0   0   0   0    0.00    0.00 7000.0 7000.0    0.94 0.000E+00 ..
    2 P           2212 102   0   0   0   0    0.00    0.00-7000.0 7000.0    0.94 0.000E+00 ..
    3 CMF            0 103   1   2   0   0    0.00    0.00    0.014000.014000.00 0.000E+00 ..
    4 GLUON         21 111   6  10   0   7    0.00    0.00  378.9  378.9    0.00 0.000E+00 ..
    5 GLUON         21 112   6   8   0   9    0.00    0.00 -223.6  223.6    0.00 0.000E+00 ..
    6 HARD           0 110   4   5   7  10    0.00    0.00  155.3  602.5  582.15 0.000E+00 ..
    7 TQRK           6 113   6   4   0   8   10.38   71.83  -18.5  190.4  175.00 0.000E+00 ..
    8 TBAR          -6 114   6   7   0   5   37.44 -107.58  215.3  300.0  175.00 0.000E+00 ..
    9 BQRK           5 114   6   5   0  10  -37.78   29.50  -67.7   83.1    4.80 0.000E+00 ..
   10 BBAR          -5 114   6   9   0   4  -10.05    6.25   26.2   29.1    4.80 0.000E+00 ..



 EVENT       2: 7000.00 GEV/C P        ON  7000.00 GEV/C P        PROCESS: 90669
 SEEDS:  1689786158 &  1462719137   STATUS:   10 ERROR:   0  WEIGHT:  3.9180E-03


                                                   ---INITIAL STATE---    

 IHEP    ID      IDPDG IST MO1 MO2 DA1 DA2   P-X     P-Y     P-Z  ENERGY    MASS     V-X   ..
    1 P           2212 101   0   0   0   0    0.00    0.00 7000.0 7000.0    0.94 0.000E+00 ..
    2 P           2212 102   0   0   0   0    0.00    0.00-7000.0 7000.0    0.94 0.000E+00 ..
    3 CMF            0 103   1   2   0   0    0.00    0.00    0.014000.014000.00 0.000E+00 ..
    4 GLUON         21 111   6   8   0   9    0.00    0.00  506.0  506.0    0.00 0.000E+00 ..
    5 GLUON         21 112   6  10   0   7    0.00    0.00 -142.1  142.1    0.00 0.000E+00 ..
    6 HARD           0 110   4   5   7  10    0.00    0.00  363.9  648.1  536.23 0.000E+00 ..
    7 TQRK           6 113   6   5   0   8  -77.38    1.14   70.1  203.8  175.00 0.000E+00 ..
    8 TBAR          -6 114   6   7   0   4   54.35  -16.24  326.2  374.5  175.00 0.000E+00 ..
    9 BQRK           5 114   6   4   0  10   -3.39   -5.11   11.9   14.2    4.80 0.000E+00 ..
   10 BBAR          -5 114   6   9   0   5   26.42   20.21  -44.3   55.6    4.80 0.000E+00 ..

          OUTPUT ON ELEMENTARY PROCESS

          N.B. NEGATIVE WEIGHTS NOT ALLOWED

          NUMBER OF EVENTS   =         100
          NUMBER OF WEIGHTS  =        1074
          MEAN VALUE OF WGT  =  6.9823E-04
          RMS SPREAD IN WGT  =  9.3117E-04
          ACTUAL MAX WEIGHT  =  5.1161E-03
          ASSUMED MAX WEIGHT =  6.9480E-03

          PROCESS CODE IPROC =       90669
          CROSS SECTION (PB) =  0.6982    
          ERROR IN C-S  (PB) =  2.8414E-02
          EFFICIENCY PERCENT =   10.05    
\end{verbatim}
}

\end{document}